\documentclass[a4paper,12pt]{article}
\usepackage{epsfig}\parskip 5pt plus 1pt
\usepackage[english]{babel}
\usepackage{amsmath,amsfonts,amssymb,latexsym}
\usepackage{cite}
\usepackage{amsmath}
\usepackage{amssymb}
\usepackage{amsfonts}
\usepackage{mathrsfs}
\usepackage{graphicx}
\usepackage{fancybox}
\usepackage{array}
\usepackage{axodraw}
\usepackage{multicol}
\usepackage{lscape}
\usepackage{pifont}
\usepackage{color}
\newcommand{\be}{\begin{equation}}
\newcommand{\ee}{\end{equation}}
\newcommand{\bea}{\begin{eqnarray}}
\newcommand{\eea}{\end{eqnarray}}
\newcommand{\onbb}{$0\nu\beta\beta$ }
\newcommand{\meff}{\langle m_{\nu}\rangle}

\newcommand {\red} {\color[rgb]{1,0,0}}

\allowdisplaybreaks[4]

\def\gsim{\raise0.3ex\hbox{$\;>$\kern-0.75em\raise-1.1ex\hbox{$\sim\;$}}}
\def\lsim{\raise0.3ex\hbox{$\;<$\kern-0.75em\raise-1.1ex\hbox{$\sim\;$}}}

\begin{document}

\begin{titlepage}
\vspace{-1cm}
\begin{flushright}
\small
IFIC/12-079,
MPP-2012-134,
STUPP-12-212
\end{flushright}
\vspace{0.2cm}
\begin{center}
{\Large \bf Systematic decomposition of the \vspace{0.2cm}\\ 
            neutrinoless double beta decay operator}
\vspace*{0.2cm}
\end{center}
\vskip0.2cm

\begin{center}
{\bf  Florian~Bonnet$^{a,}$\footnote{florian.bonnet@physik.uni-wuerzburg.de},  
Martin~Hirsch$^{b,}$\footnote{mahirsch@ific.uv.es}, 
Toshihiko~Ota$^{c,d,}$\footnote{toshi@mppmu.mpg.de}, and 
Walter~Winter$^{a,}$\footnote{winter@physik.uni-wuerzburg.de}}
\end{center}
\vskip 8pt

\begin{center}
 {\it $^{a}$ Institut f\"{u}r Theoretische Physik und Astrophysik, 
      Universit\"{a}t W\"{u}rzburg,\\
      97074 W\"{u}rzburg, Germany
 }\vspace{0.2cm}\\
 {\it $^{b}$ AHEP Group, Instituto de F\'{\i}sica Corpuscular --
      C.S.I.C./Universitat de Val{\`e}ncia \\
      Edificio de Institutos de Paterna, Apartado 22085,
      46071 Val{\`e}ncia, Spain
 }\vspace{0.2cm}\\
 {\it $^{c}$ Max-Planck-Institut f\"{u}r Physik
 (Werner-Heisenberg-Institut), 
 \\
 F\"{o}hringer Ring 6,
 80805 M\"{u}nchen, Germany
 }\vspace{0.2cm}\\
 {\it $^{d}$ Department of Physics, 
 Saitama University, \\
 Shimo-Okubo 255, 338-8570 Saitama-Sakura, Japan
 }
\end{center}

\vspace*{0.3cm}

\vglue 0.3truecm

\begin{abstract}
\vskip 3pt \noindent

We discuss the systematic decomposition of the dimension nine
neutrinoless double beta decay operator, focusing on mechanisms with
potentially small contributions to neutrino mass, while being
accessible at the LHC. 
  We first provide a ($d=9$ tree-level) complete list
of diagrams for neutrinoless double beta decay. From this list one can
easily recover all previously discussed contributions to the
neutrinoless double beta decay process, such as the celebrated mass
mechanism or ``exotics'', such as contributions from left-right symmetric 
models, R-parity violating supersymmetry and leptoquarks.  More interestingly,
however, we identify a number of new possibilities which have not been
discussed in the literature previously.  
Contact to earlier works based on a general Lorentz-invariant
parametrisation of the neutrinoless double beta decay rate is made,
which allows, in principle, to derive limits on all possible
contributions. We furthermore discuss possible signals at the LHC 
for mediators leading to the short-range part of the amplitude 
with one specific example. 
The study of such contributions would gain 
particular importance if there were a tension between different
measurements of neutrino mass such as coming from
neutrinoless double beta decay and cosmology or single
beta decay.

\end{abstract}
\end{titlepage}

\setcounter{footnote}{0}

\section{Introduction}
\label{sect:Intro}

Neutrinoless double beta (\onbb) decay is mostly known as a sensitive
probe for Majorana neutrino masses \cite{Avignone:2007fu,GomezCadenas:2011it,%
Rodejohann:2012xd,Barabash:1209.4241}. However, the mass mechanism is
only one out of many possible contributions to the \onbb decay
amplitude \cite{Rodejohann:2011mu,Deppisch:2012nb}.
The aim of the current paper is to provide a (tree-level) complete
list of all possible contributions to the neutrinoless double beta
decay dimension nine ($d=9$) operator: 
\begin{equation}
\mathcal{O} \propto \bar u \bar u  \, d d \, \bar e \bar e \, 
\label{equ:effop}
\end{equation}
From this list one can easily recover all known contributions to \onbb
decay.  More interestingly, however, we will identify a number of
new possibilities to generate \onbb decay not discussed in the
literature previously.

\begin{figure}[t]
\centering
\begin{picture}(360,140)
\ArrowLine(10,30)(70,30)
\ArrowLine(70,30)(130,30)
\Photon(70,30)(70,90){2}{5}
\ArrowLine(70,90)(130,110)
\ArrowLine(130,70)(70,90)
\ArrowLine(230,70)(290,90)
\ArrowLine(290,90)(230,110)
\Photon(290,90)(290,30){2}{5}
\ArrowLine(290,30)(230,30)
\ArrowLine(350,30)(290,30)
\EBox(130,10)(230,130)
\Text(40,25)[t]{$\nu$}
\Text(100,28)[t]{$e^-$}
\Text(65,60)[r]{$W^-$}
\Text(100,72)[t]{$d$}
\Text(100,108)[b]{$u$}
\Text(320,25)[t]{$\nu$}
\Text(260,28)[t]{$e^-$}
\Text(295,60)[l]{$W^-$}
\Text(260,72)[t]{$d$}
\Text(260,108)[b]{$u$}
\Text(180,58)[b]{\Huge $0\nu\beta\beta$}
\end{picture}
\caption{\it \label{Fig:BlackBox} Black Box diagram relating the
Majorana nature of neutrinos with \onbb decay.}
\end{figure}
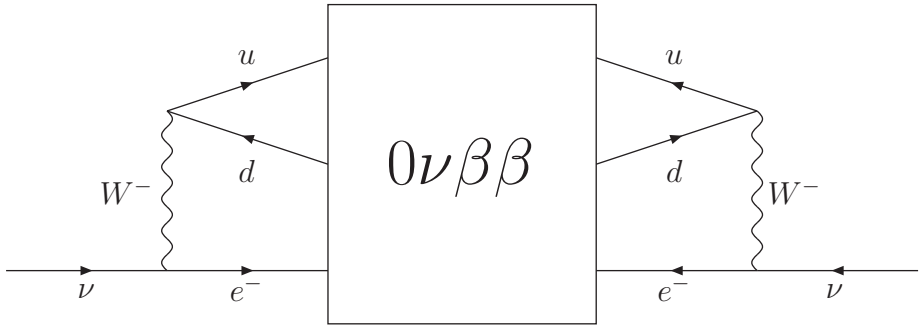

The Black Box theorem
\cite{Schechter:1981bd,Nieves:1984sn,Takasugi:1984xr} states that
since observation of \onbb indicates that lepton number is not conserved,
it proves that neutrinos must be Majorana particles.\footnote{For a
recent version of the black box theorem including lepton flavour
violation, see \cite{Hirsch:2006yk}.} Graphically the theorem can be
depicted as shown in Fig.~\ref{Fig:BlackBox}: If \onbb decay is
observed, Majorana neutrino masses are generated at least at the
4-loop order, which is a model-independent statement.  This does not 
mean that the \onbb decay contribution from Fig.~\ref{Fig:BlackBox} 
is the leading contribution to neutrino mass;
in fact, in specific models, neutrino mass is often generated at a lower
loop order. A recent
calculation \cite{Duerr:2011zd} indeed confirms that the neutrino mass generated
through this 4-loop diagram with the \onbb operator of the size of 
the current limits is roughly $m_{\nu} \simeq {\cal O}(10^{-24})$ eV. 
Obviously, this number is
too small to explain the neutrino masses observed in oscillation
experiments, and also many orders of magnitude smaller than the current
sensitivity of \onbb decay via the mass mechanism.  
 The correct
interpretation of the black box theorem thus is: If \onbb decay is
observed, neutrinos are Majorana particles, whether 
the contribution from neutrino mass dominates \onbb decay or
not.

Up to now, \onbb decay has not been observed. The best half-life
limits on \onbb decay come from experiments on two
isotopes: $^{76}$Ge and $^{136}$Xe. The Heidelberg-Moscow
collaboration gives $T^{0\nu\beta\beta}_{1/2}(^{76}{\rm Ge}) \ge 1.9
\cdot 10^{25}$ yr \cite{KlapdorKleingrothaus:2000sn},\footnote{There
is also a claim \cite{KlapdorKleingrothaus:2006ff} for the observation
of \onbb decay in $^{76}$Ge, which is, however, not supported by any
other experiment.}  while the recent 
results from EXO-200 and KamLAND-ZEN quote 
$T^{0\nu\beta\beta}_{1/2}(^{136}{\rm Xe}) \ge 1.6 \cdot 10^{25}$ yr
\cite{Auger:2012ar} and 
$T^{0\nu\beta\beta}_{1/2}(^{136}{\rm Xe}) \ge 1.9 \cdot 10^{25}$ yr
\cite{Collaboration:2012zm}, both at the  90 \% CL.
There is, however, reasonable hope that the half-lifes in
excess of $10^{26}$ yr will be probed within the next few years, since
a number of next generation \onbb experiments are under
construction or already taking data. For recent reviews and a list of
experimental references, see for example
\cite{Barabash:1209.4241,GomezCadenas:2011it}. Moreover, proposals for
ton-scale next-to-next generation \onbb experiments claim that even
sensitivities in excess $T^{0\nu\beta\beta}_{1/2} \sim 10^{27}$ yr can
be reached for $^{136}$Xe \cite{KamLANDZen:2012aa,MacLallen:2012aa}
and $^{76}$Ge \cite{Abt:2004yk,Guiseppe:2011me}. For a brief summary
of the long-term prospects for \onbb experiments, see for 
example \cite{Barabash:2011fs}.

The interpretation of these half-life limits in terms of particle
physics parameters requires assumptions, such as which contribution
dominates the \onbb decay amplitude. If neutrinos have Majorana
masses, \onbb decay can be mediated by Majorana neutrino propagation,
depending on the magnitude of the effective neutrino mass given by
$\meff=\sum_j U_{ej}^2 m_j$. This mechanism is hence forth referred to as the
{\it the mass mechanism}. The mass mechanism has attracted most of the
attention within the community. The reason for this ``bias'' is rather
straightforward: Neutrinos exist and oscillation experiments
\cite{Fukuda:1998mi,Ahmad:2002jz,Eguchi:2002dm,Abe:2011sj,Adamson:2011qu,Abe:2011fz,An:2012eh,Ahn:2012nd}
have shown that (at least two) neutrinos have non-zero masses. Thus,
if neutrinos are indeed Majorana particles, the mass mechanism is
guaranteed to give a contribution to the \onbb decay amplitude --- in
this sense the mass mechanism is the minimal possibility to generate
\onbb decay.  With the assumption of the mass mechanism 
being the dominant contribution, the current limits for the half-life
of \onbb decay \cite{KlapdorKleingrothaus:2000sn,Auger:2012ar}
correspond to the limits $\meff \lsim (0.2-0.35)$ eV [$\meff \lsim
(0.17-0.30)$ eV] for $^{76}$Ge [$^{136}$Xe] using the latest QRPA
matrix elements of \cite{Faessler:2012ku} to $\meff \lsim 0.53$ eV
[$\meff \lsim 0.34$ eV] with the matrix elements calculated within the
shell model \cite{Menendez:2008jp,Menendez:2011zza}.  Future limits of
order $T^{0\nu\beta\beta}_{1/2} \sim 10^{27}$ yr will then probe
$\meff \lsim (0.02-0.06)$ eV that is of the order of the mass scale
suggested by atmospheric neutrino oscillations, $\sqrt{\Delta m^2_{\rm
atm}} \simeq 0.05$ eV \cite{Tortola:2012te}.
If the next generation of \onbb decay experiments detects a
signal, one might expect that future cosmological data
\cite{Lesgourgues:2006nd,Hannestad:2010kz,Wong:2011ip} also provide
indications for non-zero neutrino masses. Contradicting results from
\onbb (observation) and cosmology (limit) then might point to a
non-standard explanation for \onbb decay, see \cite{Bergstrom:2011dt}
for a detailed discussion of the interplay of different data.  

Contributions to the \onbb decay rate can be divided into a
short-range~\cite{Pas:2000vn} and a long-range~\cite{Pas:1999fc}
part. The long-range contributions, which the neutrino mass mechanism
belongs to, can lead in some cases to very stringent limits on the new
physics scale $\Lambda \gsim$ $\lambda^{\mathrm{eff}}_{\rm
LNV}\times$($10^2 - 10^3$) TeV, where $\lambda^{\mathrm{eff}}_{\rm
LNV}$ is some effective Lepton Number Violating (LNV) coupling that
depends on the model under consideration.  Therefore, for some of the
``exotic'' mechanisms discussed in the literature, falling into this
category, the half-life limits from \onbb decay themselves yield the
most stringent bounds.  On the other hand, in the case of the
short-range contribution, i.e., the \onbb amplitude is mediated only
by heavy mediators with masses at a high energy scale $\Lambda$, the
effective Lagrangian describing \onbb decay is simply proportional to
$1/\Lambda^5$. In that case, the next generation \onbb decay
experiments are sensitive to new physics at scales $\Lambda \gsim$
(few) TeV.

Data from the LHC will probe physics at the TeV scale,
i.e., of a similar scale as the sensitivity of \onbb decay experiments
in case of short-range contributions. In fact, first limits on
particular models have already been published.  Just to mention one
particular example, in Left-Right (LR) symmetric models, \onbb decay
can be generated by $W_R-N-W_R$ exchange, where $N$ is a heavy
Majorana neutrino and $W_{R}$ is the charged gauge boson of the
right-handed $SU(2)$ \cite{Riazuddin:1981hz}.\footnote{ The idea to use accelerator data to test
\onbb decay contributions in left-right symmetric models has a long
history. ``Inverse'' neutrinoless double beta decay, i.e., $e^-e^- \to
W_RW_R$ has been first discussed in \cite{Rizzo:1982kn}. $W_R$
production at hadron colliders, followed by the decay $W_R\to
\mu^+\mu^+ jj$ was first studied in \cite{Keung:1983uu}.} Using the nuclear matrix
elements of \cite{Hirsch:1996qw}, the limit on the half-life
\cite{Auger:2012ar} corresponds to $\langle m_N\rangle = m_{W_R} \gsim
1.3$ TeV (assuming that the gauge coupling of the right-handed and
left-handed $SU(2)$s are equal), while the recent analysis by the ATLAS
\cite{ATLAS:2012ak} and the CMS collaborations
\cite{CMS:PAS-EXO-12-017} give $m_{W_R} \ge (2.3-2.5)$ TeV for $m_N
\lsim 1.3$ TeV.
\footnote{Here,
$(\langle m_N\rangle)^{-1}\equiv\sum_j V_{ej}^2/m_{N_j}$ is a sum
over all heavy neutrinos coupling to the electron, while in the limit
from the LHC experiments it is assumed that only one heavy neutrino
(the lightest) has a mass below the mass of $W_R$.}  Because of this
complementarity of \onbb decay and LHC, we will pay particular
attention to the short-range part of the \onbb decay amplitude.
However, our decomposition of the $d=9$ operator, which will be shown
in Sec.~\ref{Sec:decom}, is general, and the corresponding limits 
for the long-range part can be easily derived as well, using the recipes
described below and in the appendix.

Our paper is not the first work attempting a systematic analysis of
the $d=9$ \onbb decay operator, relevant earlier papers include
\cite{Pas:1999fc,Pas:2000vn,Babu:2001ex,deGouvea:2007xp,delAguila:2012nu}.
The authors of \cite{Pas:1999fc,Pas:2000vn} worked out a general
Lorentz-invariant parametrisation for the \onbb decay rate. This
approach is motivated from the nuclear physics point of view of \onbb
decay. At low energies, adequate for \onbb decay studies, any of the
\onbb diagrams in which heavy mediation fields are inserted among the
six fermions $\bar{u}\bar{u} dd \bar{e} \bar{e}$ will be reduced to a
finite set of combinations of the hadronic and the leptonic
currents corresponding to a basic set of nuclear matrix elements.
This approach leaves the LNV parameters in the \onbb decay rate
unspecified, and our current work can be understood as providing a
(tree-level) complete list of all possible ultraviolet completions
(``models'') for \onbb decay.  As expected, at low energies any
information on the particle models is reduced then to one (or
a combination of more than one) of the coefficients $\epsilon_i$ of
the effective \onbb currents presented in
\cite{Pas:1999fc,Pas:2000vn}.

Our approach also has some overlap with
\cite{Babu:2001ex,deGouvea:2007xp}.  These authors write down all
effective LNV operators from $d=5$ (the famous Weinberg operator
\cite{Weinberg:1979sa}) to $d=11$.  The main motivation of those
papers is to identify all possible Majorana neutrino mass models via
the effective LNV operators~\cite{Babu:2001ex}. This effective
operator treatment allows to estimate the scale $\Lambda$ at which new
physics appears, if these operators give neutrino masses or a \onbb
decay amplitude of the order of the current experimental
sensitivity~\cite{deGouvea:2007xp}.  However, our current work is
complementary to these papers, in that we list all possible {\em
decompositions} of a particular LNV operator, that is the $d=9$ \onbb
operator.  We also go one step further than these works in the
estimation of the bounds on the LNV operator, by making contact with
the nuclear matrix element calculation of
\cite{Pas:1999fc,Pas:2000vn}, instead of simply relying on 
dimensional arguments.\footnote{We also found a $d=9$ LNV operator
$\overline{u_{R}} \overline{u_{R}} d_{R} d_{R} \overline{e_{R}}
\overline{e_{R}}$ that was not listed in the earlier papers
\cite{Babu:2001ex,deGouvea:2007xp}. }
Finally, there is also the recent paper \cite{delAguila:2012nu}, where
the authors study \onbb decay from an effective Lagrangian point of
view.  Operators of $d>9$ are considered, in which the Standard Model
(SM) Higgs doublets are additionally inserted to the $d=9$ operator
Eq.~\eqref{equ:effop}.  Note, however, that \cite{delAguila:2012nu}
considers only the case when new physics is confined to the leptonic
part of the \onbb decay amplitude.

The rest of this paper is organised as follows.  In the next section,
as a preparation for our approach, we will first recapitulate the
characteristics of the long-range and short-range contributions to
\onbb decay. Then, in Sec.~\ref{Sec:decom} 
we will focus on the decomposition of effective
operators to find all possible models generating the $d=9$ \onbb\,
operator. Through an example, we will study the crucial role that the
LHC can play in discriminating such models in Sec.~\ref{sec:example}. 
The relations between the
list of particle models and the general decay rate of
\cite{Pas:1999fc,Pas:2000vn} are given in tabular form in the appendix
for the short-range part for the case of scalar exchange. The
corresponding relations for the other parts of the decay rate can be
easily derived from the recipes spelled out below and in the appendix.

\section{Model-independent parametrisation of the 
$\boldsymbol{0\nu\beta\beta}$ decay rate}

A general Lorentz-invariant parametrisation of new physics
contributions to \onbb has been developed in
\cite{Pas:1999fc,Pas:2000vn}. This formalism allows to derive limits
on any LNV new physics contributing to \onbb decay without recalculation 
of nuclear matrix elements. In order to make contact with
this formalism, we recapitulate the main results and
definitions of \cite{Pas:1999fc,Pas:2000vn} in this section.  
The total amplitude of \onbb is most conveniently divided
into two parts: Long-range and short-range contributions, see
Fig.~\ref{Fig:LongShort}.

\subsection{Long-range contributions}

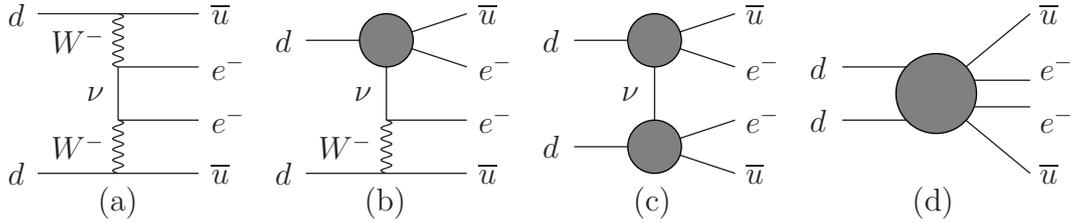
\begin{figure}[t]
\centering
\begin{picture}(400,110)
\Line(10,10)(40,10)
\Line(40,10)(70,10)
\Photon(40,10)(40,30){2}{5}
\Line(40,30)(70,30)
\Line(40,30)(40,50)
\Line(40,50)(70,50)
\Photon(40,50)(40,70){2}{5}
\Line(10,70)(40,70)
\Line(40,70)(70,70)
\Text(5,10)[r]{$d$}
\Text(5,70)[r]{$d$}
\Text(75,70)[l]{$\overline{u}$}
\Text(75,10)[l]{$\overline{u}$}
\Text(75,50)[l]{$e^-$}
\Text(75,30)[l]{$e^-$}
\Text(35,40)[r]{$\nu$}
\Text(35,60)[r]{$W^-$}
\Text(35,20)[r]{$W^-$}
\Text(40,5)[t]{(a)}
\Line(110,10)(140,10)
\Line(140,10)(170,10)
\Photon(140,10)(140,30){2}{5}
\Line(140,30)(170,30)
\Line(140,30)(140,50)
\Line(110,60)(140,60)
\Line(140,60)(170,70)
\Line(140,60)(170,50)
\GCirc(140,60){10}{0.5}
\Text(105,10)[r]{$d$}
\Text(105,60)[r]{$d$}
\Text(175,70)[l]{$\overline{u}$}
\Text(175,10)[l]{$\overline{u}$}
\Text(175,50)[l]{$e^-$}
\Text(175,30)[l]{$e^-$}
\Text(135,40)[r]{$\nu$}
\Text(135,20)[r]{$W^-$}
\Text(140,5)[t]{(b)}
\Line(210,20)(240,20)
\Line(240,20)(270,30)
\Line(240,20)(270,10)
\Line(240,30)(240,50)
\Line(210,60)(240,60)
\Line(240,60)(270,70)
\Line(240,60)(270,50)
\GCirc(240,60){10}{0.5}
\GCirc(240,20){10}{0.5}
\Text(205,20)[r]{$d$}
\Text(205,60)[r]{$d$}
\Text(275,70)[l]{$\overline{u}$}
\Text(275,10)[l]{$\overline{u}$}
\Text(275,50)[l]{$e^-$}
\Text(275,30)[l]{$e^-$}
\Text(235,40)[r]{$\nu$}
\Text(240,5)[t]{(c)}
\Line(310,30)(340,30)
\Line(310,50)(340,50)
\Line(350,45)(380,70)
\Line(350,45)(380,45)
\Line(350,35)(380,35)
\Line(350,35)(380,10)
\GCirc(345,40){15}{0.5}
\Text(305,30)[r]{$d$}
\Text(305,50)[r]{$d$}
\Text(385,70)[l]{$\overline{u}$}
\Text(385,10)[l]{$\overline{u}$}
\Text(385,50)[l]{$e^-$}
\Text(385,30)[l]{$e^-$}
\Text(345,5)[t]{(d)}
\end{picture}
\caption{\it \label{Fig:LongShort} Different contributions to \onbb:
(a)-(c) A light neutrino is exchanged between two point-like vertices, 
which are classified as ``long-range''. (d) Contributions 
mediated by heavy particles are classified as ``short-range''. 
Diagram (a) corresponds to the mass mechanism
--- the standard interpretation of \onbb with Majorana neutrino
 propagation. See main text for details.}
\end{figure}

 Consider first the long-range part.  Here,
we can sub-divide the amplitudes into parts (a)-(c) as shown in the
figure. In case (a), a massive Majorana neutrino is exchanged between 
two SM charged current vertices, while cases (b) and (c) contain 
one and two (unspecified) non-standard interactions respectively, 
indicated by the black blobs. 

At low energy, we can write the relevant part of the effective
Lagrangian with the leptonic ($j$) and hadronic ($J$) 
charged currents as 
\begin{eqnarray}\nonumber
\mathcal{L}^{\text{4-Fermi}}
&=& \mathcal{L^{\rm SM}} + \mathcal{L^{\rm LNV}} \\  
           & =& \frac{G_F}{\sqrt{2}}
           \left[
           j^{\mu}_{V-A} J_{V-A,\mu}
           +
            \hspace{-0.3cm}
           \sum_{\begin{minipage}{1.5cm}
                    {\tiny 
                      $\alpha,\beta \neq V-A$ 
                    }
                  \end{minipage}} 
            \hspace{-0.3cm}
            \epsilon_{\alpha}^{\beta} 
            \hspace{0.1cm}
            j_{\beta}J_{\alpha}\right]\,.\label{eq:defLR}
\end{eqnarray}
Here, we follow the notations of $j$ and $J$ adopted in \cite{Pas:1999fc},
which are\footnote{Note that the difference in normalisation of 
Eq.~(\ref{eq:CurrLR}) and the normal convention for $L$/$R$ in 
particle physics leads to various powers of two, see appendix, 
when relating models with the $\epsilon_{\alpha}^{\beta}$ of 
Eq.~(\ref{eq:defLR}).}
\begin{gather}\label{eq:CurrLR}
J^{\mu}_{V\pm A}
=
(J_{R/L})^{\mu}
\equiv
\overline{u}\gamma^{\mu}(1\pm\gamma_5)d\,,
\qquad
j_{V\pm A}^{\mu}
\equiv
\overline{e}\gamma^{\mu}(1\pm\gamma_5)\nu\,,
\\ \nonumber
J_{S\pm P}
=
J_{R/L}
\equiv
\overline{u}(1\pm\gamma_5)d\,,
\qquad
j_{S\pm P}\equiv\overline{e}(1\pm\gamma_5)\nu\,,
\\ \nonumber
J^{\mu\nu}_{T_{R/L}}
=
(J_{R/L})^{\mu \nu}
\equiv
\overline{u}\gamma^{\mu\nu}(1\pm\gamma_5)d\,,
\qquad
j_{T_{R/L}}^{\mu\nu} \equiv\overline{e}\gamma^{\mu\nu}(1\pm\gamma_5)\nu\ ,
\end{gather}
where the Lorenz tensor matrix $\gamma^{\mu \nu}$ is defined as
$\gamma^{\mu \nu} = \frac{\rm i}{2} [\gamma^{\mu}, \gamma^{\nu}]$.
Recall that $P_{L/R} =\frac{1}{2}(1 \mp \gamma_5)$ and we will use the
short-hand notation $L$ and $R$ for left-handed and right-handed
fermions, respectively.  The first term of Eq.~\eqref{eq:defLR}
is the SM charged current interaction, and the second term contains the new
physics contributions, which do not take the Lorenz structure of the
standard four-Fermi interaction $(V-A)(V-A)$.  The coefficients
$\epsilon_{\alpha}^{\beta}$ for the exotic four-Fermi interactions are
normalised to the SM charged current strength $G_F/\sqrt{2}$. If these
dimensionless coefficients take numbers smaller than one, diagram (c)
in Fig.~\ref{Fig:LongShort} is of order $\epsilon^2$ and becomes
immediately sub-dominant in the \onbb amplitudes.

The neutrino propagator in diagrams (a)-(c) contains two terms:
$m_{\nu} + {q\!\!\!/}$, the mass and the momentum terms. 
Since the charged current for leptons in the SM is purely
left-handed, it picks out the $m_{\nu}$ part, i.e., 
the amplitude of the (standard) mass mechanism of \onbb is proportional 
to $m_{\nu}$.
Clearly, then not all $\epsilon_{\alpha}^{\beta}$ 
can be constrained from \onbb decay, 
due to the absence of a lower bound on $m_{\nu}$. 
On the other hand, if the new physics in diagram (b)
generates a right-chiral lepton interaction 
$(j_{V+A}, j_{S+P}, j_{T_{R}})$, the
${q\!\!\!/}$-term in the neutrino propagator will enter 
the amplitude. The size of the 3-momentum $|{\vec q}|$ can be estimated 
from the typical inter-nucleon distance of two
neutrons in the nucleus to be of the order of 100 MeV. 
Therefore, the amplitudes with ${q\!\!\!/}$-terms
are highly enhanced in comparison with those with the
$m_{\nu}$-term.\footnote{%
This momentum-enhancement mechanism has also been discussed in the context of
the other LNV processes, such as 
$\mu^{-} N \rightarrow e^{+} N$~\cite{Primakoff:1981sx,Doi:1985dx}
and 
$\mu^{-} N \rightarrow \mu^{+} N$~\cite{Missimer:1994xd,Takasugi:2003ah}.
A possibility of a direct experimental test of 
the four-Fermi LNV interaction with a right-chiral lepton current 
is examined in \cite{Kanemura:2012br}. }
For this reason, the coefficients $\epsilon_{\alpha}^{\beta}$ 
with right-chiral leptonic interactions, 
are heavily constrained by \onbb decay.
The hadronic and leptonic currents 
are best defined as currents of definite chirality 
(as defined at Eq.~\eqref{eq:CurrLR}) to take care of this fact.

In Table~\ref{Tab:LimitsLong} we give the updated bounds for all
$\epsilon_{\alpha}^{\beta}$, which are taken from
\cite{Deppisch:2012nb}.  Note that the index $\beta$ for the leptonic
current in the table takes neutrino interactions with the chirality 
$R$ in the exotic four-Fermi interaction, while the hadronic currents 
can be of either $L$- or $R$-type. Note also that, while the
$\epsilon_{\alpha}^{\beta}$ are defined as dimensionless coefficients,
they scale like $\epsilon_{\alpha}^{\beta} \propto
(\lambda^{\mathrm{eff}}_{\rm LNV}/\Lambda)^2$.

\begin{table}[t]
\centering
\begin{tabular}{ccccccc}
\hline
Isotope & $|\epsilon^{V+A}_{V-A}|$ &  $|\epsilon^{V+A}_{V+A}|$ & 
          $|\epsilon^{S+P}_{S-P}|$ &  $|\epsilon^{S+P}_{S+P}|$ & 
          $|\epsilon^{TR}_{TL}|$   &  $|\epsilon^{TR}_{TR}|$   \\
\hline
$^{136}$Xe & $2.8 \cdot 10^{-9}$  & $5.6 \cdot 10^{-7}$   & 
          $6.8 \cdot 10^{-9}$    & $6.8 \cdot 10^{-9}$   & 
          $4.8 \cdot 10^{-10}$   & $8.1\cdot 10^{-10}$    \\
\hline
\end{tabular}
\caption{\it Limits on effective long-range interactions
from $T^{\text{\onbb}}_{1/2}({}^{136}\text{Xe}) \gsim 1.6 \cdot 10^{25}$ ys 
\cite{Auger:2012ar} which corresponds to approximately 
$\langle m_{\nu} \rangle \lsim 0.35 $ eV in the mass mechanism. 
These limits are taken from \cite{Deppisch:2012nb} and 
are derived assuming only one $\epsilon$ is different 
from zero at a time.}
\label{Tab:LimitsLong}
\end{table}

\subsection{Short-range contributions}

The short-range contributions encompass all processes where no light
neutrinos are exchanged, and can be understood as one $d=9$ effective
vertex diagram
as shown in diagram (d) in Fig.~\ref{Fig:LongShort}. 
In this case, one can use the basis of low energy
hadronic currents $J$ as defined in Eq.~(\ref{eq:CurrLR}), 
while for the currents $j$ of two electrons, one defines
\begin{eqnarray}\label{eq:CurrSR}
j_{L/R} &\equiv&\overline{e}(1 \mp \gamma_5)e^c\,,\\ \nonumber
(j_{L/R})^{\mu} &\equiv&\overline{e}\gamma^{\mu}(1 \mp \gamma_5)e^c , 
\end{eqnarray}
to express the effective Lagrangian for short-range \onbb\ as \cite{Pas:2000vn}
\begin{eqnarray}
\mathcal{L}^{\text{eff}}
=\frac{G_F^2}{2}m_{P}^{-1} 
 \left[    \epsilon_1 JJj+\epsilon_2 J^{\mu\nu}J_{\mu\nu}j
           +\epsilon_3 J^{\mu}J_{\mu}j+\epsilon_4 J^{\mu}J_{\mu\nu}j^{\nu}
           +\epsilon_5J^{\mu}Jj_{\mu}
 \right]\, ,
\label{eps_short}
\end{eqnarray}
where $m_{P}$ is the mass of proton. 
Here we omitted the indices for clarity. However, if chirality changes 
play a role in the value of the \onbb decay
rate,  one needs to maintain the chirality indices and define
$\epsilon_{i}=\epsilon_{i}^{xyz}$, with
$x,y,z \in \{L,R\}$; cf., App.~\ref{app:decay} for details. Since $(j_{R})^{\mu}=-(j_{L})^{\mu}$, we define
$(j)^{\mu}=(j_{R})^{\mu}=-(j_{L})^{\mu}$.

Current limits on short-range type $\epsilon_i$ are summarised in
Tab.~\ref{Tab:LimitsShort}.  Note that, different from the
$\epsilon_{\alpha}^{\beta}$ of the long-range part, here $\epsilon_i$
scale as $\epsilon_i \propto (\lambda^{\rm eff}_{\rm LNV})^4/\Lambda^5$.
For $\lambda^{\rm eff}_{\rm LNV} \simeq g_L$ the limits given 
in Tab.~\ref{Tab:LimitsShort} then correspond 
to $\Lambda \gsim (1-3)$ TeV.

\begin{table}[t]
\centering
\begin{tabular}{ccccccc}
\hline
Isotope 
& $|\epsilon_1|$ & $|\epsilon_2|$ & $|\epsilon_3^{LLz (RRz)} |$ &
$|\epsilon_3^{LRz(RLz)}|$ & $|\epsilon_4|$ & $|\epsilon_5|$ \\
\hline
$^{136}$Xe & $2.6 \cdot 10^{-7}$ & $1.4 \cdot 10^{-9}$ & $1.1 \cdot 10^{-8}$ & 
$1.7 \cdot 10^{-8}$ & $1.2 \cdot 10^{-8}$& $1.2 \cdot 10^{-7}$ \\
	\hline
\end{tabular}
\caption{\it Limits on effective short-range interactions.  These limits
are taken from \cite{Deppisch:2012nb} and are derived assuming only
one $\epsilon$ is different from zero at a time.}
\label{Tab:LimitsShort}
\end{table}

\section{General decomposition of the $\boldsymbol{d=9}$ 
$\boldsymbol{0\nu\beta\beta}$ decay operator}
\label{Sec:decom}

\begin{figure}[t]
\centering
\begin{picture}(420,110)
\Line(10,10)(50,50)
\Line(50,50)(10,90)
\ZigZag(50,50)(90,50){3}{4}
\Line(90,50)(90,90)
\Line(90,50)(130,50)
\Line(130,50)(130,90)
\ZigZag(130,50)(170,50){3}{4}
\Line(170,50)(210,10)
\Line(170,50)(210,90)
\Vertex(50,50){3}
\Vertex(90,50){3}
\Vertex(130,50){3}
\Vertex(170,50){3}
\Line(250,10)(290,50)
\Line(250,90)(290,50)
\ZigZag(290,50)(370,50){3}{8}
\ZigZag(330,50)(330,90){3}{4}
\Line(330,90)(290,130)
\Line(330,90)(370,130)
\Line(370,50)(410,10)
\Line(370,50)(410,90)
\Vertex(290,50){3}
\Vertex(330,50){3}
\Vertex(330,90){3}
\Vertex(370,50){3}
\Text(295,45)[t]{$\text v_1$}
\Text(330,45)[t]{$\text v_2$}
\Text(340,85)[t]{$\text v_3$}
\Text(370,45)[t]{$\text v_4$}
\Text(110,5)[t]{\rm{Topology I}}
\Text(330,5)[t]{\rm{Topology II}}
\Text(55,45)[t]{$\text v_1$}
\Text(90,45)[t]{$\text v_2$}
\Text(130,45)[t]{$\text v_3$}
\Text(170,45)[t]{$\text v_4$}
\end{picture}
\caption{\it \label{Fig:0nbbTopologies}The two basic tree-level
topologies realizing a $d=9$ \onbb\, operator. External lines are
fermions; internal lines can be fermions (solid), or scalars or
vectors (zig zag).}
\end{figure}
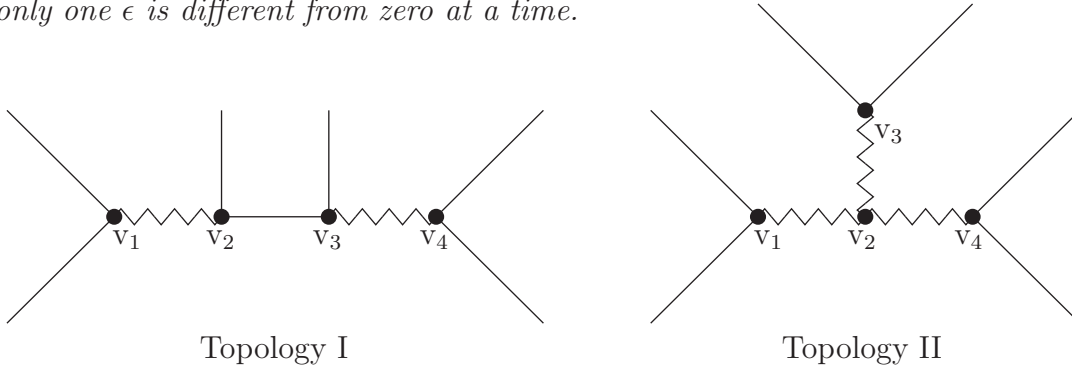

The $d=9$ effective operator generating \onbb decay at the quark level
can be written as in Eq.~\eqref{equ:effop}. In this section, 
we decompose this operator, following the techniques developed in
\cite{Gavela:2008ra,Bonnet:2009ej,Bonnet:2011yx,Bonnet:2012kz}, 
in terms of the SM quantum numbers of the mediators.
Note that the results obtained in this section are valid for both
short- and long-range contributions, while the quantitative impact of
the contribution to \onbb\ depends on the Lorentz nature. Details 
are left for the appendix, where we tabulate all possibilities for 
the short-range part mediated by fermions and scalars.

At  tree-level, there are only two possible topologies for this operator, 
which are shown in Fig.~\ref{Fig:0nbbTopologies}.
For topology~I (T-I) the internal particles between vertices 
$\text{v}_1-\text{v}_2$ and $\text{v}_3-\text{v}_4$ can be either 
scalars (S) or vectors (V), while the particle between 
$\text{v}_2-\text{v}_3$ must be a fermion (F).
The topology thus contains three classes of diagrams: VFV, SFS and SFV. 
The inner particles for topology~II (T-II), on the other hand, must all 
be either scalars or vectors, all possible combination in principle 
can occur (SSS, VVV, VVS and SSV). 

One has multiple choices for assigning the fermions to the outer legs. 
Once a particular assignment is chosen, the electric charge and the
colour of the internal particles are fixed (the latter up to a
two-fold ambiguity), but not the $U(1)_{Y}$ hypercharge. 
The assignments of hypercharge are fixed as well, 
once the chiralities of the six outer fermions are determined.  
We will come back to this point later in the discussion, and more
details are given in the appendix. We will now discuss the general
decompositions of T-I and T-II in turn.

\subsection{Decomposition of topology~I}

\begin{table}[p]
\vspace*{-0.5cm}
\small{
\begin{center}
\begin{tabular}{ccccccl}
\hline \hline 
&&Long&
\multicolumn{3}{c}{Mediator $(U(1)_{\rm em}, SU(3)_{c})$}
\\
\# & Decomposition & Range?  & $S$ or  $V_{\rho}$ &  $\psi$ & $S'$ or $V'_{\rho}$ 
& Models/Refs./Comments \\
\hline 
1-i 
&
$(\bar{u} d) (\bar{e}) (\bar{e}) (\bar{u} d)$
&
(a)
&
$(+1, {\bf 1})$
&
$(0, {\bf 1}) $
&
$(-1, {\bf 1})$
&
Mass mechan., 
RPV~\cite{Mohapatra:1986su,Hirsch:1995zi,Hirsch:1995ek},
\\
& & & & & & 
LR-symmetric models~\cite{Riazuddin:1981hz},  
\\
& & & & & & 
Mass mechanism with $\nu_{S}$~\cite{Goswami:2005ng},
\\
& & & & & & 
TeV scale seesaw, e.g.,~\cite{Blennow:2010th,Ibarra:2010xw}
\\
&
&
&
$(+1, {\bf 8})$
&
$(0, {\bf 8}) $
&
$(-1, {\bf 8})$
&
\cite{Choubey:2012ux}
\\
1-ii-a 
&
$(\bar{u} d) (\bar{u}) (d) (\bar{e} \bar{e})$
&
&
$ (+1, {\bf 1}) $
&
$ (+ 5/3, {\bf 3})$
&
$ (+2, {\bf 1})$
\\
&
&
&
$ (+1, {\bf 8}) $
&
$ (+ 5/3, {\bf 3})$
&
$ (+2, {\bf 1})$
\\
1-ii-b 
&
$(\bar{u} d) (d) (\bar{u}) (\bar{e} \bar{e})$
&
&
$(+1, {\bf 1}) $
&
$(+4/3, \overline{\bf 3})$ 
&
$(+2, {\bf 1})$
\\
&
&
&
$(+1, {\bf 8}) $
&
$(+4/3, \overline{\bf 3})$ 
&
$(+2, {\bf 1})$
\\
\hline
2-i-a
&
$(\bar{u} d) (d) (\bar{e}) (\bar{u} \bar{e})$
&
&
$(+1, {\bf 1}) $
&
$(+4/3, \overline{\bf 3})$ 
&
$(+1/3, \overline{\bf 3})$
\\
&
&
&
$(+1, {\bf 8}) $
&
$(+4/3, \overline{\bf 3})$ 
&
$(+1/3, \overline{\bf 3})$
\\
2-i-b
&
$(\bar{u} d) (\bar{e}) (d) (\bar{u} \bar{e})$
&
(b)
&
$(+1, {\bf 1})$
&
$ (0,{\bf 1})$
&
$ (+1/3, \overline{\bf 3})$
&
RPV~\cite{Mohapatra:1986su,Hirsch:1995zi,Hirsch:1995ek}, LQ ~\cite{Hirsch:1996qy,Hirsch:1996ye}
\\
&
&
&
$(+1, {\bf 8})$
&
$ (0,{\bf 8})$
&
$ (+1/3, \overline{\bf 3})$
&
\\
2-ii-a
&
$(\bar{u} d) (\bar{u}) (\bar{e}) (d \bar{e})$
&
&
$(+1, {\bf 1})$
&
$ (+5/3,{\bf 3})$
&
$ (+2/3, {\bf 3})$ 
\\
&
&
&
$(+1, {\bf 8})$
&
$ (+5/3,{\bf 3})$
&
$ (+2/3, {\bf 3})$ 
\\
2-ii-b
&
$(\bar{u} d) (\bar{e}) (\bar{u}) (d \bar{e})$
&
(b)
&
$(+ 1,{\bf 1})$
&
$ (0,{\bf 1})$
&
$ (+ 2/3, {\bf 3})$ 
&
RPV~\cite{Mohapatra:1986su,Hirsch:1995zi,Hirsch:1995ek}, LQ ~\cite{Hirsch:1996qy,Hirsch:1996ye}
\\
&
&
&
$(+ 1,{\bf 8})$
&
$ (0,{\bf 8})$
&
$ (+ 2/3, {\bf 3})$ 
&
\\
2-iii-a
&
$(d \bar{e}) (\bar{u}) (d) (\bar{u} \bar{e})$
&
(c)
&
$ (- 2/3, \overline{\bf 3})$  
&
$ (0, {\bf 1})$
&
$ (+ 1/3, \overline{\bf 3})$
&

RPV~\cite{Mohapatra:1986su,Hirsch:1995zi,Hirsch:1995ek}
\\
&
&
&
$ (- 2/3, \overline{\bf 3})$  
&
$ (0, {\bf 8})$
&
$ (+ 1/3, \overline{\bf 3})$
&
RPV~\cite{Mohapatra:1986su,Hirsch:1995zi,Hirsch:1995ek}
\\
2-iii-b
&
$(d \bar{e}) (d) (\bar{u}) (\bar{u} \bar{e})$
&
&
$ (- 2/3, \overline{\bf 3})$ 
&
$ (-1/3, {\bf 3}) $
&
$ (+ 1/3, \overline{\bf 3})$
\\
&
&
&
$ (- 2/3, \overline{\bf 3})$ 
&
$ (-1/3, \overline{\bf 6}) $
&
$ (+ 1/3, \overline{\bf 3})$
\\
\hline
3-i
&
$(\bar{u} \bar{u}) (\bar{e})(\bar{e}) (dd)$
&
&
$ (+ 4/3, \overline{\bf 3}) $
&
$ (+1/3, \overline{\bf 3}) $
&
$(- 2/3, \overline{\bf 3})$  
&
only with $V_{\rho}$ and $V'_{\rho}$
\\
&
&
&
$ (+ 4/3, {\bf 6}) $
&
$ (+1/3, {\bf 6}) $
&
$(- 2/3, {\bf 6})$  
\\
3-ii
&
$(\bar{u} \bar{u}) (d) (d) (\bar{e} \bar{e})$
&
&
$(+ 4/3, \overline{\bf 3}) $ 
&
$ (+5/3, {\bf 3})$  
&
$(+2, {\bf 1}) $ 
&
only with $V_{\rho}$
\\
&
&
&
$(+ 4/3, {\bf 6}) $ 
&
$ (+5/3, {\bf 3})$  
&
$(+2, {\bf 1}) $ 
\\
3-iii
&
$(dd) (\bar{u}) (\bar{u}) (\bar{e} \bar{e})$
&
&
$ (+ 2/3, {\bf 3}) $  
&
$ (+4/3, \overline{\bf 3}) $ 
&
$ (+ 2, {\bf 1}) $ 
&
only with $V_{\rho}$
\\
&
&
&
$ (+ 2/3, \overline{\bf 6}) $  
&
$ (+4/3, \overline{\bf 3}) $ 
&
$ (+ 2, {\bf 1}) $ 
\\
\hline
4-i
&
$(d \bar{e}) (\bar{u}) (\bar{u}) (d \bar{e})$
&
(c)
&
$(- 2/3, \overline{\bf 3})$  
&
$( 0, {\bf 1}) $ 
&
$ (+ 2/3, {\bf 3}) $  
&
RPV~\cite{Mohapatra:1986su,Hirsch:1995zi,Hirsch:1995ek}
\\
&

&&
$(- 2/3, \overline{\bf 3})$  
&
$( 0, {\bf 8}) $ 
&
$ (+ 2/3, {\bf 3}) $  
&
RPV~\cite{Mohapatra:1986su,Hirsch:1995zi,Hirsch:1995ek}
\\
4-ii-a
&
$(\bar{u} \bar{u}) (d) (\bar{e}) (d \bar{e})$
&
&
$(+ 4/3, \overline{\bf 3}) $ 
&
$ (+5/3, {\bf 3})$ 
&
$ (+ 2/3, {\bf 3}) $  
&
only with $V_{\rho}$
\\
&
&
&
$(+ 4/3, {\bf 6}) $ 
&
$ (+5/3, {\bf 3})$ 
&
$ (+ 2/3, {\bf 3}) $  
&
see Sec.~\ref{sec:example} (this work)
\\
4-ii-b
&
$(\bar{u} \bar{u}) (\bar{e}) (d) (d \bar{e})$
&
&
$ (+ 4/3, \overline{\bf 3}) $
&
$ (+1/3, \overline{\bf 3}) $
&
$ (+ 2/3, {\bf 3}) $  
&
only with $V_{\rho}$
\\
&
&
&
$ (+ 4/3, {\bf 6}) $
&
$ (+1/3, {\bf 6}) $
&
$ (+ 2/3, {\bf 3}) $  
\\
\hline
5-i
&
$(\bar{u} \bar{e}) (d) (d) (\bar{u} \bar{e})$
&
(c)
&
$ (- 1/3, {\bf 3}) $
&
$(0, {\bf 1}) $
&
$ (+ 1/3, \overline{\bf 3}) $
&
RPV~\cite{Mohapatra:1986su,Hirsch:1995zi,Hirsch:1995ek}
\\
&
&
&
$ (- 1/3, {\bf 3}) $
&
$(0, {\bf 8}) $
&
$ (+ 1/3, \overline{\bf 3}) $
&
RPV~\cite{Mohapatra:1986su,Hirsch:1995zi,Hirsch:1995ek}
\\
5-ii-a
&
$(\bar{u} \bar{e}) (\bar{u}) (\bar{e}) (dd)$
&
&
$ (- 1/3, {\bf 3}) $
&
$ (+1/3, \overline{\bf 3}) $
&
$ (- 2/3, \overline{\bf 3}) $  
&
only with $V'_{\rho}$
\\
&
&
&
$ (- 1/3, {\bf 3}) $
&
$ (+1/3, {\bf 6}) $
&
$ (- 2/3, {\bf 6}) $  
\\
5-ii-b
&
$(\bar{u} \bar{e}) (\bar{e}) (\bar{u}) (dd)$
&
&
$ (- 1/3, {\bf 3}) $
&
$ (-4/3, {\bf 3})$ 
&
$(- 2/3, \overline{\bf 3}) $ 
&
only with $V'_{\rho}$ 
\\
&
&
&
$ (- 1/3, {\bf 3}) $
&
$ (-4/3, {\bf 3})$ 
&
$(- 2/3, {\bf 6}) $
\\  
\hline \hline
\end{tabular}
\end{center}
} 
\caption{\it \label{Tab:TopoI} General decomposition of the $d=9$
operator ${\bar u} {\bar u} dd{\bar e}{\bar e}$ for topology~I.  Here
we do not specify the chirality of outer fermions, and the mediators
are given with the charge of electromagnetic $U(1)_{\rm
em}$ and that of colour $SU(3)_{c}$. The symbols $S$ and $S'$ denote
scalars, $V_\rho$ and $V_\rho'$ vectors, and $\psi$ a fermion.  The
column ``Long Range?'' indicates if and which type of long-range diagram in
Fig.~\ref{Fig:LongShort} can be constructed, apart from the
short-range diagram (d). The column ``Models/Refs./Comments'' lists possible
models, discussed previously in the literature, and references, and comments on
possible limitations for the mediators. Here
``RPV'' stands for R-parity violating SUSY models, and ``LQ'' for
``leptoquarks''. }
\end{table}

In Tab.~\ref{Tab:TopoI} we list the general decompositions for the
\onbb decay operator for T-I. The chiralities of the outer fermions
are left unspecified here for a more compact presentation.  In total there
are 18 (times 2 for the choice of colour) possibilities for 
realizing T-I.  Because of the $SU(3)$ multiplication rules, 
${\bf \bar{3}} \otimes {\bf \bar{3}} = {\bf 3}_a \oplus {\bf \bar{6}}_s$ 
and ${\bf 3} \otimes {\bf \bar{3}} = {\bf 1} \oplus {\bf 8}$, 
there are always two possible
colour assignments for the internal particles. The table is valid for
all three possible classes of diagrams (VFV, SFS and SFV).
In some of the cases listed in Tab.~\ref{Tab:TopoI}, only VFV or 
SFV exchange is possible, because in the decomposition 
${\bf \bar{3}} \otimes {\bf \bar{3}} = {\bf 3}_a \oplus {\bf \bar{6}}_s$ 
the coupling of a scalar to two identical quarks (two ${\bf \bar{3}}$) 
vanishes for 
the ${\bf 3}_a$. These affects all the cases in T-I-3, 
T-I-4-ii and T-I-5-ii, see table for nomenclature.

Each of the models listed in this table leads to an effective
operator with a different Lorentz structure, which needs to be
projected onto the basis shown in Eqs. (\ref{eq:defLR}) and
(\ref{eps_short}), once the chiralities are fixed.  The projection can
be done by Fierz transformations and the transformations of the SM
gauge group indices.  This allows to identify immediately if a model
gives important contributions to \onbb decay. More details and results
of this procedure are shown in the appendix.

From this table we can identify all (T-I) contributions discussed in
the literature, once the nature of the internal bosons and the
chirality of the outer fermions are chosen. For example, the mass
mechanism corresponds to T-I-1-i, with the bosons being vectors 
($W^{\pm}$) 
and all the outer fermions being left-handed. In this case, 
the quantum numbers of the internal fermion $\psi$ 
are equal to those of a (light) neutrino, and the mass term 
is picked out from the propagator.

Other examples can be identified as easily. To list a few more, 
the afore-mentioned $W_R-N-W_R$ exchange diagram is also contained 
in T-I-1-i with vectors ($W_{R}^{\pm}$), 
all outer fermions now being right-handed. Since $N$ 
must be heavy, however, this is now a short-range contribution. 
Concrete models can lead to the occurrence of more than one of the 
operators listed, an example is provided by (trilinear) R-parity 
violating (RPV) supersymmetry (SUSY). The six short-range diagrams 
\cite{Hirsch:1995zi,Hirsch:1995ek} for the RPV SUSY mechanism 
of \onbb decay \cite{Mohapatra:1986su}, correspond to class SFS 
and are identified as 
T-I-1-i (${\tilde e}-\tilde{\chi}^{0}-{\tilde e}$ diagram), 
T-I-2-i-b (${\tilde e}-\tilde{\chi}^{0}-{\tilde d}$), 
T-I-2-ii-b (${\tilde e}-\tilde{\chi}^{0}-{\tilde u}$), 
T-I-2-iii-a (${\tilde u}-\tilde{\chi}^{0}/\tilde{g}-{\tilde d}$), 
T-I-4-i (${\tilde u}-\tilde{\chi}^{0}/\tilde{g}-{\tilde u}$) 
and  
T-I-5-i (${\tilde d}-\tilde{\chi}^{0}/\tilde{g}-{\tilde d}$). 
In those diagrams,
the neutralino $\tilde{\chi}^{0}$ corresponds to $\psi(0,{\bf 1})$, 
while the gluino $\tilde{g}$ corresponds to  $\psi(0,{\bf 8})$. 
Finally, the leptoquark (LQ) mechanism of
\cite{Hirsch:1996qy,Hirsch:1996ye} is 
a long-range contribution of the class SFV with
T-I-2-i-b and T-I-2-ii-b. 
The internal fermion $\psi(0,{\bf 1})$ is again identified 
with a light neutrino.

A few more comments on Tab.~\ref{Tab:TopoI} might be in order. 
There are a total of six possibilities in which the intermediate fermion
transforms $\psi(0,{\bf 1})$ under the gauge symmetries 
$(U(1)_{\rm em}, SU(3)_{c})$. 
Only they can lead to long-range contributions, 
all the other models in the list are necessarily of the
short-range type.  Among those six, only the cases marked (a) or (b) 
in the column ``Long Range?'' can lead to interesting constraints, 
since the
remaining three cases marked as (c) are suppressed with $\epsilon^2$. 
Note, however, that all of these cases can also be of short-range type.
There are 12 (times two) cases listed, which require that the internal
fermion has a fractional electric charge. As far as we know, none of them have
been discussed in the literature before. All of these new ``models''
not only require fractionally charged fermions, but also exotic
bosons.  The latter can be doubly-charged 
bileptons~\cite{Cuypers:1996ia}, diquarks, or leptoquarks~\cite{Buchmuller:1986zs}.

The \onbb decay process violates lepton number $L$. 
We can easily identify the different possibilities for LNV 
from Tab.~\ref{Tab:TopoI} (cf., Fig.~\ref{Fig:0nbbTopologies}, left panel). 
If the internal fermion is neutral, it can have a Majorana mass 
and a mass insertion leads then to $\Delta(L)=2$. This is the case, for 
example, in the mass mechanism and in the $W_R-N-W_R$ diagram of 
LR-symmetric models. The other possibility is to have LNV vertices. 
For the cases with a doubly charged bilepton $S'(+2, {\bf 1})$, 
for example, one can have one $\Delta(L)=2$ vertex. 
And, finally, it is possible to have models with two $\Delta(L)=1$
vertices. An example is trilinear RPV SUSY, e.g., 
from the superpotential 
$\mathcal{W} = \lambda' {\widehat L}{\widehat Q}{\widehat D}^c$. 

Most of the new models we find are of the short-range type and thus
should be testable at the LHC.  We will discuss one particular example
in greater detail in Sec.~\ref{sec:example}.

\subsection{Decomposition of topology~II}

\begin{table}[t]
\begin{center}
\begin{tabular}{cccccl}
\hline \hline 
&& 
\multicolumn{3}{c}{Mediator $(Q_{\rm em}, Q_{\text{colour}})$} \\
\# & Decomposition & $S$ or $V_{\rho}$ & $S'$  or $V'_{\rho}$ & $S'' $ or $V''_{\rho}$ 
& Models/Refs./Comments \\
\hline 
1 
&
$(\bar{u} d) (\bar{u} d) (\bar{e} \bar{e})$
&
$ (+1,{\bf 1})$
&
$ (+1,{\bf 1})$
&
$ (-2, {\bf 1})$
&
Addl. triplet scalar~\cite{Mohapatra:1981pm} \\
& & & & & LR-symmetric models~\cite{Rizzo:1982kn,Hirsch:1996qw}
\\
&
&
$ (+1,{\bf 8})$
&
$ (+1,{\bf 8})$
&
$ (-2, {\bf 1})$
\\
2
&
$(\bar{u} d) (\bar{u} \bar{e}) (\bar{e} d)$
&
$(+1,{\bf 1})$
&
$(-1/3,{\bf 3})$
&
$(-2/3,\overline{\bf 3})$
\\
&
&
$(+1,{\bf 8})$
&
$(-1/3,{\bf 3})$
&
$(-2/3,\overline{\bf 3})$
\\
3
&
$(\bar{u} \bar{u}) (dd) (\bar{e} \bar{e})$
&
$(+4/3,\overline{\bf 3})$
&
$(+2/3,{\bf 3})$
&
$(-2,{\bf 1})$
&
only with $V_{\rho}$ and $V'_{\rho}$
\\
&
&
$(+4/3,{\bf 6})$
&
$(+2/3,\overline{\bf 6})$
&
$(-2,{\bf 1})$
\\
4
&
$(\bar{u} \bar{u}) (\bar{e} d) (\bar{e} d)$
&
$ (+4/3, \overline{\bf 3})$
&
$ (-2/3, \overline{\bf 3})$
&
$ (-2/3, \overline{\bf 3})$
&
only with $V_{\rho}$
\\
&
&
$ (+4/3, {\bf 6})$
&
$ (-2/3, \overline{\bf 3})$
&
$ (-2/3, \overline{\bf 3})$
\\
5
&
$(\bar{u} \bar{e}) (\bar{u} \bar{e}) (d d)$
&
$ (-1/3,{\bf 3})$
&
$ (-1/3,{\bf 3})$
&
$ (+2/3, {\bf 3})$
&
only with $V''_{\rho}$
\\
&
&
$ (-1/3,{\bf 3})$
&
$ (-1/3,{\bf 3})$
&
$ (+2/3, \overline{\bf 6})$
& \cite{Gu:2011ak,Kohda:2012sr}
\\
\hline \hline
\end{tabular}
\end{center}
\caption{\it \label{Tab:TopoII}
Decomposition of topology~II. As in Tab~\ref{Tab:TopoI}, we do 
only give electric and colour charges of the internal bosons here. 
The mediators can be either scalars ($S$, $S'$, $S''$) or
vectors ($V_\rho$, $V_\rho'$, $V_\rho''$).
All listed possibilities give short-range contributions.}
\end{table}

There are a total of five (again times two due to colour) possibilities 
to assign the outer fermions to T-II. These are listed together 
with the electric and colour charges of the possible mediators in 
Tab.~\ref{Tab:TopoII}. As before, this table is valid for both 
V and S intermediate states and we do not specify the chiralities 
of the outer fermions here. 
In the appendix, we give a table for the operator decompositions 
with fixed chiralities. 

Much fewer models with T-II have been studied in the literature 
than for T-I. In fact, we have found only the cases 
T-II-1 and T-II-5 (with a scalar ${\bf\bar 6}$) have been considered 
previously. The best known is the case T-II-1 with bosons transforming 
as $V(+1,{\bf 1})$. This diagram can be produced by adding an 
$SU(2)_L$ triplet scalar to the SM particle content 
\cite{Mohapatra:1981pm} (case SSS).\footnote{However, it was shown in 
\cite{Schechter:1981bd,Wolfenstein:1982bf,Haxton:1982ff} that this 
contribution is always sub-dominant.} The same diagram can occur 
in LR-symmetric models \cite{Rizzo:1982kn} (as VVS). In this case, the 
outer fermions are again all right-handed, and 
this possibility has been considered in \cite{Hirsch:1996qw}. 
Note that in all these models, there is  always an additional contribution 
from T-I,  
which is not necessarily present for all T-II models. 
In the case of the LR-symmetric model, T-II can be comparable with T-1, 
but can never completely dominate the contributions to \onbb decay.

Finally there is the recent paper \cite{Gu:2011ak}, in which the
author constructs a model with the particle content corresponding to
the scalars in T-II-5 (with ${\bf\bar 6}$ representation under
$SU(3)_{c}$). In this model neutrinos are pseudo-Dirac, such that
T-II-5 gives the dominant contribution to \onbb decay.  This is the
only example in the literature, which we are aware of that a T-II
contribution dominates the \onbb decay amplitude.\footnote{A similar
diagram can be found in Fig.~2 of~\cite{Kohda:2012sr}.  The model
therein was constructed to describe neutrino mass at two-loop
order. 
Note that neutrino mass may be obtained
at a lower than 4-loop order (as postulated by the black box diagram),
in a specific model. That, however, necessarily requires that the effective 
operator can be promoted to an $SU(2)$ invariant operator  with 
one or two lepton doublets, which cannot be done for every model.
The $SU(2)$ invariant effective operator
with one lepton doublet and a right-handed electron may induce neutrino
mass at the 3-loop level (as our example in Sec.~\ref{sec:example}), and the
 operator with two lepton doublets
may induce neutrino mass at the 2-loop level. If one assumes that  
the SM gauge invariant effective operator is constrained by the 
current limit of $0\nu\beta\beta$ decay,
the 2- and 3-loop-induced neutrino masses via the associated operators $SU(2)_{L}$
should  be sub-dominant.}
Note, however,
that our Tab.~\ref{Tab:TopoII} allows to construct a number of
additional models with this property.

\section{An example for short-range $\boldsymbol{0\nu\beta\beta}$ decay, 
and its test at the LHC}
\label{sec:example}

As mentioned above, one expects that exotic models with LNV, which 
lead to short-range contributions for \onbb decay, yield testable 
phenomenology at the LHC. The classical example is the LR-symmetric 
model, and there is also a recent paper that has made a study for trilinear 
RPV SUSY and \onbb decay~\cite{Allanach:2009iv}. 
In this section we will discuss 
basic LHC phenomenology of one particular example in our
Tab.~\ref{Tab:TopoI}, based on the decomposition T-I-4-ii-a. We have 
chosen this particular case basically for two reasons: (i) it belongs 
to the class of models, which have not been studied in the literature 
before, and (ii) it leads to richer phenomenology at the LHC than 
either the LR-symmetric model or RPV SUSY. By using the different signals 
which we will discuss in the following, one could distinguish this model 
from the other possibilities, such as the LR-symmetric model and RPV SUSY.

\begin{figure}[t]
\centering
\begin{picture}(250,130)
\Line(10,10)(60,60)
\Line(10,120)(60,60)
\DashLine(60,60)(100,60){5}
\Line(100,60)(150,60)
\DashLine(150,60)(190,60){5}
\Line(190,60)(240,120)
\Line(190,60)(240,10)
\Line(100,60)(100,120)
\Line(150,60)(150,120)
\Text(5,10)[r]{$\overline{u_{R}}$}
\Text(5,120)[r]{$\overline{u_{R}}$}
\Text(100,125)[b]{$d_{L}$}
\Text(150,125)[b]{$\overline{e_{R}}$}
\Text(245,10)[l]{$d_{R}$}
\Text(245,120)[l]{$\overline{e_{L}}$}
\Text(80,55)[t]{$S_{\rm DQ}^{4/3}$}
\Text(125,55)[t]{$\Psi^{5/3}$}
\Text(170,55)[t]{$S^{2/3}_{\rm LQ}$}
\end{picture}
\caption{\it The diagram of \onbb for the example T-I-4-ii-a
with two scalar mediators (SFS type diagram).}
\label{Fig:0nbbExample}
\end{figure}
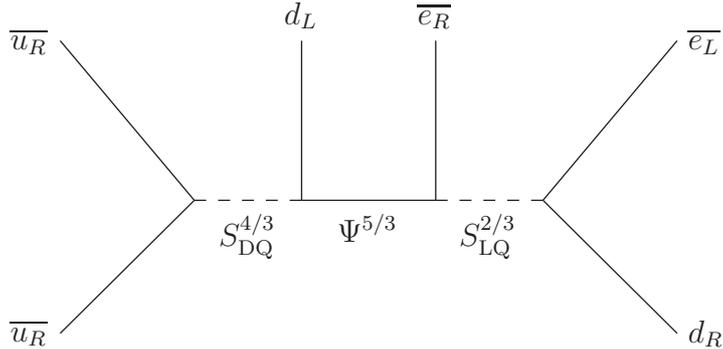

For the decomposition T-I-4-ii-a one needs to 
introduce three new particles to the SM particle content, 
which are 
(a) a diquark, which is either a vector $V_{{\rm DQ}}^{4/3}$ 
or a scalar $S_{\rm DQ}^{4/3}$, 
(b) an exotic coloured (${\bf 3}$) vector-like fermion $\Psi^{5/3}$ 
with electric charge ${5/3}$,
and (c) a leptoquark, which is again either a vector 
$V_{{\rm LQ}}^{2/3}$ or a scalar $S_{\rm LQ}^{2/3}$. 
We will concentrate on the case where both the diquark and the
leptoquark are scalars. The vector case is qualitatively similar from
the point of view of LHC phenomenology.\footnote{Of course, some
numerical factors are different in the vector cases from the scalar
case.} The \onbb decay is generated through the diagram shown 
in Fig.~\ref{Fig:0nbbExample}.

So far, the chiralities of the outer fermions are not determined.
When taking into account the fact that the effective operator
should be a component of a SM gauge invariant operator, 
the number of the choices is limited. 
There are three possibilities of the SM gauge invariant operators 
that can contain the T-I-4-ii-a operator,\footnote{%
The chiralities of the outer fermions on the vertices $\text{v}_{2}$
and $\text{v}_{3}$ are chosen so that the mass term in the propagator 
of the fermion mediator $\Psi^{5/3}$ is picked out. See appendix for
more details.} which are
$(\overline{Q} \overline{Q}) (d_{R}) (\overline{L}) (\overline{L} d_{R})$,
$(\overline{u_{R}} \overline{u_{R}}) (Q) (\overline{e_{R}})
(\overline{L} d_{R})$,
and 
$(\overline{u_{R}} \overline{u_{R}}) (d_{R}) (\overline{L})
(\overline{e_{R}} Q)$.
We note in passing that they correspond to the 
effective operators (\#11, \#20, and also \#20 respectively)
shown in \cite{Babu:2001ex}.
Here, we take the second one for our example,
and consequently the chiralities of the outer fermions
are fixed as 
$(\overline{u_{R}} \overline{u_{R}}) (d_{L}) (\overline{e_{R}})
(\overline{e_{L}} d_{R})$. Note that in this
example, neutrino mass can be generated at the 3-loop order,
with one effective vertex given by the \onbb\ operator. 
This contribution should be negligible by similar arguments as in
Ref.~\cite{Duerr:2011zd}.

Diquarks at the LHC have been studied recently in \cite{Han:2010rf}.
Among them, the relevant one for our \onbb example is
a scalar colour-sextet ({\bf 6}) diquark
$S_{\rm DQ}^{+4/3} ({\bf 6}, {\bf 1})_{+4/3}$
which is an 
$SU(2)_{L}$ singlet $(\bf 1)$ and has +4/3 $U(1)_{Y}$-hypercharge.\footnote{%
Here we use the notation $(SU(3)_{c}, SU(2)_{L})_{U(1)_{Y}}$ from the appendix.}
It interacts with two right-handed up-quarks as
\begin{equation}\label{eq:DQ}
\mathcal{L}_{\rm{DQ}}
=
\left[
\lambda_{\rm DQ}^{\alpha \beta} 
(\overline{u_{\alpha R}})^{I a} 
(T_{\bf \bar{6}})_{IJ}^{X} 
({u_{\beta R}}^{c})^{J}_{a}  
(S_{\rm DQ}^{+4/3})_{X}
+ 
h.c. 
\right]
- 
m_{\rm DQ}^2 
(S_{\rm DQ}^{-4/3})^{X}
(S_{\rm DQ}^{+4/3})_{X}.
\end{equation}
where the indices $\alpha$ and $\beta$ indicate the generations of the
up-type quarks, $I$ and $J$ label the fundamental representations
(${\bf 3}$ and ${\bf \bar{3}}$) of $SU(3)_{c}$ ($I,J=1,2,3$), and $a$ is
the index for the 2-component left-handed (=conjugate of right-handed)
spinor.  Here, the matrices $(T_{\bf \bar{6}})_{IJ}^{X}$ ($X=$1-6)
provides $SU(3)_c$ Clebsch-Gordan coefficients, which are symmetric
under the exchange of $I$ and $J$.  The concrete form of $T_{\bf
\bar{6}}$ is given in App.~\ref{app:fcol}.  If the diquark is chosen as a
colour ${\bf \bar{3}}$ (i.e., $\epsilon_{IJK} (\overline{u_{\alpha R}})^{I}
({u_{\beta R}}^{c})^{J} (S_{\rm DQ}^{+4/3})^{K}$) instead of a {\bf 6}, the
coefficient $\lambda_{\rm DQ}^{\alpha\beta}$ must be antisymmetric in
the indices, due to the transformation properties of the ${\bf 3}$
that is made from an antisymmetric combination in colour indices of
two ${\bf \bar{3}}$ (two $\bar{u}$).  For \onbb decay, only the choice $\alpha
= \beta = u$ is relevant.  Thus, the scalar diquark $S_{\rm DQ}$ must
be a colour ${\bf 6}$ representation.\footnote{This is strictly true only for unmixed (valence) quarks. 
Note also that for vector diquarks in principle both, the 
${\bf 6}_{s}$ and the ${\bf \bar{3}}_{a}$ can contribute.} 

A diquark with couplings as in Eq.~(\ref{eq:DQ}) will be copiously
produced at the LHC. In \cite{Han:2010rf}, the authors evaluated 
the production cross sections $\sigma$, which are as large as 
$\sigma/(\lambda_{\rm DQ}^2 \rm{BR}_{jj}) = 400$ $(1)$ pb for
$m_{\rm DQ}= 1$ (3) TeV. Here, $\rm{BR}_{jj}$ is the branching ratio for
the diquark decaying to two jets. Due to the s-channel resonance
in the cross section, $\sigma$ scales approximately as $\sigma \propto
\lambda_{\rm DQ}^2$ (in the narrow width approximation).  
Recently, the CMS \cite{Chatrchyan:2011ns} and ATLAS
\cite{Aad:2012xz,Aad:2012yz} collaborations have searched for
resonances in the dijet mass spectrum and upper limits 
on $\sigma \times \rm{BR}_{jj}\times {\cal A}$ have been derived 
as a function of the invariant dijet mass. 
Here, ${\cal A}$ is the acceptance, which is
estimated to be ${\cal A} \simeq 0.6$ for isotropic 
decays~\cite{Chatrchyan:2011ns}. The experimental upper limits range 
from $\sigma\times \rm{BR}_{jj}\times {\cal A} \simeq 1$ ($0.01$) pb 
for $m_{\rm DQ} = 1$ (3) TeV. These limits get stronger for larger values 
of $m_{\rm DQ}$, because of the larger QCD background for smaller 
invariant masses. These limits, together with the theoretical calculation of 
the cross sections \cite{Han:2010rf}, 
imply upper limits on $\lambda_{\rm DQ}^{uu}$ of the order of roughly 
$\lambda_{\rm DQ}^{uu} \lsim 0.2$ over the whole mass range explored 
($m_{\rm DQ} \sim (1-4)$ TeV).

Note that a scalar diquark has been proposed 
\cite{Dorsner:2009mq,Dorsner:2011ai,Kosnik:2011jr} as a possible 
explanation for the unexpectedly larger $t{\bar t}$ asymmetry 
observed at the Tevatron. However, these papers consider only a scalar 
${\bf 3}_{a}$, which will not contribute to \onbb decay, as explained 
above. A ${\bf {\bar 6}}_{s}$ would probably be able to give a similar 
enhancement, but a recent paper by the ATLAS collaboration claims that 
most of the parameter space of  \cite{Dorsner:2011ai} is now ruled 
out by LHC data \cite{Aad:2012em}. We will therefore not enter into 
a detailed discussion of this possibility.

The mediator $\Psi^{5/3}$ is a heavy vector-like coloured (${\bf 3}$)
fermion, aka Vector-like Quark (VLQ).  The LHC phenomenology of such
states has been recently studied by a number of
authors~\cite{Cacciapaglia:2010vn,Cacciapaglia:2011fx,Okada:2012gy,Cacciapaglia:2012dd}.
From the SM gauge invariance, this exotic fermion should be a
component field of an $SU(2)_{L}$ doublet
$\Psi=(\Psi^{5/3},\Psi^{2/3})^{T}$ with hypercharge $7/6$.
Current limits from pair production
have been summarised recently in \cite{Okada:2012gy}. For the
$\Psi^{5/3}$ the ATLAS search for pair-produced heavy quarks decaying
to $WqWq$ gives $m_{\Psi^{5/3}} \gsim 350$ GeV \cite{Okada:2012gy}.
Note that vector-like quarks have received a lot of
attention recently \cite{Azatov:2012rj,Bonne:2012im,Batell:2012ca,%
Bertuzzo:2012bt,McKeen:2012av,Bae:2012ir,Davoudiasl:2012ig,Batell:2012mj,%
An:2012vp,Martin:2012dg,Wang:2012gm,Iwamoto:2012hh,Endo:2011xq} 
as a possibility to explain the larger than expected event rate in
$h\to \gamma\gamma$ observed by the ATLAS \cite{Aad:2012gk} and CMS
\cite{CMS:2012gu} collaborations.

The other scalar mediator $S^{2/3}_{\rm LQ}$,
which interacts with $d_{R}$ and $L$, can be identified 
as  so-called a first generation Leptoquark (LQ) which interacts 
only with the first generation fermions
and it comes from the $SU(2)_{L}$ doublet with hypercharge,
$1/6$, $S_{\rm LQ} = (S_{\rm LQ}^{2/3}, S_{\rm
LQ}^{-1/3})^{T}$~\cite{Buchmuller:1986zs}. 
The Lagrangian relevant 
for generating the \onbb decay diagram of Fig.~\ref{Fig:0nbbExample} 
contains
\begin{equation}
\mathcal{L}_{\rm LQ} 
= 
\left[
\lambda_{\rm LQ} 
(\overline{L})^{i}_{\dot{a}}
(d_{R})_{I}^{\dot{a}}
({\rm i} \tau^{2})_{ij}
(S_{\rm LQ}^{*})^{I j} 
+
\rm{h.c.}
\right]
-
m_{\rm LQ}^{2}
(S_{\rm LQ}^{\dagger})^{I i} 
(S_{\rm LQ})_{I i}\,,
\label{eq:LQ-int}
\end{equation}
where $({\rm i}\tau^{2})$ is an antisymmetric tensor for the $SU(2)_{L}$
indices.  At the LHC, the first generation LQs are studied through
pair production via the strong interaction. The produced LQs can then
decay into $eq$ or $\nu q$ pairs through the interaction shown in
Eq.~\eqref{eq:LQ-int}.  This allows to derive absolute bounds on the
LQ mass (nearly independent of $\lambda_{\rm LQ}$).  The current
bounds~\cite{Aad:2011ch} from ATLAS are $m_{\rm LQ}>660\,\rm{GeV}$.
The CMS searches for the LQ give the limits which range from 830 GeV
to 640 GeV for first generation LQs with Br$_{eq}=1$ to Br$_{eq}=0.5$.
The HERA experiment \cite{Abramowicz:2012tg} has also searched for
LQs, but via single LQ production.  This leads to limits in the
parameter plane $\lambda_{\rm LQ}$-$m_{\rm LQ}$. However, practically
all the HERA-excluded combinations are now superseded by the recent
LHC limits.

For the diagram generating \onbb shown in Fig.~\ref{Fig:0nbbExample},
two more interaction terms among the different mediators are needed:
\begin{align}\label{eq:exaLNV}
\mathcal{L}_{\Psi}
=&
\lambda_{{\rm DQ}\Psi}^{\alpha} 
(\overline{{Q_{\alpha}}^{c}})_{I i}^{a}
(T_{\bf 6})^{IJ}_{X}
({\rm i} \tau^{2})^{ij}
(\Psi_{L})_{J j a}
(S_{\rm DQ}^{-4/3})^{X}
+
\lambda_{{\rm LQ}\Psi}
(\overline{\Psi_{R}})^{I i a}
({e_{R}}^{c})_{a}
(S_{\rm LQ})_{Ii}
+
h.c. 
\end{align}
Note that Eq.~(\ref{eq:exaLNV}), together with the previously
specified pieces of Lagrangians, necessarily violates lepton (but not
baryon) number, as is necessary for the generation of a finite \onbb
decay amplitude. No constraints on 
$\lambda_{{\rm DQ}\Psi}$ and $\lambda_{{\rm LQ}\Psi}$ exist 
in the literature up to now.

After integrating out all the heavy fields, 
the LNV $d=9$ effective Lagrangian for \onbb decay 
is given as 
\begin{equation}\label{eq:Lageff}
{\cal L}^{\rm eff} 
= 
\frac{
\lambda_{\rm DQ} 
\lambda_{{\rm DQ}\Psi} 
\lambda_{{\rm LQ}\Psi} 
\lambda_{\rm LQ}
}
{
m_{\rm DQ}^{2} m_{\Psi} m_{\rm LQ}^{2}
}
\left[
(\overline{u_{R}})^{I' a}
(T_{\bf \bar{6}})_{I'J'}^{X}
({u_{R}}^{c})^{J'}_{a} 
\right]
\left[
(\overline{{d_{L}}^{c}})_{I}^{b}
(T_{\bf 6})^{IJ}_{X}
({e_{R}}^{c})_{b}
\right]
\left[
(\overline{e_{L}})_{\dot{c}}
(d_{R})_{J}^{\dot{c}}
\right]
+
h.c.
\end{equation}
As demonstrated in App.~\ref{app:fcol}, we arrive at a linear combination 
of the basis operators --- the effective current description 
of \cite{Pas:2000vn} 
--- after Fierz and the colour-index
transformation, which is, 
\begin{equation}\label{eq:Fierzexa}
\mathcal{L}^{\text{eff}}
=
\frac{
\lambda_{\rm DQ} 
\lambda_{{\rm DQ}\Psi} 
\lambda_{{\rm LQ}\Psi} 
\lambda_{\rm LQ}
}
{
m_{\rm DQ}^{2} m_{\Psi} m_{\rm LQ}^{2}
}
\frac{1}{32}
\left[
{\rm i}
(\mathcal{O}_{4})_{LR}
-
(\mathcal{O}_{5})_{LR}
\right] \equiv 
 \mathcal{C}_4 \, (\mathcal{O}_{4})_{LR}
-
 \mathcal{C}_5 \, (\mathcal{O}_{5})_{LR}
\, 
\end{equation}
with correspondingly defined coefficients $\mathcal{C}_i$ ($\propto
\Lambda^{-5}$)  and $|\mathcal{C}_4| = |\mathcal{C}_5|$ in this
particular model.
The basis operators are defined with the chirality indices $L$ and $R$
as described in App.~\ref{app:decay}, cf., Eqs.~(\ref{equ:o4}) and
(\ref{equ:o5}).  Note that the transition from Eq.~(\ref{eq:Lageff})
to the basis in Eq.~(\ref{eq:Fierzexa}) can be directly read off from
our tables in App.~\ref{app:topI}. This example corresponds to the
second line of T1-4-ii-2 in Tab.~\ref{Tab:Decom-345}, where
Eq.~(\ref{eq:Fierzexa}) can be read off from the last
column. Therefore, this example serves to illustrate how to use the
tables in our appendix.

The general formula to calculate the half-life time is shown in
\cite{Pas:2000vn}, cf., Eq.~(\ref{eq:Tinv}) in App.~\ref{app:decay},
and is given with the normalised (mass dimensionless) coefficients
$\epsilon_{i} \equiv 2m_{P} \mathcal{C}_{i}/G_{F}^{2}$. The relevant
part is
\begin{align}
(T_{1/2}^{\text{\onbb}})^{-1}
=
G_{2}
\left|
\sum_{i=4}^{5}
\epsilon_{i} \mathcal{M}_{i}
\right|^{2}
\label{eq:T-half-in-example}
\end{align}
in this example.
Here $G_{2}$ [yr$^{-1}$] is a phase space factor,
and $\mathcal{M}_{i}$ are the 
nuclear matrix element parts of the total amplitude,
which are normalised to 
be mass dimensionless.

The half-life Eq.~\eqref{eq:T-half-in-example} 
is dominated by the $\epsilon_{4}$ contribution,
because of $\mathcal{M}_{4} \gg \mathcal{M}_{5}$.\footnote{%
In fact, given our tables in the appendix,
one does not need to rely on the assumption of only one 
dominating NME since the coefficients are explicitely given, 
and one can directly translate the half-life time
into the model constraints for a specific model. In this specific example,
we use this dominance for the sake of simplicity.
}
Using the experimental bounds to $\epsilon_{4}$
listed in Tab.~\ref{Tab:LimitsShort},
we obtain the bounds for the masses of the heavy particles 
as a function of the couplings involved:
\begin{equation}
|\mathcal{C}_4 | = \frac{
\left|
\lambda_{\rm DQ} 
\lambda_{{\rm DQ}\Psi} 
\lambda_{{\rm LQ}\Psi} 
\lambda_{\rm LQ}
\right|
}
{
m_{\rm DQ}^{2} m_{\Psi} m_{\rm LQ}^{2}
}
\frac{1}{32} = 
\frac{G_F^2}{2 m_{P}} \epsilon_{4} 
<\frac{G_F^2}{2 m_{P}} 1.2 \cdot 10^{-8} \, 
\end{equation}
which, assuming that all masses are of order $\Lambda$, 
leads to
\begin{equation}
\Lambda \gtrsim  2.0 
\lambda_{\text{eff}}^{\frac{4}{5}} \, 
\text{ TeV} ,
\end{equation}
where $\lambda_{\text{eff}}\equiv(\lambda_{\rm DQ}\lambda_{{\rm DQ} \Psi}
\lambda_{{\rm LQ} \Psi}\lambda_{\rm LQ})^{(1/4)}$.

While the individual masses involved in these expressions have been
constrained from the searches at colliders, which were discussed
above, the most direct test of this model --- a signal directly
related to the diagram shown in Fig.~\ref{Fig:0nbbExample} --- can be
done at the LHC in the following way.  In the case where $m_{\rm LQ} <
m_{\Psi} < m_{\rm DQ}$, once a diquark $S^{4/3}_{\rm DQ}$ is produced,
it will decay to the VLQ $\Psi^{5/3}$ with a branching ratio of
roughly (neglecting kinematical factors) ${\rm Br}\sim \frac{\Gamma
(S_{\rm DQ} \rightarrow \Psi Q)}{ \Gamma (S_{\rm DQ} \rightarrow \Psi
Q) + \Gamma (S_{\rm DQ} \rightarrow u_{R} u_{R})} \sim
\frac{\lambda_{\rm DQ\Psi}^{2}} {\lambda_{\rm DQ\Psi}^2 + \lambda_{\rm
DQ}^2}$, i.e., ${\rm Br} \sim 1/2$ for $\lambda_{{\rm DQ}
\Psi}=\lambda_{\rm DQ}$, and the VLQ $\Psi$ will then further decay to
the LQ $S^{2/3}_{\rm LQ}$ plus a lepton.  This decay channel will
usually dominate over the 3-body decay $\Psi^{5/3} \to 3 j$ via an
off-shell diquark.  The total signal for this decay chain is then
$e^+e^+jj$.\footnote{ If $m_{\Psi}< m_{\rm LQ}$ both the $3j$ and the
$eej$ final states of the $\Psi$ decay will occur.}  This signal is
the same as the process searched for by the ATLAS \cite{ATLAS:2012ak}
and the CMS collaborations \cite{CMS:PAS-EXO-12-017} in the context of
the LR-symmetric model, and thus the result of the search can already
be used to derive limits on the parameter space of the mediator 
fields, DQ, VLQ, and LQ.  The upper limit on this channel reported in
\cite{CMS:PAS-EXO-12-017} is around 2.5 fb with an assumption of
$m_{W_{R}}=3$ TeV in the LR symmetric model. This bound corresponds
then to roughly $\lambda_{{\rm DQ}\Psi} = \lambda_{\rm DQ} \lsim 0.07$
for $m_{\rm DQ}=3$ TeV (with an assumption on the mass difference
$m_{\rm DQ}-m_{\Psi} \gsim 100$ GeV due to the experimental cuts).
While this limit provides already some interesting constraints on this
example ``model'' shown in Fig.~\ref{Fig:0nbbExample}, a more detailed
analysis is required, before it is ruled out as the dominant mechanism
of \onbb decay process. We expect, of course, that much more stringent
limits will be provided by the forth-coming LHC run with $\sqrt{s}=14$
TeV.

\section{Summary and conclusions}

We have systematically decomposed the \onbb operator with mass
dimension nine ($d=9$), resulting in a tree-level complete list of
possible contributions to \onbb decay. Our main results are summarised
in Tables~\ref{Tab:TopoI} and~\ref{Tab:TopoII}.

Our list encompasses all previously discussed contributions to \onbb
decay and, more interestingly, demonstrates that actually most cases
have not been discussed yet.  The new options typically require not
only fractionally charged fermions, but also exotic bosons.  The
latter can be doubly-charged bileptons, diquarks, or
leptoquarks. For topology~II (cf., right panel of
Fig.~\ref{Fig:0nbbTopologies}), we have also found a possibility with
integer charges and scalars (or vectors) only. In fact, almost all of
the topology~II possibilities have not been discussed in the
literature before as leading contribution to \onbb decay. 

The $d=9$ \onbb decay operator is genuinely suppressed by $1/\Lambda^5$ if
the lightest mediator is heavier than a few GeV.  
On the other hand, the \onbb decay
operator mediated by a light neutrino with right-chiral interactions
(long-range contribution) is weaker suppressed, and therefore, the new
physics scale $\Lambda$ could be quite high, unreachable in current
collider experiments. \onbb decay itself does very likely give the strongest
constraint on such mediators.  In
the short-range case, $\Lambda$ points towards the TeV scale, if a
signal is detected at the next generation \onbb decay experiments, which
means that the mediators leading to \onbb decay may be constrained at the
LHC.  We have therefore focused on the short-range case for a more
detailed analysis.

In order to translate the bound on the half-life
$T_{1/2}^{\text{\onbb}}$ into a bound for the masses and couplings of
a particular model, the Lorentz structure of the effective \onbb decay
operator is important, since each model will be sensitive to a distinct
combination of Nuclear Matrix Elements (NME). For the short-range case
mediated by scalars and a fermion, we have
therefore expanded all models in terms of  the
chiralities of their effective low-energy operators. The results 
are given in tabular form, and can be used to directly translate
new bounds or re-computed NMEs into a mass and coupling limit
for a specific model.
From our lists and the recipes discussed in
the appendix, also the corresponding vector cases  can be derived in a
straightforward manner.

We have also worked out one example which can be tested at the LHC in
greater detail. This example requires a diquark, an exotic
colour-triplet vector-like fermion with electric charge $5/3$, and a
leptoquark.  While the individual mediators can be produced and tested
at the LHC, not all couplings needed for \onbb decay are directly
accessible  in all kinematically possible configurations. 
However, in case the exotic fermion is 
lighter than the diquark, the \onbb decay diagram can be
directly tested by processes with two like-sign leptons and two jets
in the final state.  While our example only serves as a prototype, we
expect that a more systematic study of all short-range \onbb decay
contributions for the LHC is feasible.

In conclusion, a discovery of \onbb\ guarantees physics beyond the 
Standard Model. Whether this new physics is due to the $d=5$ Weinberg 
operator which implies heavy mediators, such as heavy right-handed neutrinos 
in the famous (type I) seesaw mechanism, or some other mechanism,
is an open question. There are many different possibilities to mediate 
\onbb decay without Majorana neutrinos, even at tree level. 
We have discussed that many of these options have in common that the 
mediators should be found at the LHC, or that LHC will provide very 
stringent constraints. Finally, the most interesting case may be that 
\onbb decay is discovered in conflict with neutrino mass bounds from tritium 
endpoint experiments~\cite{Osipowicz:2001sq} or cosmology
\cite{Lesgourgues:2006nd,Hannestad:2010kz,Wong:2011ip} which would 
point towards one of our exotic mechanisms.

\section*{Acknowledgements}

M.H. acknowledges support from the Spanish MICINN grants
FPA2011-22975, MULTIDARK CSD2009-00064 and by the Generalitat
Valenciana grant Prometeo/2009/091 and the EU~Network grant UNILHC
PITN-GA-2009-237920.  F. B. and W. W. acknowledge support from DFG
grants WI 2639/3-1 and WI 2639/4-1. 
T.O acknowledges support from 
Grant-in-Aid for Scientific Research No. 24340044 by
Japan Society for the Promotion of Science.
This work has been also supported by the EU FP7 project 
``In$\nu$isibles'' (Marie Curie Actions,
PITN-GA-2011-289442).

\appendix
\section{From effective Lagrangians to the decay rate}

Our general decomposition of the \onbb decay operator is given in
terms of the quark (and the lepton) currents.  However, \onbb is a
low-energy process in which we must treat hadronic currents ---
neutrons are converted into protons in a nucleus.  Therefore, the
derivation of the decay rate for any model leading to an effective
Lagrangian of the form Eqs.~\eqref{eq:defLR} and \eqref{eps_short},
involves a number of steps.  However, since this derivation has been
studied several times in the literature, we summarise only the
relevant definitions necessary for making contact with the general
Lorentz-invariant description of \cite{Pas:1999fc,Pas:2000vn} in this
appendix.

\subsection{Decay rate}
\label{app:decay}

Although we have already given the effective Lagrangian for 
short-range contributions in Eq.~\eqref{eps_short}, 
here we re-define them with the chiralities:
\begin{align}
\mathcal{L}^{\text{eff}}
=&
\frac{G_{F}^{2}}{2} m_{P}^{-1}
\left[
\sum_{i=1}^{3}
\epsilon_{i}^{\{XY\}Z}
(\mathcal{O}_{i})_{\{XY\}Z}
+
\sum_{i=4}^{5}
\epsilon_{i}^{XY} (\mathcal{O}_{i})_{XY}
\right],
\label{eq:short-w-chirality}
\end{align}
where the effective operators are described as
\begin{align}
(\mathcal{O}_{1})_{\{XY\}Z}
\equiv&
J_{X} J_{Y} j_{Z},
\\
(\mathcal{O}_{2})_{\{XY\}Z}
\equiv&
(J_{X})^{\mu \nu} (J_{Y})_{\mu \nu} j_{Z},
\\
(\mathcal{O}_{3})_{\{XY\}Z}
\equiv&
(J_{X})^{\mu} (J_{Y})_{\mu} j_{Z},
\\
(\mathcal{O}_{4})_{XY}
\equiv&
(J_{X})^{\mu \nu} (J_{Y})_{\mu} (j)_{\nu},
\label{equ:o4}
\\
(\mathcal{O}_{5})_{XY}
\equiv&
J_{X} (J_{Y})^{\mu} (j)_{\mu}.
\label{equ:o5}
\end{align} 
The following formula directly relates the inverse half-life 
with the effective Lagrangian Eq.~\eqref{eq:short-w-chirality}:
\footnote{ The formulae for the long-range contributions are given in
\cite{Pas:1999fc}. Note that, when one decomposes
the amplitudes into a long-range and a short range parts, one assumes
implicitly that there is no new physics with a mass scale similar to
the nuclear Fermi scale, i.e., ${\cal O}(100)$ MeV.}
\begin{align}\label{eq:Tinv}
\left(T^{0\nu\beta\beta}_{1/2}\right)^{-1} 
=& 
G_{1} 
\left|
\sum_{i=1}^{3} 
\epsilon_i \mathcal{M}_i
\right|^2
+
G_{2}
\left|
\sum_{i=4}^{5} 
\epsilon_i \mathcal{M}_i
\right|^2
+
G_{3}
{\rm Re}
\left[
\left(
\sum_{i=1}^{3} 
\epsilon_i \mathcal{M}_i
\right)
\left(
\sum_{i=4}^{5} 
\epsilon_i \mathcal{M}_i
\right)^{*}
\right],
\end{align}
Eq. (\ref{eq:Tinv}) contains the product  
of three distinct factors, $G_{i}$, $\mathcal{M}_{i}$ and 
$\epsilon_{i}$.
Here, $G_{i\in\{1, 2, 3\}}$ are the leptonic phase space integrals, which
can be calculated accurately, e.g., \cite{Doi:1985dx}.
The nuclear Matrix Elements (NME) 
${\cal M}_{i\in\{1\text{-}5\}}$ are different for different short-range 
contributions $\epsilon_{i\in\{1\text{-}5\}}$, detailed definitions 
can be found in \cite{Pas:2000vn}, numerical values are given in 
\cite{Pas:2000vn,Deppisch:2012nb}.
The contribution from the mass mechanism 
can be expressed  as
\begin{align}\label{eq:Tinv2}
\left(T^{0\nu\beta\beta}_{1/2}\right)^{-1} 
=& 
G_{1} 
\left|
\frac{\langle m_{\nu} \rangle}{m_{e}}
\left[
\mathcal{M}_{\rm GT} 
- 
\frac{g_{V}^{2}}{g_{A}^{2}}
\mathcal{M}_{\rm F}
\right]
\right|^{2},
\end{align}
where $\mathcal{M}_{\rm F}$ and $\mathcal{M}_{\rm GT}$ are
the  standard Fermi and Gamow-Teller transition matrix elements,
$g_{V}$ and $g_{A}$ are the vector and axial-vector couplings
of the hadron current, and $m_{e}$ is the mass of electron.%
\footnote{The NME for the mass mechanism have been calculated in a 
number of papers in the literature. Unfortunately, however, some of 
the NME of the general decay rate have so far been calculated only in 
\cite{Pas:1999fc,Pas:2000vn,Deppisch:2012nb}.}

Given the numerical values of $G_{i}$ and $\mathcal{M}_{i}$, 
once the combination of $\epsilon_i$'s for a given model have
been identified (see the tables below), the derivation of 
the limits from $T_{1/2}^{\text{\onbb}}$ becomes straightforward.

\subsection{Fierz and colour-index transformations}
\label{app:fcol}

The general effective Lagrangians given in Eqs.~(\ref{eq:defLR}) and
(\ref{eps_short}) are described with {\it the standard form $J$ of 
the quark current}, which 
(i) is a singlet under the colour $SU(3)_{c}$ 
and 
(ii) takes the bi-linear form $({\bar u} \Gamma d)$ in terms of 
the Lorentz structure, where 
$\Gamma\in\{1\pm\gamma^{5}, \gamma^{\mu} (1 \pm \gamma^{5}),
\gamma^{\mu \nu} (1 \pm \gamma^{5})\}$.
It is suitable for the calculation 
of the hadron transition amplitudes 
of the type $\langle P(p)|({\bar u} \Gamma d) |N(p')\rangle$, to 
describe the conversion of neutrons to protons. 
On the other hand, the decomposed effective operators which 
are listed in Tabs.~\ref{Tab:TopoI} and \ref{Tab:TopoII}
do not take this standard form (except for T-I-1 and T-II-1).
In order to bring them to the standard form shown as 
the effective Lagrangian Eq.~\eqref{eps_short},
we need to transform the Lorentz and the colour indices in the 
effective operators.

First, we discuss the treatment of colour indices.
In the operator decomposition,
we introduce the antisymmetric tensors $\epsilon_{IJK}$ 
and $\epsilon^{IJK}$ ($I,J,K =1\text{-}3$),
the symmetric matrices 
$(T_{\bf 6})_{X}^{IJ}$ and
$(T_{\bf \bar{6}})^{X}_{IJ}$ ($X=1\text{-}6$)
under the exchange of $I$ and $J$,
and the Gell-Mann matrices ${(\lambda^{A})_{I}}^{J}$ 
($A=1\text{-}8$), where
the lower $I$-index is for ${\bf 3}$ representation ($d_{I}$),
while the upper one for ${\bf \bar{3}}$ ($\bar{u}^{I}$).
Here, the matrices $T_{\bf 6}$ and $T_{\bf \bar{6}}$ are explicitly 
defined as
\begin{gather}
(T_{\bf 6})_{1}^{IJ}
=
(T_{{\bf \bar{6}}})^{1}_{IJ}
=
\begin{pmatrix}
1 && \\
&0&\\
&&0
\end{pmatrix},
\qquad
(T_{\bf 6})_{2}^{IJ}
=
(T_{{\bf \bar{6}}})^{2}_{IJ}
=
\begin{pmatrix}
0 &1/\sqrt{2}& \\
1/\sqrt{2}&0&\\
&&0
\end{pmatrix},
\nonumber 
\\
(T_{\bf 6})_{3}^{IJ}
=
(T_{{\bf \bar{6}}})^{3}_{IJ}
=
\begin{pmatrix}
0 && \\
&1&\\
&&0
\end{pmatrix},
\qquad
(T_{\bf 6})_{4}^{IJ}
=
(T_{{\bf \bar{6}}})^{4}_{IJ}
=
\begin{pmatrix}
0 && 1/\sqrt{2} \\
&0&\\
1/\sqrt{2} &&0
\end{pmatrix},
\nonumber
\\
(T_{\bf 6})_{5}^{IJ}
=
(T_{{\bf \bar{6}}})^{5}_{IJ}
=
\begin{pmatrix}
0 && \\
&0& 1/\sqrt{2}\\
&1/\sqrt{2}&0
\end{pmatrix},
\qquad
(T_{\bf 6})_{6}^{IJ}
=
(T_{{\bf \bar{6}}})^{6}_{IJ}
=
\begin{pmatrix}
0 && \\
&0&\\
&&1
\end{pmatrix}.
\end{gather}
The transformation rules relevant to our work are summarised as
\begin{align}
\epsilon^{IJK} \epsilon_{KI'J'}
=&
\delta_{I'}^{I} \delta_{J'}^{J}
-
\delta_{I'}^{J} \delta_{J'}^{I},
\\
(T_{\bf 6})_{X}^{IJ}
(T_{\bf \bar{6}})^{X}_{I'J'}
=&
\frac{1}{2}
\left[
\delta_{I'}^{I} \delta_{J'}^{J}
+
\delta_{I'}^{J} \delta_{J'}^{I}
\right],
\label{eq:T6T6-dd}
\\
{(\lambda^{A})_{I'}}^{I}
{(\lambda^{A})_{J'}}^{J}
=&
-
\frac{2}{3}
\delta_{I'}^{I} \delta_{J'}^{J}
+
2 \delta_{I'}^{J} \delta_{J'}^{I}.
\end{align}

Next, we transform the spinor indices.  General formulas for Fierz
transformations can be found in the literature in many references, see
for example \cite{Dreiner:2008tw} for 2-component spinor
representations.  In some decompositions, we must Fierz-transform all
the six fermions, to which the well-known transformation rules for
four fermions are not applicable.  However, after appropriate
successive transformations with the formulae shown in
\cite{Dreiner:2008tw}, one can reach the standard form.

Here, we demonstrate the procedure of the operator projection
with an operator as an example, which is we saw in
Sec.~\ref{sec:example},
$(\overline{u_{R}} \overline{u_{R}})
(d_{L})(\overline{e_{R}})
(\overline{e_{L}} d_{R})$.
Writing down the operator again with all the indices explicitly,
we have
\begin{align}
\mathcal{O}_{\text{example}}
\equiv&
\left[
(\overline{u_{R}})^{I' a}
(T_{\bf \bar{6}})^{X}_{I'J'}
({u_{R}}^{c})^{J'}_{a}
\right]
\left[
(\overline{{d_{L}}^{c}})_{I}^{b}
(T_{\bf 6})_{X}^{IJ}
({e_{R}}^{c})_{b}
\right]
\left[
(\overline{e_{L}})_{\dot{c}}
(d_{R})_{J}^{\dot{c}}
\right].
\end{align}
Applying Eq.~\eqref{eq:T6T6-dd}, we obtain
the following colour-singlet $\bar{u}d$ combinations,
that however do not form standard quark currents yet:
\begin{align}
\mathcal{O}_{\text{example}}
=&
\frac{1}{2}
\left[
\delta_{I'}^{I} 
\delta_{J'}^{J}
+
\delta_{I'}^{J} 
\delta_{J'}^{I}
\right]
(\overline{u_{R}})^{I' a}
({u_{R}}^{c})^{J'}_{a}
(\overline{{d_{L}}^{c}})_{I}^{b}
({e_{R}}^{c})_{b}
(\overline{e_{L}})_{\dot{c}}
(d_{R})_{J}^{\dot{c}}
\nonumber 
\\
=&
(\overline{u_{R}})^{I a}
(\overline{{d_{L}}^{c}})_{I}^{b}
({u_{R}}^{c})^{J}_{a}
(d_{R})_{J}^{\dot{c}}
(\overline{e_{L}})_{\dot{c}}
({e_{R}}^{c})_{b}.
\label{eq:Operator-Projection-example-Nr1}
\end{align}
In order to obtain the standard form,
we transform also the spinor indices.
The formulae that are necessary for this transformation 
will be shown later. 
\begin{align}
&\text{(RHS) of Eq.~\eqref{eq:Operator-Projection-example-Nr1}}
\nonumber 
\\
=&
\delta_{b}^{d}
\delta_{\dot{f}}^{\dot{c}}
(\overline{u_{R}})^{I a}
(\overline{{d_{L}}^{c}})_{I}^{b}
({u_{R}}^{c})^{J}_{a}
(d_{R})_{J}^{\dot{f}}
(\overline{e_{L}})_{\dot{c}}
({e_{R}}^{c})_{d}
\nonumber 
\\
=&
-
\frac{1}{32}
\left[
J_{L}
(J_{R})^{\mu}
(j)_{\mu}
+
\frac{1}{{\rm i}}
(J_{L})^{\mu \nu}
(J_{R})_{\mu}
(j)_{\nu}
\right].
\label{eq:Operator-Projection-example-Nr2}
\end{align} 
Here, Fierz transformations are carried out in the 2-component
representation (as in \cite{Dreiner:2008tw}),
which is related to the 4-component representation
in the following manner.
We take so-called chiral representation 
for a 4-component spinor, 
i.e., the Lorentz vector matrices $\sigma^{\mu}\equiv (1, \sigma^{a})$
and $\overline{\sigma}^{\mu} \equiv (1, -\sigma^{a})$ 
for 2-component spinors
are introduced, which are given as the components of 
the $\gamma^{\mu}$ matrices as
\begin{align}
\gamma^{\mu} 
= \begin{pmatrix}
   & (\sigma^{\mu})_{a\dot{b}} \\
  (\overline{\sigma}^{\mu})^{\dot{a} b} &
  \end{pmatrix}.
\end{align}
The Lorentz tensor matrices $\sigma^{\mu \nu}$ 
for 2-component spinors, which appear in 
Eq.~\eqref{eq:Operator-Projection-example-Nr2},
are defined with $\sigma^{\mu}$ and $\overline{\sigma}^{\mu}$
as\footnote{Note that 
the definition of 2-component tensor matrices here
is different from \cite{Dreiner:2008tw} 
by an imaginary unit {\rm i}.}
\begin{align}
{(\sigma^{\mu \nu})_{a}}^{b}
\equiv&
\frac{1}{4}
\left[
(\sigma^{\mu})_{a\dot{a}}
(\overline{\sigma}^{\nu})^{\dot{a} b}
-
(\sigma^{\nu})_{a\dot{a}}
(\overline{\sigma}^{\mu})^{\dot{a} b}
\right],
\\
{(\overline{\sigma}^{\mu \nu})^{\dot{a}}}_{\dot{b}}
\equiv&
\frac{1}{4}
\left[
(\overline{\sigma}^{\mu})^{\dot{a} a}
(\sigma^{\nu})_{a \dot{b}}
-
(\overline{\sigma}^{\nu})^{\dot{a} a}
(\sigma^{\mu})_{a \dot{b}}
\right],
\end{align}
which are related to the matrices $\gamma^{\mu \nu}$ 
for 4-component spinors as
\begin{align}
\gamma^{\mu \nu}
=
2 {\rm i}
\begin{pmatrix}
{(\sigma^{\mu \nu})_{a}}^{b} & \\
& 
{(\overline{\sigma}^{\mu \nu})^{\dot{a}}}_{\dot{b}}
\end{pmatrix}.
\end{align}
The relations between the currents in the 2-component representation 
and those (defined at Eq.~\eqref{eq:CurrLR})
in the 4-component representation
are explicitly described as
\begin{align}
(\overline{u_{R}})^{I a} 
(d_{L})_{I a} 
=& 
\frac{1}{2} J_{L},
\\
(\overline{u_{R}})^{I a}
(\sigma^{\mu})_{a \dot{b}} 
(d_{R})_{I}^{\dot{b}}
=&
\frac{1}{2} (J_{R})^{\mu}, 
\\
(\overline{u_{R}})^{I a}
{(\sigma^{\mu \nu})_{a}}^{b}
(d_{L})_{I b}
=&
\frac{1}{4 {\rm i}} (J_{L})^{\mu \nu},
\\
(\overline{e_{L}})_{\dot{a}}
(\overline{\sigma}^{\mu})^{\dot{a} b}
({e_{R}}^{c})_{b}
=&
\frac{1}{2} (j_{L})^{\mu}.
\end{align}
In the steps of the transformation shown 
in Eq.~\eqref{eq:Operator-Projection-example-Nr2},
we applied the following formulae of Fierz-transformation, 
\begin{align}
\delta_{a}^{d} \delta_{\dot{b}}^{\dot{c}}
=&
\frac{1}{2}
(\sigma^{\mu})_{a \dot{b}}
(\overline{\sigma}^{\mu})^{\dot{c} d},
\\
\epsilon_{ac} \epsilon^{bd}
=&
-
\frac{1}{2}
\left[
\delta_{a}^{b} \delta_{c}^{d}
+
{(\sigma^{\rho \sigma})_{a}}^{b}
{(\sigma_{\rho \sigma})_{c}}^{d}
\right],
\end{align}
and the nature of the sigma matrices,
\begin{align}
{(\sigma^{\mu \nu})_{a}}^{b} (\sigma^{\rho})_{b \dot{b}}
=&
\frac{1}{2}
\left[
(\sigma^{\mu})_{a \dot{b}}
g^{\nu \rho}
-
(\sigma^{\nu})_{a \dot{b}}
g^{\mu \rho}
+
{\rm i} 
\epsilon^{\mu \nu \rho \sigma}
(\sigma_{\sigma})_{a \dot{b}}
\right],
\\
{\rm i} \epsilon^{\mu \nu \rho \sigma} 
{(\sigma_{\rho \sigma})_{a}}^{b}
=&
-2
{(\sigma^{\mu \nu})_{a}}^{b}.
\end{align}

\subsection{Full decompositions for the short-range 
scalar-mediated topology I}
\label{app:topI}

In case of topology I the intermediate states contain a fermion, 
leading to a different treatment in case of long-range and short-range 
contributions.  For the long-range part of the amplitude, in 
order to derive interesting constraints, one has to pick out the 
neutrino momentum $q_{\nu}$ from the propagator, 
because of $q_{\nu} \gg m_{\nu}$.
On the other hand, in the short-range case, 
the relevant amplitude must take the mass part $m_{\psi}$ 
in the propagator of the fermion mediator $\psi$,
because of $q_{\psi} \ll m_{\psi}$. 
Thus, not all possibilities to assign chiralities of the six 
outer fermions lead to interesting models. 
When the decomposition is symbolically written as $(a b)(c)(d)(e f)$ 
where each parenthesis corresponds to each vertex $\text{v}_{i}$ in 
the left panel of Fig.~\ref{Fig:0nbbTopologies},
i.e., $(ab)$ corresponds to the outer fermions on the vertex 
$\text{v}_{1}$, and $(c)$ does to $\text{v}_{2}$, and so on,
in order to pick out $m_{\psi}$ from the propagator, 
only the combinations $(\overline{{c_L}^{c}})(d_L)$ and 
$(\overline{{c_R}^{c}})(d_R)$ need to be 
considered. Note, that $\overline{a_R}\equiv \overline{a}P_L$.
Since we concentrate on the SFS case, in which
both the mediators between the vertices $\text{v}_{1}$ and $\text{v}_{2}$
and that between $\text{v}_{3}$ and $\text{v}_{4}$ are scalars, 
the chirality structures of the outer fermions on 
$\text{v}_{1}$ and $\text{v}_{4}$ are also restricted
to be
$(ab) \in \{(\overline{{a_{L}}^{c}}b_{L}),(\overline{{a_{R}}^{c}}b_{R})\}$
and 
$(ef) \in \{
(\overline{{e_{L}}^{c}}f_{L}),
(\overline{{e_{R}}^{c}}f_{R})\}$.
At this stage, there are eight choices 
for the combinations of chiralities of the six outer fermions.
However, some of them are not generated from the 
$d=9$ SM gauge invariant operators
which are listed in \cite{Babu:2001ex,deGouvea:2007xp}.
We exclude the chirality choices which require an additional Higgs 
doublet(s) (i.e., those originated from the operators of $d>9$),
and list all the possibilities inspired from
the $d=9$ SM gauge invariant operators.

The results are summarised in Tabs.~\ref{Tab:Decom-1-i}-\ref{Tab:Decom-345}. 
The id-numbers indicated in the column ``\#'' correspond
to those in Tab.~\ref{Tab:TopoI}.
One can find the chirality choices explicitly in the column ``Operators''.
The id-numbers in the column ``BL'' tells the 
correspondence to the $d=9$ SM gauge invariant operators
listed by Babu and Leung~\cite{Babu:2001ex}.
The SM charges of all possible mediators are fully identified in
the column ``Mediators''.\footnote{%
Here, the charges are fixed, following the charge flow 
$\text{v}_{1} \leftarrow \text{v}_{2} 
\leftarrow \text{v}_{3} \leftarrow \text{v}_{4}$.
}
The basis operators, cf. Eq.~(\ref{eps_short}), resulting 
after Fierz and the colour-index transformations
are given in the column ``Basis op.''. 
Some of the decompositions appear necessarily with different ones 
at the same time. 
These associated decompositions are listed in the column ``Appears with''. 

The use and application of our tables is illustrated with the example
in Sec.~\ref{sec:example}, see discussion after
Eq.~(\ref{eq:Lageff}). Once a specific model is chosen, the basis
operators with the corresponding coefficients can be directly read off
from the table. These basis operators are directly related with the
NMEs, i.e., the table lists which NMEs specific models are testing.
As a consequence, the lifetime bounds can be translated into
constraints on masses and couplings for a specific model --- as it is
illustrated in our example.

\begin{table}[tp]
\begin{center}
{\small
\begin{tabular}{ccccccc}
\hline \hline 
&
&
&
\multicolumn{3}{c}{Mediators $(SU(3)_{c}, SU(2)_{L})_{U(1)_{Y}}$}
&
\\
\# & Operators & BL
& $S$ & $\psi$ & $S'$ & Basis op.
\\ 
\hline
1-i
&$(\overline{u_{L}} d_{R}) (\overline{e_{L}})
 (\overline{e_{L}}) (\overline{u_{L}} d_{R})$
& \#11
& $({\bf 1},{\bf 2})_{+1/2}$
& $({\bf 1}, {\bf 1})_{0}$
& $({\bf 1},{\bf 2})_{-1/2}$
& $\frac{1}{8} (\mathcal{O}_{1})_{\{RR \} R}$ 
\\
&
&
& $({\bf 1},{\bf 2})_{+1/2}$
& $({\bf 1}, {\bf 3})_{0}$
& $({\bf 1},{\bf 2})_{-1/2}$
& s.a.a
\\
&
 
&
& $({\bf 8},{\bf 2})_{+1/2}$
& $({\bf 8}, {\bf 1})_{0}$
& $({\bf 8},{\bf 2})_{-1/2}$
&
$ - \frac{5}{24} (\mathcal{O}_{1})_{\{RR\}R}$ 
$ - \frac{1}{32} (\mathcal{O}_{2})_{\{RR\}R}$  
\\
&
&
& $({\bf 8},{\bf 2})_{+1/2}$
& $({\bf 8}, {\bf 3})_{0}$
& $({\bf 8},{\bf 2})_{-1/2}$
& s.a.a
\\
& 
$(\overline{u_{L}} d_{R}) (\overline{e_{L}})
 (\overline{e_{L}}) (\overline{u_{R}} d_{L})$
& \#14
& $({\bf 1},{\bf 2})_{+1/2}$
& $({\bf 1}, {\bf 1})_{0}$
& $({\bf 1},{\bf 2})_{-1/2}$
& $\frac{1}{8} (\mathcal{O}_{1})_{\{LR \} R}$ 
\\
&
&
& $({\bf 1},{\bf 2})_{+1/2}$
& $({\bf 1}, {\bf 3})_{0}$
& $({\bf 1},{\bf 2})_{-1/2}$
& s.a.a
\\ 
& 
& 
& $({\bf 8},{\bf 2})_{+1/2}$
& $({\bf 8}, {\bf 1})_{0}$
& $({\bf 8},{\bf 2})_{-1/2}$
&
  $ -\frac{1}{12} (\mathcal{O}_{1})_{\{LR\}R} $
  $ - \frac{1}{8} (\mathcal{O}_{3})_{\{LR\}R} $ 
\\
&
&
& $({\bf 8},{\bf 2})_{+1/2}$
& $({\bf 8}, {\bf 3})_{0}$
& $({\bf 8},{\bf 2})_{-1/2}$
& s.a.a
\\ 
&
$(\overline{u_{R}} d_{L}) (\overline{e_{L}})
 (\overline{e_{L}}) (\overline{u_{R}} d_{L})$
& \#12
& $({\bf 1},{\bf 2})_{+1/2}$
& $({\bf 1}, {\bf 1})_{0}$
& $({\bf 1},{\bf 2})_{-1/2}$
& $\frac{1}{8} (\mathcal{O}_{1})_{\{LL \} R}$  
\\
&
&
& $({\bf 1},{\bf 2})_{+1/2}$
& $({\bf 1}, {\bf 3})_{0}$
& $({\bf 1},{\bf 2})_{-1/2}$
& s.a.a
\\
&
& 
& $({\bf 8},{\bf 2})_{+1/2}$
& $({\bf 8}, {\bf 1})_{0}$
& $({\bf 8},{\bf 2})_{-1/2}$
& $ - \frac{5}{24} (\mathcal{O}_{1})_{\{LL\}R} $
  $ - \frac{1}{32} (\mathcal{O}_{2})_{\{LL\} R} $
\\
&
&
& $({\bf 8},{\bf 2})_{+1/2}$
& $({\bf 8}, {\bf 3})_{0}$
& $({\bf 8},{\bf 2})_{-1/2}$
& s.a.a
\\
\hline
1-ii-a 
&$(\overline{u_{L}} d_{R}) (\overline{u_{L}})
 (d_{R}) (\overline{e_{L}} \overline{e_{L}})$
& \#11
& $({\bf 1},{\bf 2})_{+1/2}$
& $({\bf 3}, {\bf 3})_{+2/3}$
& $({\bf 1},{\bf 3})_{+1}$
& $\frac{1}{8} (\mathcal{O}_{1})_{\{RR \} R}$  
\\
&
& 
& $({\bf 8},{\bf 2})_{+1/2}$
& $({\bf 3}, {\bf 3})_{+2/3}$
& $({\bf 1},{\bf 3})_{+1}$
& $-\frac{5}{24} (\mathcal{O}_{1})_{\{RR\} R} $ 
  $-\frac{1}{32} (\mathcal{O}_{2})_{\{RR\} R} $
\\
&$(\overline{u_{L}} d_{R}) (\overline{u_{R}})
 (d_{L}) (\overline{e_{L}} \overline{e_{L}})$
& \#14
& $({\bf 1},{\bf 2})_{+1/2}$
& $({\bf 3}, {\bf 2})_{+7/6}$
& $({\bf 1},{\bf 3})_{+1}$
& $\frac{1}{8} (\mathcal{O}_{1})_{\{LR \} R}$  
\\
&
& 
& $({\bf 8},{\bf 2})_{+1/2}$
& $({\bf 3}, {\bf 2})_{+7/6}$
& $({\bf 1},{\bf 3})_{+1}$
& $ - \frac{1}{12} (\mathcal{O}_{1})_{\{LR\}R}$
  $ - \frac{1}{8} (\mathcal{O}_{3})_{\{LR\} R}$
\\
&$(\overline{u_{R}} d_{L}) (\overline{u_{L}})
 (d_{R}) (\overline{e_{L}} \overline{e_{L}})$
& \#14
& $({\bf 1},{\bf 2})_{+1/2}$
& $({\bf 3}, {\bf 3})_{+2/3}$
& $({\bf 1},{\bf 3})_{+1}$
& $\frac{1}{8} (\mathcal{O}_{1})_{\{LR \} R}$  
\\
&
&
& $({\bf 8},{\bf 2})_{+1/2}$
& $({\bf 3}, {\bf 3})_{+2/3}$
& $({\bf 1},{\bf 3})_{+1}$
& $-\frac{1}{12} (\mathcal{O}_{1})_{\{LR\}R}$ 
  $- \frac{1}{8} (\mathcal{O}_{3})_{\{LR\}R}$
\\
&$(\overline{u_{R}} d_{L}) (\overline{u_{R}})
 (d_{L}) (\overline{e_{L}} \overline{e_{L}})$
& \#12
& $({\bf 1},{\bf 2})_{+1/2}$
& $({\bf 3}, {\bf 2})_{+7/6}$
& $({\bf 1},{\bf 3})_{+1}$
& $\frac{1}{8} (\mathcal{O}_{1})_{\{LL \} R}$ 
\\
&
&
& $({\bf 8},{\bf 2})_{+1/2}$
& $({\bf 3}, {\bf 2})_{+7/6}$
& $({\bf 1},{\bf 3})_{+1}$
& $-\frac{5}{24} (\mathcal{O}_{1})_{\{LL\}R} $
  $- \frac{1}{32} (\mathcal{O}_{2})_{\{LL\}R}$
\\
\hline
1-ii-b 
&$(\overline{u_{L}} d_{R}) 
 (d_{L})
 (\overline{u_{R}})
 (\overline{e_{L}} \overline{e_{L}})$
& \#14
& $({\bf 1},{\bf 2})_{+1/2}$
& $(\overline{\bf 3}, {\bf 3})_{+1/3}$
& $({\bf 1},{\bf 3})_{+1}$
&  $\frac{1}{8} (\mathcal{O}_{1})_{\{LR \} R}$ 
\\
&
& 
& $({\bf 8},{\bf 2})_{+1/2}$
& $(\overline{\bf 3}, {\bf 3})_{+1/3}$
& $({\bf 1},{\bf 3})_{+1}$
& $ - \frac{1}{12} (\mathcal{O}_{1})_{\{LR\}R}$  
  $ - \frac{1}{8} (\mathcal{O}_{3})_{\{LR\}R}$
\\
&$(\overline{u_{L}} d_{R}) 
 (d_{R})
 (\overline{u_{L}})
 (\overline{e_{L}} \overline{e_{L}})$
& \#11
& $({\bf 1},{\bf 2})_{+1/2}$
& $(\overline{\bf 3}, {\bf 2})_{+5/6}$
& $({\bf 1},{\bf 3})_{+1}$
&  $\frac{1}{8} (\mathcal{O}_{1})_{\{RR \} R}$  
\\
&
&
& $({\bf 8},{\bf 2})_{+1/2}$
& $(\overline{\bf 3}, {\bf 2})_{+5/6}$
& $({\bf 1},{\bf 3})_{+1}$
& $ - \frac{5}{24} (\mathcal{O}_{1})_{\{RR\}R} $
  $ - \frac{1}{32} (\mathcal{O}_{2})_{\{RR\}R} $
\\
&$(\overline{u_{R}} d_{L}) 
 (d_{L})
 (\overline{u_{R}})
 (\overline{e_{L}} \overline{e_{L}})$
& \#12
& $({\bf 1},{\bf 2})_{+1/2}$
& $(\overline{\bf 3}, {\bf 3})_{+1/3}$
& $({\bf 1},{\bf 3})_{+1}$
&  $\frac{1}{8} (\mathcal{O}_{1})_{\{LL \} R}$ 
\\
&
& 
& $({\bf 8},{\bf 2})_{+1/2}$
& $(\overline{\bf 3}, {\bf 3})_{+1/3}$
& $({\bf 1},{\bf 3})_{+1}$
& $- \frac{5}{24} (\mathcal{O}_{1})_{\{LL\}R} $
  $- \frac{1}{32} (\mathcal{O}_{2})_{\{LL\}R} $
\\
&$(\overline{u_{R}} d_{L}) 
 (d_{R})
 (\overline{u_{L}})
 (\overline{e_{L}} \overline{e_{L}})$
& \#14
& $({\bf 1},{\bf 2})_{+1/2}$
& $(\overline{\bf 3}, {\bf 2})_{+5/6}$
& $({\bf 1},{\bf 3})_{+1}$
& $\frac{1}{8} (\mathcal{O}_{1})_{\{LR \} R}$ 
\\
&
&
& $({\bf 8},{\bf 2})_{+1/2}$
& $(\overline{\bf 3}, {\bf 2})_{+5/6}$
& $({\bf 1},{\bf 3})_{+1}$
& $-\frac{1}{12} (\mathcal{O}_{1})_{\{LR\}R}$  
  $- \frac{1}{8} (\mathcal{O}_{3})_{\{LR\}R}$
\\
\hline \hline
\end{tabular}
} 
\end{center}
\caption{\it The results of decomposition and projection of the
operators categorised to \#1.  We also show the ID-number of the
lepton number violating operators listed by Babu and Leung
(BL)~\cite{Babu:2001ex} (see also Ref.~\cite{deGouvea:2007xp}). 
For T-I-i-1, there are only three independent choices of chiralities.
Here, we assume that the effective operators are originated from the
SM gauge ($SU(3)_{c} \times SU(2)_{L} \times U(1)_{Y}$) 
invariant $d=9$ operators.  Note that the scalar
mediators $S$ and $S'$ of \#1-i-(2) take interactions with different
combinations of quarks, although their SM charges are the same. 
The abbreviation ``s.a.a'' in ``Basis op.'' column 
means ``same as above''. The hypercharge $Y$ is defined
as $Y \equiv Q_{\rm em} - I_3$.
}
\label{Tab:Decom-1-i}
\end{table}

\begin{table}[tp]
\begin{center}
\small{
\begin{tabular}{ccccccp{2.5cm}c}
\hline \hline
&
&
&
\multicolumn{3}{c}{Mediators $(SU(3)_{c}, SU(2)_{L})_{U(1)_{Y}}$}
&
& Appears
\\ 
\# & Operators 
& BL
& $S$ & $\psi$ & $S'$ & Basis op.
& with
\\ 
\hline
2-i-a 
&$(\overline{u_{L}} d_{R}) 
 (d_{R})
 (\overline{e_{L}})
 (\overline{u_{L}} \overline{e_{L}})$
& \#11
& $({\bf 1},{\bf 2})_{+1/2}$
& $(\overline{\bf 3}, {\bf 2})_{+5/6}$
& $(\overline{\bf 3},{\bf 1})_{+1/3}$
& $ - \frac{1}{16} (\mathcal{O}_{1})_{\{ RR \}R}$
\\
&
&
& $({\bf 1},{\bf 2})_{+1/2}$
& $(\overline{\bf 3}, {\bf 2})_{+5/6}$
& $(\overline{\bf 3},{\bf 3})_{+1/3}$
& s.a.a
\\
&
&
& $({\bf 8},{\bf 2})_{+1/2}$
& $(\overline{\bf 3}, {\bf 2})_{+5/6}$
& $(\overline{\bf 3},{\bf 1})_{+1/3}$
& $\frac{5}{48} (\mathcal{O}_{1})_{\{RR\}R} $ 
  $ + \frac{1}{64} (\mathcal{O}_{2})_{\{RR\}R}$ 
\\
&
&
& $({\bf 8},{\bf 2})_{+1/2}$
& $(\overline{\bf 3}, {\bf 2})_{+5/6}$
& $(\overline{\bf 3},{\bf 3})_{+1/3}$
& s.a.a
\\
&$(\overline{u_{L}} d_{R}) 
 (d_{R})
 (\overline{e_{L}})
 (\overline{u_{R}} \overline{e_{R}})$
& \#19
& $({\bf 1},{\bf 2})_{+1/2}$
& $(\overline{\bf 3}, {\bf 2})_{+5/6}$
& $(\overline{\bf 3},{\bf 1})_{+1/3}$
& $\frac{1}{16} (\mathcal{O}_{5})_{RR} $
\\
&
& 
& $({\bf 8},{\bf 2})_{+1/2}$
& $(\overline{\bf 3}, {\bf 2})_{+5/6}$
& $(\overline{\bf 3},{\bf 1})_{+1/3}$
& $ - \frac{1}{16{\rm i}} (\mathcal{O}_{4})_{RR}$
  $ - \frac{5}{48} (\mathcal{O}_{5})_{RR} $
\\
&$(\overline{u_{R}} d_{L}) 
 (d_{R})
 (\overline{e_{L}})
 (\overline{u_{L}} \overline{e_{L}})$
& \#14
& $({\bf 1},{\bf 2})_{+1/2}$
& $(\overline{\bf 3}, {\bf 2})_{+5/6}$
& $(\overline{\bf 3},{\bf 1})_{+1/3}$
& $- \frac{1}{16} (\mathcal{O}_{1})_{\{ LR \} R} $
\\
&
&
& $({\bf 1},{\bf 2})_{+1/2}$
& $(\overline{\bf 3}, {\bf 2})_{+5/6}$
& $(\overline{\bf 3},{\bf 3})_{+1/3}$
& s.a.a
\\
&
&
& $({\bf 8},{\bf 2})_{+1/2}$
& $(\overline{\bf 3}, {\bf 2})_{+5/6}$
& $(\overline{\bf 3},{\bf 1})_{+1/3}$
& $ \frac{1}{24} (\mathcal{O}_{1})_{\{LR\}R} $
  $ + \frac{1}{16} (\mathcal{O}_{3})_{\{LR\}R}$
\\
&
&
& $({\bf 8},{\bf 2})_{+1/2}$
& $(\overline{\bf 3}, {\bf 2})_{+5/6}$
& $(\overline{\bf 3},{\bf 3})_{+1/3}$
& s.a.a
\\
&$(\overline{u_{R}} d_{L}) 
 (d_{R})
 (\overline{e_{L}})
 (\overline{u_{R}} \overline{e_{R}})$
& \#20
& $({\bf 1},{\bf 2})_{+1/2}$
& $(\overline{\bf 3}, {\bf 2})_{+5/6}$
& $(\overline{\bf 3},{\bf 1})_{+1/3}$
& $ \frac{1}{16} (\mathcal{O}_{5})_{LR}$
\\
&
& 
& $({\bf 8},{\bf 2})_{+1/2}$
& $(\overline{\bf 3}, {\bf 2})_{+5/6}$
& $(\overline{\bf 3},{\bf 1})_{+1/3}$
& $  \frac{1}{16{\rm i}} (\mathcal{O}_{4})_{LR} $
  $ - \frac{5}{48} (\mathcal{O}_{5})_{LR}  $
\\ 
\hline 
2-i-b 
&$(\overline{u_{L}} d_{R}) 
      (\overline{e_{L}})
     (d_{R})
 (\overline{u_{L}} \overline{e_{L}})$
& \#11
& $({\bf 1},{\bf 2})_{+1/2}$
& $({\bf 1}, {\bf 1})_{0}$
& $(\overline{\bf 3},{\bf 1})_{+1/3}$
& $ -  \frac{1}{16} (\mathcal{O}_{1})_{\{ RR \} R} $
& 1-i \& 5-i
\\
&
&
& $({\bf 1},{\bf 2})_{+1/2}$
& $({\bf 1}, {\bf 3})_{0}$
& $(\overline{\bf 3},{\bf 3})_{+1/3}$
& s.a.a
& 1-i \& 5-i
\\
&
& 
& $({\bf 8},{\bf 2})_{+1/2}$
& $({\bf 8}, {\bf 1})_{0}$
& $(\overline{\bf 3},{\bf 1})_{+1/3}$
& $ \frac{5}{48}  (\mathcal{O}_{1})_{\{RR\}R} $ 
  $ + \frac{1}{64}  (\mathcal{O}_{2})_{\{RR\}R}$ 
& 1-i \& 5-i
\\
&
&
& $({\bf 8},{\bf 2})_{+1/2}$
& $({\bf 8}, {\bf 3})_{0}$
& $(\overline{\bf 3},{\bf 3})_{+1/3}$
& s.a.a 
&1-i \& 5-i
\\
&$(\overline{u_{L}} d_{R}) 
      (\overline{e_{L}})
 (d_{R})
 (\overline{u_{R}} \overline{e_{R}})$
& \#19
& $({\bf 1},{\bf 2})_{+1/2}$
& $({\bf 1}, {\bf 1})_{0}$
& $(\overline{\bf 3},{\bf 1})_{+1/3}$
& $ \frac{1}{16} (\mathcal{O}_{5})_{\{ RR \}} $
& 1-i \& 5-i
\\
&
&
& $({\bf 8},{\bf 2})_{+1/2}$
& $({\bf 8}, {\bf 1})_{0}$
& $(\overline{\bf 3},{\bf 1})_{+1/3}$
& $- \frac{1}{16{\rm i}} (\mathcal{O}_{4})_{RR}$
  $- \frac{5}{48} (\mathcal{O}_{5})_{RR} 
  $
& 1-i \& 5-i
\\
&$(\overline{u_{R}} d_{L}) 
 (\overline{e_{L}})
 (d_{R})
 (\overline{u_{L}} \overline{e_{L}})$
& \#14
& $({\bf 1},{\bf 2})_{+1/2}$
& $({\bf 1}, {\bf 1})_{0}$
& $(\overline{\bf 3},{\bf 1})_{+1/3}$
& $- \frac{1}{16} (\mathcal{O}_{1})_{\{ LR \} R}$
& 1-i \& 5-i
\\
&
&
& $({\bf 1},{\bf 2})_{+1/2}$
& $({\bf 1}, {\bf 3})_{0}$
& $(\overline{\bf 3},{\bf 3})_{+1/3}$
& s.a.a
& 1-i \& 5-i
\\
&
&
& $({\bf 8},{\bf 2})_{+1/2}$
& $({\bf 8}, {\bf 1})_{0}$
& $(\overline{\bf 3},{\bf 1})_{+1/3}$
& $ \frac{1}{24} (\mathcal{O}_{1})_{\{LR\}R} $
  $ + \frac{1}{16} (\mathcal{O}_{3})_{\{LR\}R}$
& 1-i \& 5-i
\\
&
&
& $({\bf 8},{\bf 2})_{+1/2}$
& $({\bf 8}, {\bf 3})_{0}$
& $(\overline{\bf 3},{\bf 3})_{+1/3}$
& s.a.a
& 1-i \& 5-i
\\
&$(\overline{u_{R}} d_{L}) 
 (\overline{e_{L}})
 (d_{R})
 (\overline{u_{R}} \overline{e_{R}})$
& \#20
& $({\bf 1},{\bf 2})_{+1/2}$
& $({\bf 1}, {\bf 1})_{0}$
& $(\overline{\bf 3},{\bf 1})_{+1/3}$
& $ \frac{1}{16} (\mathcal{O}_{5})_{LR} $
& 1-i \& 5-i
\\
&
&
& $({\bf 8},{\bf 2})_{+1/2}$
& $({\bf 8}, {\bf 1})_{0}$
& $(\overline{\bf 3},{\bf 1})_{+1/3}$
& $ \frac{1}{16{\rm i}}(\mathcal{O}_{4})_{LR} $
  $ - \frac{5}{48} (\mathcal{O}_{5})_{LR}$
& 1-i \& 5-i
\\ 
\hline \hline
\end{tabular}
} 
\end{center}
\caption{\it Decomposition \#2-i. 
In case of decomposition \#2-i-b, we will have
not only \#2-i-b, but also \#1-i and \#5-i.}
\end{table}

\begin{table}[tp]
\begin{center}
\small{
\begin{tabular}{ccccccp{2.5cm}c}
\hline \hline
&
&
&
\multicolumn{3}{c}{Mediators $(SU(3)_{c}, SU(2)_{L})_{U(1)_{Y}}$}
&
& Appears
\\ 
\# & Operators 
& BL
& $S$ & $\psi$ & $S'$ & Basis op.
& with
\\ 
\hline
2-ii-a &
$(\overline{u_{L}} d_{R}) (\overline{u_{L}}) 
(\overline{e_{L}})(d_{R} \overline{e_{L}})$
& \#11
& $({\bf 1}, {\bf 2})_{+1/2}$
& $({\bf 3}, {\bf 3})_{+2/3}$
& $({\bf 3}, {\bf 2})_{+1/6}$
& $ -\frac{1}{16} (\mathcal{O}_{1})_{\{ RR \} R} $
\\
&
& 
& $({\bf 8}, {\bf 2})_{+1/2}$
& $({\bf 3}, {\bf 3})_{+2/3}$
& $({\bf 3}, {\bf 2})_{+1/6}$
& $\frac{5}{48}  (\mathcal{O}_{1})_{\{RR\}R} $
  $ + \frac{1}{64} (\mathcal{O}_{2})_{\{RR\}R}$
\\
&
$(\overline{u_{L}} d_{R}) (\overline{u_{R}}) 
(\overline{e_{R}}) (d_{R} \overline{e_{L}})$
& \#19
& $({\bf 1}, {\bf 2})_{+1/2}$
& $({\bf 3}, {\bf 2})_{+7/6}$ 
& $({\bf 3}, {\bf 2})_{+1/6}$
& $ \frac{1}{16} (\mathcal{O}_{5})_{RR} $
\\
&
&
& $({\bf 8}, {\bf 2})_{+1/2}$
& $({\bf 3}, {\bf 2})_{+7/6}$ 
& $({\bf 3}, {\bf 2})_{+1/6}$
& $- \frac{1}{16{\rm i}} (\mathcal{O}_{4})_{RR} $ 
  $-\frac{5}{48} (\mathcal{O}_{5})_{RR}$  
\\
&
$(\overline{u_{R}} d_{L}) (\overline{u_{L}})
(\overline{e_{L}}) (d_{R} \overline{e_{L}})$
& \#14
& $({\bf 1}, {\bf 2})_{+1/2}$
& $({\bf 3}, {\bf 3})_{+2/3}$
& $({\bf 3}, {\bf 2})_{+1/6}$
& $ -\frac{1}{16} (\mathcal{O}_{1})_{\{ LR \} R} $
\\
&
& 
& $({\bf 8}, {\bf 2})_{+1/2}$
& $({\bf 3}, {\bf 3})_{+2/3}$
& $({\bf 3}, {\bf 2})_{+1/6}$
& $ \frac{1}{24} (\mathcal{O}_{1})_{\{LR\}R}$  
  $ + 
    \frac{1}{16} (\mathcal{O}_{3})_{\{LR\}R}$
\\
&
$(\overline{u_{R}} d_{L}) (\overline{u_{R}})
(\overline{e_{R}}) (d_{R} \overline{e_{L}})$
& \#20
& $({\bf 1}, {\bf 2})_{+1/2}$
& $({\bf 3}, {\bf 2})_{+7/6}$
& $({\bf 3}, {\bf 2})_{+1/6}$
& $ \frac{1}{16} (\mathcal{O}_{5})_{LR}$
\\
&
&
& $({\bf 8}, {\bf 2})_{+1/2}$
& $({\bf 3}, {\bf 2})_{+7/6}$
& $({\bf 3}, {\bf 2})_{+1/6}$
& $ \frac{1}{16{\rm i}} (\mathcal{O}_{4})_{LR} $
  $ - \frac{5}{48} (\mathcal{O}_{5})_{LR} $
\\
\hline
2-ii-b 
&
$(\overline{u_{L}} d_{R}) 
(\overline{e_{L}}) (\overline{u_{L}}) 
(d_{R} \overline{e_{L}})$
& \#11
& $({\bf 1}, {\bf 2})_{+1/2}$
& $({\bf 1}, {\bf 1})_{0}$
& $({\bf 3}, {\bf 2})_{+1/6}$
& $ -\frac{1}{16} (\mathcal{O}_{1})_{\{ RR \}R} $
& 1-i \& 4-i
\\
&
&
& $({\bf 1}, {\bf 2})_{+1/2}$
& $({\bf 1}, {\bf 3})_{0}$
& $({\bf 3}, {\bf 2})_{+1/6}$
& s.a.a 
& 1-i \& 4-i
\\
&
& 
& $({\bf 8}, {\bf 2})_{+1/2}$
& $({\bf 8}, {\bf 1})_{0}$
& $({\bf 3}, {\bf 2})_{+1/6}$
& $\frac{5}{48} (\mathcal{O}_{1})_{\{RR\}R} $
  $ +  \frac{1}{64} (\mathcal{O}_{2})_{\{RR\}R}$
& 1-i \& 4-i
\\
&
&
& $({\bf 8}, {\bf 2})_{+1/2}$
& $({\bf 8}, {\bf 3})_{0}$
& $({\bf 3}, {\bf 2})_{+1/6}$
& s.a.a
& 1-i \& 4-i
\\
&
$(\overline{u_{L}} d_{R}) 
 (\overline{e_{R}}) (\overline{u_{R}}) 
 (d_{R} \overline{e_{L}})$
& \#19
& $({\bf 1}, {\bf 2})_{+1/2}$
& $({\bf 1}, {\bf 2})_{-1/2}$
& $({\bf 3}, {\bf 2})_{+1/6}$
& $ \frac{1}{16} (\mathcal{O}_{5})_{RR} $ 
\\
&
& 
& $({\bf 8}, {\bf 2})_{+1/2}$
& $({\bf 8}, {\bf 2})_{-1/2}$
& $({\bf 3}, {\bf 2})_{+1/6}$
& $ - \frac{1}{16{\rm i}} (\mathcal{O}_{4})_{RR} $
  $ -\frac{5}{48} (\mathcal{O}_{5})_{RR}  $
\\
&
$(\overline{u_{R}} d_{L}) 
(\overline{e_{L}}) (\overline{u_{L}}) 
(d_{R} \overline{e_{L}})$
& \#14
& $({\bf 1}, {\bf 2})_{+1/2}$
& $({\bf 1}, {\bf 1})_{0}$
& $({\bf 3}, {\bf 2})_{+1/6}$
& $ -\frac{1}{16} (\mathcal{O}_{1})_{\{ LR \} R} $
& 1-i \& 4-i
\\
&
&
& $({\bf 1}, {\bf 2})_{+1/2}$
& $({\bf 1}, {\bf 3})_{0}$
& $({\bf 3}, {\bf 2})_{+1/6}$
& s.a.a
& 1-i \& 4-i
\\
&
&
& $({\bf 8}, {\bf 2})_{+1/2}$
& $({\bf 8}, {\bf 1})_{0}$
& $({\bf 3}, {\bf 2})_{+1/6}$
& $ \frac{1}{24} (\mathcal{O}_{1})_{\{LR\}R} $ 
  $ + 
    \frac{1}{16} (\mathcal{O}_{3})_{\{LR\}R} $
& 1-i \& 4-i
\\
&
&
& $({\bf 8}, {\bf 2})_{+1/2}$
& $({\bf 8}, {\bf 3})_{0}$
& $({\bf 3}, {\bf 2})_{+1/6}$
& s.a.a
& 1-i \& 4-i
\\
&
$(\overline{u_{R}} d_{L}) 
(\overline{e_{R}}) (\overline{u_{R}}) 
(d_{R} \overline{e_{L}})$
& \#20
& $({\bf 1}, {\bf 2})_{+1/2}$
& $({\bf 1}, {\bf 2})_{-1/2}$
& $({\bf 3}, {\bf 2})_{+1/6}$
& $ \frac{1}{16} (\mathcal{O}_{5})_{LR} $
\\
&
&
& $({\bf 8}, {\bf 2})_{+1/2}$
& $({\bf 8}, {\bf 2})_{-1/2}$
& $({\bf 3}, {\bf 2})_{+1/6}$
& $  \frac{1}{16{\rm i}} (\mathcal{O}_{4})_{LR}$
  $  - \frac{5}{48} (\mathcal{O}_{5})_{LR}$
\\
\hline \hline
\end{tabular}
} 
\end{center}
\caption{\it Decomposition \#2-ii. 
In case of decomposition \#2-ii-b, we will have
not only \#2-ii-b, but also \#1-i and \#4-i.}
\end{table}

\begin{table}[tp]
\begin{center}
{\small
\begin{tabular}{ccccccp{2.5cm}c}
\hline \hline
&
&
&
\multicolumn{3}{c}{Mediators $(SU(3)_{c}, SU(2)_{L})_{U(1)_{Y}}$}
&
& Appears
\\ 
\# & Operators 
& BL
& $S$ & $\psi$ & $S'$ & Basis op.
& with
\\ 
\hline
2-iii-a 
&$(d_{R} \overline{e_{L}}) 
      (\overline{u_{L}})
     (d_{R})
 (\overline{u_{L}} \overline{e_{L}})$
& \#11
& $(\overline{\bf 3},{\bf 2})_{-1/6}$
& $({\bf 1}, {\bf 1})_{0}$
& $(\overline{\bf 3},{\bf 1})_{+1/3}$
& $ \frac{1}{32} (\mathcal{O}_{1})_{\{ RR \} R}$  
  $+ \frac{1}{128} (\mathcal{O}_{2})_{\{ RR \}R} $
&
4-i \& 5-i
\\
&
& 
& $(\overline{\bf 3},{\bf 2})_{-1/6}$
& $({\bf 1}, {\bf 3})_{0}$
& $(\overline{\bf 3},{\bf 3})_{+1/3}$
& s.a.a
& 4-i \& 5-i 
\\
&
&
& $(\overline{\bf 3},{\bf 2})_{-1/6}$
& $({\bf 8}, {\bf 1})_{0}$
& $(\overline{\bf 3},{\bf 1})_{+1/3}$
& $- \frac{7}{48} (\mathcal{O}_{1})_{\{RR\}R} $
  $ - \frac{1}{192} (\mathcal{O}_{2})_{\{RR\}R}$
&
4-i \& 5-i
\\
&
& 
& $(\overline{\bf 3},{\bf 2})_{-1/6}$
& $({\bf 8}, {\bf 3})_{0}$
& $(\overline{\bf 3},{\bf 3})_{+1/3}$
& s.a.a
&
4-i \& 5-i
\\
&$(d_{R} \overline{e_{L}}) 
      (\overline{u_{L}})
     (d_{R})
 (\overline{u_{R}} \overline{e_{R}})$
& \#19
& $(\overline{\bf 3},{\bf 2})_{-1/6}$
& $({\bf 1}, {\bf 1})_{0}$
& $(\overline{\bf 3},{\bf 1})_{+1/3}$
& $- \frac{1}{32 {\rm i}} (\mathcal{O}_{4})_{RR}$
  $- \frac{1}{32} (\mathcal{O}_{5})_{RR} $
&
4-i \& 5-i
\\
&
&
& $(\overline{\bf 3},{\bf 2})_{-1/6}$
& $({\bf 8}, {\bf 1})_{0}$
& $(\overline{\bf 3},{\bf 1})_{+1/3}$
& 
$ + \frac{1}{48{\rm i}} (\mathcal{O}_{4})_{RR}$
$ +\frac{7}{48} (\mathcal{O}_{5})_{RR} $
&
4-i \& 5-i
\\
&$(d_{R} \overline{e_{L}}) 
      (\overline{u_{R}})
     (d_{L})
 (\overline{u_{L}} \overline{e_{L}})$
& \#14
& $(\overline{\bf 3},{\bf 2})_{-1/6}$
& $({\bf 1}, {\bf 2})_{+1/2}$
& $(\overline{\bf 3},{\bf 1})_{+1/3}$
& $\frac{1}{32} (\mathcal{O}_{3})_{\{ LR \} R} $
\\
&
&
& $(\overline{\bf 3},{\bf 2})_{-1/6}$
& $({\bf 1}, {\bf 2})_{+1/2}$
& $(\overline{\bf 3},{\bf 3})_{+1/3}$
& s.a.a
\\
&
&
& $(\overline{\bf 3},{\bf 2})_{-1/6}$
& $({\bf 8}, {\bf 2})_{+1/2}$
& $(\overline{\bf 3},{\bf 1})_{+1/3}$
& $ - \frac{1}{8} (\mathcal{O}_{1})_{\{LR\}R} $
  $ - \frac{1}{48}(\mathcal{O}_{3})_{\{LR\}R} $
\\
&
&
& $(\overline{\bf 3},{\bf 2})_{-1/6}$
& $({\bf 8}, {\bf 2})_{+1/2}$
& $(\overline{\bf 3},{\bf 3})_{+1/3}$
& s.a.a
\\
&$(d_{R} \overline{e_{L}}) 
      (\overline{u_{R}})
     (d_{L})
 (\overline{u_{R}} \overline{e_{R}})$
& \#20
& $(\overline{\bf 3},{\bf 2})_{-1/6}$
& $({\bf 1}, {\bf 2})_{+1/2}$
& $(\overline{\bf 3},{\bf 1})_{+1/3}$
& 
$ \frac{1}{32 {\rm i}} (\mathcal{O}_{4})_{LR} $
$  - \frac{1}{32} (\mathcal{O}_{5})_{LR} $
\\
&
&
& $(\overline{\bf 3},{\bf 2})_{-1/6}$
& $({\bf 8}, {\bf 2})_{+1/2}$
& $(\overline{\bf 3},{\bf 1})_{+1/3}$
& $ - \frac{1}{48{\rm i}} (\mathcal{O}_{4})_{LR} $
  $ + \frac{7}{48} (\mathcal{O}_{5})_{LR} $
\\
\hline
2-iii-b &
$(d_{R} \overline{e_{L}}) 
(d_{L}) (\overline{u_{R}})
(\overline{u_{L}} \overline{e_{L}})$
& \#14
& $(\overline{\bf 3}, {\bf 2})_{-1/6}$
& $({\bf 3}, {\bf 1})_{-1/3}$
& $(\overline{\bf 3}, {\bf 1})_{+1/3}$
&  $  \frac{1}{16}
     (\mathcal{O}_{1})_{\{LR\}R} 
   $
   $ + 
      \frac{1}{32} 
     (\mathcal{O}_{3})_{\{LR\}R}
   $
\\
&
&
& $(\overline{\bf 3}, {\bf 2})_{-1/6}$
& $({\bf 3}, {\bf 3})_{-1/3}$
& $(\overline{\bf 3}, {\bf 3})_{+1/3}$
& s.a.a
\\
&&
& $(\overline{\bf 3}, {\bf 2})_{-1/6}$
& $(\overline{\bf 6}, {\bf 1})_{-1/3}$
& $(\overline{\bf 3}, {\bf 1})_{+1/3}$
& $ - 
   \frac{1}{32} 
   (\mathcal{O}_{1})_{\{LR\}R} 
   $
   $+
    \frac{1}{64}
    (\mathcal{O}_{3})_{\{LR\}R}
    $
\\
&
&
& $(\overline{\bf 3}, {\bf 2})_{-1/6}$
& $(\overline{\bf 6}, {\bf 3})_{-1/3}$
& $(\overline{\bf 3}, {\bf 3})_{+1/3}$
& s.a.a
\\
&
$(d_{R} \overline{e_{L}}) 
(d_{L}) (\overline{u_{R}})
(\overline{u_{R}} \overline{e_{R}})$
& \#20
& $(\overline{\bf 3}, {\bf 2})_{-1/6}$
& $({\bf 3}, {\bf 1})_{-1/3}$
& $(\overline{\bf 3}, {\bf 1})_{+1/3}$
& $ \frac{1}{32{\rm i}}
       (\mathcal{O}_{4})_{LR}
  $
  $ - \frac{3}{32} 
      (\mathcal{O}_{5})_{LR}
    $
\\
&&
& $(\overline{\bf 3}, {\bf 2})_{-1/6}$
& $(\overline{\bf 6}, {\bf 1})_{-1/3}$
& $(\overline{\bf 3}, {\bf 1})_{+1/3}$
& $ \frac{1}{64{\rm i}} 
    (\mathcal{O}_{4})_{LR}
  $
  $ +
    \frac{1}{64} 
    (\mathcal{O}_{5})_{LR} 
    $
\\
&
$(d_{R} \overline{e_{L}}) 
(d_{R}) (\overline{u_{L}})
(\overline{u_{L}} \overline{e_{L}})$
& \#11
& $(\overline{\bf 3}, {\bf 2})_{-1/6}$
& $({\bf 3}, {\bf 2})_{+1/6}$
& $(\overline{\bf 3}, {\bf 1})_{+1/3}$
& $ \frac{3}{32}
    (\mathcal{O}_{1})_{\{RR\}R}$   
  $+ 
    \frac{1}{128}
    (\mathcal{O}_{2})_{\{RR\}R} $
\\
&
&
& $(\overline{\bf 3}, {\bf 2})_{-1/6}$
& $({\bf 3}, {\bf 2})_{+1/6}$
& $(\overline{\bf 3}, {\bf 3})_{+1/3}$
& s.a.a
\\
&&
& $(\overline{\bf 3}, {\bf 2})_{-1/6}$
& $(\overline{\bf 6}, {\bf 2})_{+1/6}$
& $(\overline{\bf 3}, {\bf 1})_{+1/3}$
& $
   -\frac{1}{64} 
    (\mathcal{O}_{1})_{\{RR\}R}$
   $+
    \frac{1}{256} 
    (\mathcal{O}_{2})_{\{RR\}R}$
\\
&
&
& $(\overline{\bf 3}, {\bf 2})_{-1/6}$
& $(\overline{\bf 6}, {\bf 2})_{+1/6}$
& $(\overline{\bf 3}, {\bf 3})_{+1/3}$
& s.a.a
\\
&
$(d_{R} \overline{e_{L}}) 
(d_{R}) (\overline{u_{L}})
(\overline{u_{R}} \overline{e_{R}})$
& \#19
& $(\overline{\bf 3}, {\bf 2})_{-1/6}$
& $({\bf 3}, {\bf 2})_{+1/6}$
& $(\overline{\bf 3}, {\bf 1})_{+1/3}$
& $
   -
   \frac{1}{32{\rm i}} (\mathcal{O}_{4})_{RR}$
   $-
   \frac{3}{32} 
   (\mathcal{O}_{5})_{RR}
    $
\\
&&
& $(\overline{\bf 3}, {\bf 2})_{-1/6}$
& $(\overline{\bf 6}, {\bf 2})_{+1/6}$
& $(\overline{\bf 3}, {\bf 1})_{+1/3}$
& $ - 
    \frac{1}{64{\rm i}}
    (\mathcal{O}_{4})_{RR}$
   $+
    \frac{1}{64}
    (\mathcal{O}_{5})_{RR}
   $
\\
\hline \hline
\end{tabular}
} 
\end{center}
\caption{\it Decomposition \#2-iii. 
In case of decomposition \#2-iii-a, we will have
not only \#2-iii-a, but also \#4-i and \#5-i.}
\label{Tab:Decom-2-iii}
\end{table}

\begin{table}[pt]
\begin{center}
\small{
\begin{tabular}{ccccccp{5cm}}
\hline \hline
&
&
&
\multicolumn{3}{c}{Mediators $(SU(3)_{c}, SU(2)_{L})_{U(1)_{Y}}$}
&
\\ 
\# & Operators 
& BL
& $S$ & $\psi$ & $S'$ & Basis op.
\\ 
\hline
3-i
&
$(\overline{u_{L}} \overline{u_{L}})
(\overline{e_{L}}) (\overline{e_{L}})
(d_{R} d_{R})$
& \#11
& $({\bf 6}, {\bf 3})_{+1/3}$
& $({\bf 6}, {\bf 2})_{-1/6}$
& $({\bf 6}, {\bf 1})_{-2/3}$
& 
$ - \frac{1}{16} (\mathcal{O}_{1})_{\{RR\}R} 
  + \frac{1}{64} (\mathcal{O}_{2})_{\{RR\}R}$
\\
&
$(\overline{u_{R}} \overline{u_{R}})
(\overline{e_{L}}) (\overline{e_{L}})
(d_{L} d_{L})$
& \#12
& $({\bf 6}, {\bf 1})_{+4/3}$
& $({\bf 6}, {\bf 2})_{+5/6}$
& $({\bf 6}, {\bf 3})_{+1/3}$
& 
$- \frac{1}{16} (\mathcal{O}_{1})_{\{LL\}R}  
 + \frac{1}{64} (\mathcal{O}_{2})_{\{LL\}R}$
\\
& 
$(\overline{u_{R}} \overline{u_{R}})
(\overline{e_{R}}) (\overline{e_{R}})
(d_{R} d_{R})$
& ---
& $({\bf 6}, {\bf 1})_{+4/3}$
& $({\bf 6}, {\bf 1})_{+1/3}$
& $({\bf 6}, {\bf 1})_{-2/3}$
& 
$\frac{1}{16} (\mathcal{O}_{3})_{\{RR\}L} $
\\
\hline
3-ii
&
$(\overline{u_{L}} \overline{u_{L}})
(d_{R}) (d_{R})
(\overline{e_{L}} \overline{e_{L}})$
& \#11
& $({\bf 6}, {\bf 3})_{+1/3}$
& $({\bf 3}, {\bf 3})_{+2/3}$
& $({\bf 1}, {\bf 3})_{+1}$
& 
$ - \frac{1}{16} (\mathcal{O}_{1})_{\{RR\}R}
  + \frac{1}{64} (\mathcal{O}_{2})_{\{RR\}R}
$
\\
&
$(\overline{u_{R}} \overline{u_{R}})
(d_{L}) (d_{L})
(\overline{e_{L}} \overline{e_{L}})$
& \#12
& $({\bf 6}, {\bf 1})_{+4/3}$
& $({\bf 3}, {\bf 2})_{+7/6}$
& $({\bf 1}, {\bf 3})_{+1}$
& 
$- \frac{1}{16} (\mathcal{O}_{1})_{\{LL\}R}
 + \frac{1}{64} (\mathcal{O}_{2})_{\{LL\}R}
$
\\
&
$(\overline{u_{R}} \overline{u_{R}})
(d_{R}) (d_{R})
(\overline{e_{R}} \overline{e_{R}})$
& ---
& $({\bf 6}, {\bf 1})_{+4/3}$
& $({\bf 3}, {\bf 1})_{+5/3}$
& $({\bf 1}, {\bf 1})_{+2}$
& 
$ \frac{1}{16} (\mathcal{O}_{3})_{\{RR\}L} $
\\
\hline 
3-iii
& 
$(d_{L} d_{L}) 
(\overline{u_{R}})
(\overline{u_{R}})
(\overline{e_{L}} \overline{e_{L}})$
& \#12
& $(\overline{\bf 6}, {\bf 3})_{-1/3}$
& $(\overline{\bf 3}, {\bf 3})_{+1/3}$
& $({\bf 1}, {\bf 3})_{+1}$
& 
$ - \frac{1}{16} (\mathcal{O}_{1})_{\{LL\}R}  
  + \frac{1}{64} (\mathcal{O}_{2})_{\{LL\}R}
$
\\
&
$(d_{R} d_{R}) 
(\overline{u_{L}})
(\overline{u_{L}})
(\overline{e_{L}} \overline{e_{L}})$
& \#11
& $(\overline{\bf 6}, {\bf 1})_{+2/3}$
& $(\overline{\bf 3}, {\bf 2})_{+5/6}$
& $({\bf 1}, {\bf 3})_{+1}$
& $- \frac{1}{16} (\mathcal{O}_{1})_{\{RR\}R} 
   + \frac{1}{64} (\mathcal{O}_{2})_{\{RR\}R} $
\\
&
$(d_{R} d_{R}) 
(\overline{u_{R}})
(\overline{u_{R}})
(\overline{e_{R}} \overline{e_{R}})$
& ---
& $(\overline{\bf 6}, {\bf 1})_{+2/3}$
& $(\overline{\bf 3}, {\bf 1})_{+4/3}$
& $({\bf 1}, {\bf 1})_{+2}$
& $ \frac{1}{16} (\mathcal{O}_{3})_{\{RR\}L} $
\\
\hline
4-i 
&
$(d_{L} \overline{e_{R}})
(\overline{u_{R}}) (\overline{u_{R}})
(d_{R} \overline{e_{L}})$
& \#20
& $(\overline{\bf 3}, {\bf 2})_{-7/6}$
& $({\bf 1}, {\bf 2})_{-1/2}$
& $({\bf 3}, {\bf 2})_{+1/6}$
& $ - \frac{1}{32 {\rm i}} (\mathcal{O}_{4})_{LR}
    - \frac{1}{32} (\mathcal{O}_{5})_{LR} 
 $
  \\
&
& 
& $(\overline{\bf 3}, {\bf 2})_{-7/6}$
& $({\bf 8}, {\bf 2})_{-1/2}$
& $({\bf 3}, {\bf 2})_{+1/6}$
& $ - \frac{1}{24{\rm i}} (\mathcal{O}_{4})_{LR}
    - \frac{1}{24} (\mathcal{O}_{5})_{LR}
   $
  \\
&
$(d_{R} \overline{e_{L}})
(\overline{u_{L}}) (\overline{u_{L}})
(d_{R} \overline{e_{L}})$
& \#11
& $(\overline{\bf 3}, {\bf 2})_{-1/6}$
& $({\bf 1}, {\bf 1})_{0}$
& $({\bf 3}, {\bf 2})_{+1/6}$
& $ \frac{1}{32} (\mathcal{O}_{1})_{\{ RR \} R}  
   - \frac{1}{128} (\mathcal{O}_{2})_{\{ RR \} R}$
\\
&
& 
& $(\overline{\bf 3}, {\bf 2})_{-1/6}$
& $({\bf 1}, {\bf 3})_{0}$
& $({\bf 3}, {\bf 2})_{+1/6}$
& s.a.a
\\
&
& 
& $(\overline{\bf 3}, {\bf 2})_{-1/6}$
& $({\bf 8}, {\bf 1})_{0}$
& $({\bf 3}, {\bf 2})_{+1/6}$
& $ \frac{1}{24} (\mathcal{O}_{1})_{\{RR\}R}
    -
    \frac{1}{96} 
    (\mathcal{O}_{2})_{\{RR\}R}$
\\
&
& 
& $(\overline{\bf 3}, {\bf 2})_{-1/6}$
& $({\bf 8}, {\bf 3})_{0}$
& $({\bf 3}, {\bf 2})_{+1/6}$
& s.a.a
\\
\hline
4-ii-a
& 
$(\overline{u_{L}}\overline{u_{L}})
(d_{R})(\overline{e_{L}})
(\overline{e_{L}} d_{R})$
& \#11
& $({\bf 6},{\bf 3})_{+1/3}$
& $({\bf 3}, {\bf 3})_{+2/3}$
& $({\bf 3}, {\bf 2})_{+1/6}$
& $\frac{1}{32} (\mathcal{O}_{1})_{\{RR\}R} 
  -
  \frac{1}{128} (\mathcal{O}_{2})_{\{RR\}R}$
\\
& 
$(\overline{u_{R}}\overline{u_{R}})
(d_{L})(\overline{e_{R}})
(\overline{e_{L}} d_{R})$
& \#20
& $({\bf 6},{\bf 1})_{+4/3}$
& $({\bf 3}, {\bf 2})_{+7/6}$
& $({\bf 3}, {\bf 2})_{+1/6}$
& $
 - \frac{1}{32{\rm i}} (\mathcal{O}_{4})_{LR}
 - \frac{1}{32} (\mathcal{O}_{5})_{LR}  
 $
\\
& $(\overline{u_{R}}\overline{u_{R}})
(d_{R})(\overline{e_{L}})
(\overline{e_{R}} d_{L})$
& \#20
& $({\bf 6},{\bf 1})_{+4/3}$
& $({\bf 3}, {\bf 1})_{+5/3}$
& $({\bf 3}, {\bf 2})_{+7/6}$
& $
   - \frac{1}{32 {\rm i}} (\mathcal{O}_{4})_{LR}
   - \frac{1}{32} (\mathcal{O}_{5})_{LR}
  $
\\
\hline 
4-ii-b&
$(\overline{u_{L}}\overline{u_{L}})
(\overline{e_{L}}) (d_{R})
(\overline{e_{L}} d_{R})$
& \#11
& $({\bf 6},{\bf 3})_{+1/3}$
& $({\bf 6},{\bf 2})_{-1/6}$
& $({\bf 3}, {\bf 2})_{+1/6}$
& $\frac{1}{32} (\mathcal{O}_{1})_{\{RR\}R}  
   - 
   \frac{1}{128} (\mathcal{O}_{2})_{\{RR\}R}$
\\
&
$(\overline{u_{R}}\overline{u_{R}})
(\overline{e_{R}})(d_{L})
(\overline{e_{L}} d_{R})$
& \#20
& $({\bf 6},{\bf 1})_{+4/3}$
& $({\bf 6}, {\bf 1})_{+1/3}$
& $({\bf 3}, {\bf 2})_{+1/6}$
& $ 
    - \frac{1}{32 {\rm i}} (\mathcal{O}_{4})_{LR} 
    - \frac{1}{32} (\mathcal{O}_{5})_{LR}
  $
  \\
& $(\overline{u_{R}}\overline{u_{R}})
(\overline{e_{L}})(d_{R})
(\overline{e_{R}} d_{L})$
& \#20
& $({\bf 6},{\bf 1})_{+4/3}$
& $({\bf 6}, {\bf 2})_{+5/6}$
& $({\bf 3}, {\bf 2})_{+7/6}$
& $- \frac{1}{32} (\mathcal{O}_{5})_{LR}
   - \frac{1}{32 {\rm i}} (\mathcal{O}_{4})_{LR} 
  $
\\
\hline
5-i 
& 
$(\overline{u_{L}} \overline{e_{L}})
(d_{R}) (d_{R})
(\overline{u_{L}} \overline{e_{L}})
$
& \#11
& $({\bf 3}, {\bf 1})_{-1/3}$
& $({\bf 1}, {\bf 1})_{0}$
& $(\overline{\bf 3}, {\bf 1})_{+1/3}$
&  $ \frac{1}{32} (\mathcal{O}_{1})_{\{ RR \} R} 
   - \frac{1}{128} (\mathcal{O}_{2})_{\{ RR \} R}$
\\
&
&
& $({\bf 3}, {\bf 3})_{-1/3}$
& $({\bf 1}, {\bf 3})_{0}$
& $(\overline{\bf 3}, {\bf 3})_{+1/3}$
& s.a.a
\\
& 
& 
& $({\bf 3}, {\bf 1})_{-1/3}$
& $({\bf 8}, {\bf 1})_{0}$
& $(\overline{\bf 3}, {\bf 1})_{+1/3}$
& $\frac{1}{24} (\mathcal{O}_{1})_{\{RR\}R} 
   -
   \frac{1}{96} 
   (\mathcal{O}_{2})_{\{RR\}R}$
\\
&
&
& $({\bf 3}, {\bf 3})_{-1/3}$
& $({\bf 8}, {\bf 3})_{0}$
& $(\overline{\bf 3}, {\bf 3})_{+1/3}$
& s.a.a 
\\
&
$(\overline{u_{R}} \overline{e_{R}})
(d_{R}) (d_{R})
(\overline{u_{L}} \overline{e_{L}})$
& \#19
& $({\bf 3}, {\bf 1})_{-1/3}$
& $({\bf 1}, {\bf 1})_{0}$
& $(\overline{\bf 3}, {\bf 1})_{+1/3}$
& $ \frac{1}{32 {\rm i}} (\mathcal{O}_{4})_{RR}
    - \frac{1}{32} (\mathcal{O}_{5})_{RR} 
  $
  \\
&
& 
& $({\bf 3}, {\bf 1})_{-1/3}$
& $({\bf 8}, {\bf 1})_{0}$
& $(\overline{\bf 3}, {\bf 1})_{+1/3}$
& $ \frac{1}{24{\rm i}} (\mathcal{O}_{4})_{RR}
    - 
    \frac{1}{24} 
    (\mathcal{O}_{5})_{RR} 
    $
  \\
&
$(\overline{u_{R}} \overline{e_{R}})
(d_{R}) (d_{R}) 
(\overline{u_{R}} \overline{e_{R}})$
& ---
& $({\bf 3}, {\bf 1})_{-1/3}$
& $({\bf 1}, {\bf 1})_{0}$
& $(\overline{\bf 3}, {\bf 1})_{+1/3}$
& $ - \frac{1}{32} (\mathcal{O}_{3})_{\{ RR \}L}$
\\
&
& 
& $({\bf 3}, {\bf 1})_{-1/3}$
& $({\bf 8}, {\bf 1})_{0}$
& $(\overline{\bf 3}, {\bf 1})_{+1/3}$
& $ - \frac{1}{24} 
      (\mathcal{O}_{3})_{\{RR\}L} $
\\
\hline
5-ii-a
&
$(\overline{u_{L}} \overline{e_{L}})
(\overline{u_{L}})(\overline{e_{L}})
(d_{R} d_{R})$
& \#11
& $({\bf 3}, {\bf 1})_{-1/3}$
& $({\bf 6}, {\bf 2})_{-1/6}$
& $({\bf 6}, {\bf 1})_{-2/3}$
& $\frac{1}{32} (\mathcal{O}_{1})_{\{RR\}R}  
   -  
   \frac{1}{128}
   (\mathcal{O}_{2})_{\{RR\}R}
   $
\\
&
&
& $({\bf 3}, {\bf 3})_{-1/3}$
& $({\bf 6}, {\bf 2})_{-1/6}$
& $({\bf 6}, {\bf 1})_{-2/3}$
& s.a.a
\\
&
$(\overline{u_{L}} \overline{e_{L}})
(\overline{u_{R}})(\overline{e_{R}})
(d_{R} d_{R})$
& \#19
& $({\bf 3}, {\bf 1})_{-1/3}$
& $({\bf 6}, {\bf 1})_{+1/3}$
& $({\bf 6}, {\bf 1})_{-2/3}$
& $  \frac{1}{32{\rm i}} (\mathcal{O}_{4})_{RR} 
     - \frac{1}{32} (\mathcal{O}_{5})_{RR}
  $
\\
&
$(\overline{u_{R}} \overline{e_{R}})
(\overline{u_{L}}) (\overline{e_{L}})
(d_{R} d_{R})$
& \#19
& $({\bf 3}, {\bf 1})_{-1/3}$
& $({\bf 6}, {\bf 2})_{-1/6}$
& $({\bf 6}, {\bf 1})_{-2/3}$
&  $   \frac{1}{32 {\rm i}} (\mathcal{O}_{4})_{RR} 
     - \frac{1}{32}
       (\mathcal{O}_{5})_{RR} 
   $
  \\
&
$(\overline{u_{R}} \overline{e_{R}})
(\overline{u_{R}}) (\overline{e_{R}})
(d_{R} d_{R})$
& ---
& $({\bf 3}, {\bf 1})_{-1/3}$
& $({\bf 6}, {\bf 1})_{+1/3}$
& $({\bf 6}, {\bf 1})_{-2/3}$
& 
$ -\frac{1}{32} 
 (\mathcal{O}_{3})_{\{RR\}L}
$
\\
\hline
5-ii-b
&
$(\overline{u_{L}} \overline{e_{L}})
(\overline{e_{L}})(\overline{u_{L}})
(d_{R} d_{R})$
& \#11
& $({\bf 3}, {\bf 1})_{-1/3}$
& $({\bf 3}, {\bf 2})_{-5/6}$
& $({\bf 6}, {\bf 1})_{-2/3}$
& $ \frac{1}{32} (\mathcal{O}_{1})_{\{RR\}R}  
    - 
    \frac{1}{128} (\mathcal{O}_{2})_{\{RR\}R}$
\\
&
&
& $({\bf 3}, {\bf 3})_{-1/3}$
& $({\bf 3}, {\bf 2})_{-5/6}$
& $({\bf 6}, {\bf 1})_{-2/3}$
& s.a.a
\\
&
$(\overline{u_{L}} \overline{e_{L}})
(\overline{e_{R}})(\overline{u_{R}})
(d_{R} d_{R})$
& \#19
& $({\bf 3}, {\bf 1})_{-1/3}$
& $({\bf 3}, {\bf 1})_{-4/3}$
& $({\bf 6}, {\bf 1})_{-2/3}$
& $   \frac{1}{32{\rm i}}(\mathcal{O}_{4})_{RR} 
      -\frac{1}{32} (\mathcal{O}_{5})_{RR}$
\\
&
$(\overline{u_{R}} \overline{e_{R}})
(\overline{e_{L}})(\overline{u_{L}})(d_{R} d_{R})$
& \#19
& $({\bf 3}, {\bf 1})_{-1/3}$
& $({\bf 3}, {\bf 2})_{-5/6}$
& $({\bf 6}, {\bf 1})_{-2/3}$
& $ \frac{1}{32{\rm i}}
    (\mathcal{O}_{4})_{RR} 
    -  
    \frac{1}{32} (\mathcal{O}_{5})_{RR} 
  $
  \\
&
$(\overline{u_{R}} \overline{e_{R}})
(\overline{e_{R}})(\overline{u_{R}})
(d_{R} d_{R})$
& ---
& $({\bf 3}, {\bf 1})_{-1/3}$
& $({\bf 3}, {\bf 1})_{-4/3}$
& $({\bf 6}, {\bf 1})_{-2/3}$
& $- \frac{1}{32} (\mathcal{O}_{3})_{\{RR\}L}$
\\
\hline \hline
\end{tabular}
} 
\end{center}
\caption{\it Decompositions \#3, \#4, and \#5. 
The operators with two $\overline{e_{R}}$'s are not listed in the 
paper by  Babu and Leung (BL)~\cite{Babu:2001ex}.}
\label{Tab:Decom-345}
\end{table}

\subsection{Topology II}

\begin{figure}
\centering
 \includegraphics[scale=0.75]{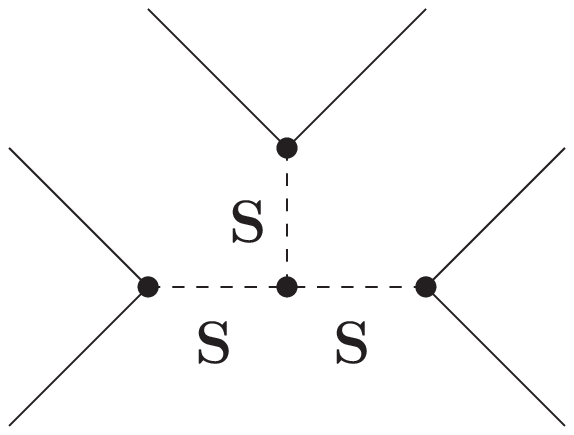}
\includegraphics[scale=0.75]{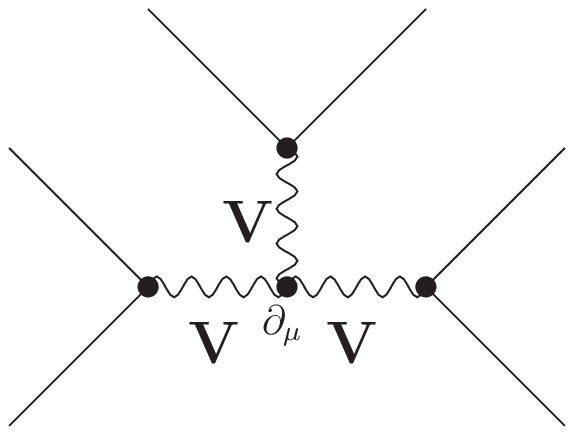}\\
\includegraphics[scale=0.75]{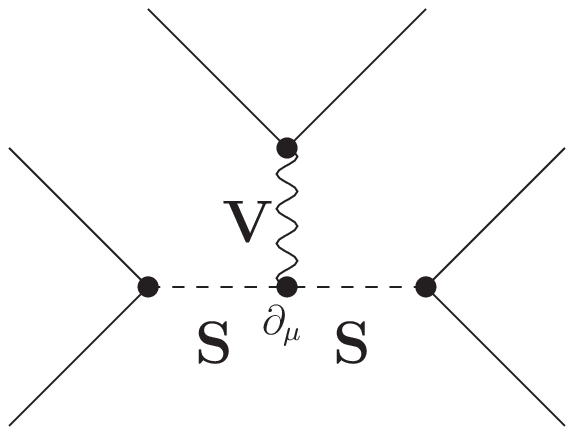}
\includegraphics[scale=0.75]{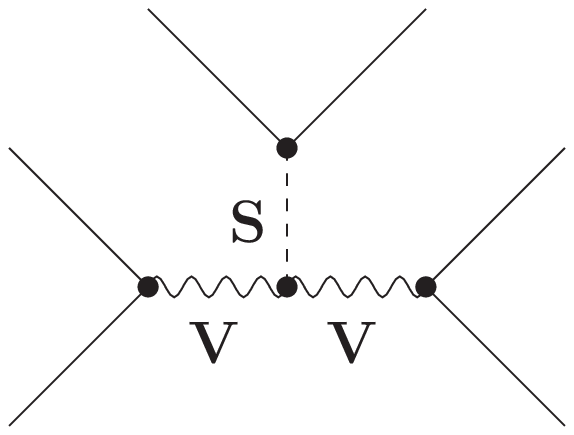}
\caption{\label{fig:split}The four different insertions for topology-II: 
SSS, VVV, SSV and VVS. Note, that both VVV and SSV have necessarily 
derivative couplings. For a discussion see text.}
\end{figure}

We can decompose the \onbb operator with T-II in practically 
the same way as T-I. However, there are also some differences, 
which we will briefly discuss. As mentioned above, 
and shown in Fig.~\ref{fig:split}, there are four possible combinations 
of scalars and vectors to complete the T-II decomposition. 
As shown in the figure, both VVV and SSV necessarily involve derivative 
couplings. They lead to an effective $d=10$ operator, 
\begin{align}
\mathcal{O}_{d=10}
\propto
\frac{1}{\Lambda^{6}} 
\partial_{\rho} 
(\bar{u} \bar{u} dd \bar{e} \bar{e})^{\rho}
=
\frac{q}{\Lambda} \frac{1}{\Lambda^{5}}
(\bar{u} \bar{u} dd \bar{e} \bar{e}),
\end{align}
where $q$ is a typical momentum of the \onbb process.
Therefore, they are suppressed by a factor of $q/\Lambda$ in
comparison with decompositions of 
the $d=9$ operators without derivatives and can be 
safely neglected.

Diagrams of the type SVV come from vectors being gauge bosons, 
i.e., from the covariant derivative of the scalar field 
$\mathcal{S}$:
\begin{align}
\mathcal{L}_{\rm gauge}
= (D_{\mu} \mathcal{S})^{\dagger}(D^\mu \mathcal{S}) 
\supset g^2 \mathcal{S}^{\dagger} \mathcal{S} V_{\mu}V^{\mu}
\supset g^2 \langle \mathcal{S} \rangle 
S V_{\mu}V^{\mu},
\end{align}
if the scalar $\mathcal{S}$ can take a vacuum expectation 
value $\langle \mathcal{S} \rangle$.
Here, $S$ is a fluctuation around the vev 
$\mathcal{S} = \langle \mathcal{S} \rangle + S$,
 which would be 
a scalar mediator.
If this vev breaks $SU(2)_L$, this leads to a suppression order
$v/\Lambda$, but, as can be seen from the example of LR symmetry, 
an SM singlet vev can produce a coupling whose order is of $\Lambda$, 
such that the total amplitude for T-II is again proportional 
to $\Lambda^{-5}$. 
Similarly, for diagrams of the type SSS, the coupling has a 
dimension of mass, leading potentially to a $\Lambda^{-5}$ total 
factor for the diagram.

Since new vectors require an extension of the gauge group, we consider
the SSS case to be the more easily motivated choice.  In our detailed
decomposition of T-II, presented in Tab.~\ref{Tab:Top2-3S}, we
therefore concentrate on the case of SSS. The results for SVV can be 
derived easily from the recipes discussed above.

\begin{table}[tp]
\begin{center}
\small{
\begin{tabular}{ccccccp{5cm}}
\hline \hline 
&
&
&
\multicolumn{3}{c}{Mediators $(SU(3)_{c}, SU(2)_{L})_{U(1)_{Y}}$}
&
\\
\# & Operators & BL
& $S$ & $S'$ & $S''$ & Basis op.
\\ 
\hline
1 
& 
$(\overline{u_{L}} d_{R})
(\overline{u_{L}} d_{R})
(\overline{e_{L}} \overline{e_{L}})$ 
& \#11 
& $({\bf 1}, {\bf 2})_{+1/2}$
& $({\bf 1}, {\bf 2})_{+1/2}$
& $({\bf 1}, {\bf 3})_{-1}$
& $\frac{1}{8} (\mathcal{O}_{1})_{\{ RR \}R}$
\\
& 
&  
& $({\bf 8}, {\bf 2})_{+1/2}$
& $({\bf 8}, {\bf 2})_{+1/2}$
& $({\bf 1}, {\bf 3})_{-1}$
& $- 
   \frac{5}{24} (\mathcal{O}_{1})_{\{RR\}R} 
   -
   \frac{1}{32}
   (\mathcal{O}_{2})_{\{RR\}R}$
\\
& 
$(\overline{u_{R}} d_{L})
(\overline{u_{L}} d_{R})
(\overline{e_{L}} \overline{e_{L}})$ 
& \#14
& $({\bf 1}, {\bf 2})_{+1/2}$
& $({\bf 1}, {\bf 2})_{+1/2}$
& $({\bf 1}, {\bf 3})_{-1}$
& $\frac{1}{8} (\mathcal{O}_{1})_{\{ LR \}R}$
\\
& 
& 
& $({\bf 8}, {\bf 2})_{+1/2}$
& $({\bf 8}, {\bf 2})_{+1/2}$
& $({\bf 1}, {\bf 3})_{-1}$
& $ - \frac{1}{12} 
      (\mathcal{O}_{1})_{\{LR\}R} 
    - 
    \frac{1}{8} (\mathcal{O}_{3})_{\{LR\} R}
    $
\\
& 
$(\overline{u_{R}} d_{L})
(\overline{u_{R}} d_{L})
(\overline{e_{L}} \overline{e_{L}})$ 
& \#12
& $({\bf 1}, {\bf 2})_{+1/2}$
& $({\bf 1}, {\bf 2})_{+1/2}$
& $({\bf 1}, {\bf 3})_{-1}$
& $\frac{1}{8} (\mathcal{O}_{1})_{\{ LL \}R}$
\\
& 
& 
& $({\bf 8}, {\bf 2})_{+1/2}$
& $({\bf 8}, {\bf 2})_{+1/2}$
& $({\bf 1}, {\bf 3})_{-1}$
& $- \frac{5}{24} 
     (\mathcal{O}_{1})_{\{LL\}R} 
   - 
   \frac{1}{32} 
   (\mathcal{O}_{2})_{\{LL\}R}$
\\
\hline
2 
& $(\overline{u_{L}} d_{R})
(\overline{u_{L}}\overline{e_{L}})
(d_{R} \overline{e_{L}})$
& \#11
& $({\bf 1}, {\bf 2})_{+1/2}$
& $({\bf 3}, {\bf 1})_{-1/3}$
& $(\overline{\bf 3}, {\bf 2})_{-1/6}$
& $ - \frac{1}{16} (\mathcal{O}_{1})_{\{ RR \}R}$
\\
&
& 
& $({\bf 1}, {\bf 2})_{+1/2}$
& $({\bf 3}, {\bf 3})_{-1/3}$
& $(\overline{\bf 3}, {\bf 2})_{-1/6}$
& s.a.a
\\
&
& 
& $({\bf 8}, {\bf 2})_{+1/2}$
& $({\bf 3}, {\bf 1})_{-1/3}$
& $(\overline{\bf 3}, {\bf 2})_{-1/6}$
& $\frac{5}{48} (\mathcal{O}_{1})_{\{RR\}R} 
   +
   \frac{1}{64} 
   (\mathcal{O}_{2})_{\{RR\}R}$
\\
&
& 
& $({\bf 8}, {\bf 2})_{+1/2}$
& $({\bf 3}, {\bf 3})_{-1/3}$
& $(\overline{\bf 3}, {\bf 2})_{-1/6}$
& s.a.a
\\
& $(\overline{u_{L}} d_{R})
(\overline{u_{R}}\overline{e_{R}})
(d_{R} \overline{e_{L}})$
& \#19
& $({\bf 1}, {\bf 2})_{+1/2}$
& $({\bf 3}, {\bf 1})_{-1/3}$
& $(\overline{\bf 3}, {\bf 2})_{-1/6}$
& $ \frac{1}{16} (\mathcal{O}_{5})_{RR}$
\\
&
& 
& $({\bf 8}, {\bf 2})_{+1/2}$
& $({\bf 3}, {\bf 1})_{-1/3}$
& $(\overline{\bf 3}, {\bf 2})_{-1/6}$
& $  - \frac{1}{16{\rm i}} (\mathcal{O}_{4})_{RR}
     - \frac{5}{48} (\mathcal{O}_{5})_{RR}
  $
\\
& $(\overline{u_{R}} d_{L})
(\overline{u_{L}}\overline{e_{L}})
(d_{R} \overline{e_{L}})$
& \#14
& $({\bf 1}, {\bf 2})_{+1/2}$
& $({\bf 3}, {\bf 1})_{-1/3}$
& $(\overline{\bf 3}, {\bf 2})_{-1/6}$
& $ - \frac{1}{16} (\mathcal{O}_{1})_{\{ LR \} R}$
\\
&
&
& $({\bf 1}, {\bf 2})_{+1/2}$
& $({\bf 3}, {\bf 3})_{-1/3}$
& $(\overline{\bf 3}, {\bf 2})_{-1/6}$
& s.a.a
\\
&
& 
& $({\bf 8}, {\bf 2})_{+1/2}$
& $({\bf 3}, {\bf 1})_{-1/3}$
& $(\overline{\bf 3}, {\bf 2})_{-1/6}$
& $ \frac{1}{24} 
    (\mathcal{O}_{1})_{\{LR\}R} 
    + 
    \frac{1}{16} 
    (\mathcal{O}_{3})_{\{LR\}R}$
\\
&
&
& $({\bf 8}, {\bf 2})_{+1/2}$
& $({\bf 3}, {\bf 3})_{-1/3}$
& $(\overline{\bf 3}, {\bf 2})_{-1/6}$
& s.a.a
\\
& $(\overline{u_{R}} d_{L})
(\overline{u_{R}}\overline{e_{R}})
(d_{R} \overline{e_{L}})$
& \#20
& $({\bf 1}, {\bf 2})_{+1/2}$
& $({\bf 3}, {\bf 1})_{-1/3}$
& $(\overline{\bf 3}, {\bf 2})_{-1/6}$
& $ \frac{1}{16} (\mathcal{O}_{5})_{LR} $
\\
&
& 
& $({\bf 8}, {\bf 2})_{+1/2}$
& $({\bf 3}, {\bf 1})_{-1/3}$
& $(\overline{\bf 3}, {\bf 2})_{-1/6}$
& $  \frac{1}{16{\rm i}} (\mathcal{O}_{4})_{LR}
     - \frac{5}{48} (\mathcal{O}_{5})_{LR} 
  $
\\
\hline
3 
&
$(\overline{u_{L}} \overline{u_{L}})
(d_{R} d_{R})
(\overline{e_{L}} \overline{e_{L}})$
& \#11
& $({\bf 6}, {\bf 3})_{+1/3}$
& $(\overline{\bf 6}, {\bf 1})_{+2/3}$
& $({\bf 1}, {\bf 3})_{-1}$
& $ 
 - \frac{1}{24} (\mathcal{O}_{1})_{\{ RR \} R} 
 + \frac{1}{96} (\mathcal{O}_{2})_{\{ RR \}R}$
\\
&
$(\overline{u_{R}} \overline{u_{R}})
(d_{L} d_{L})
(\overline{e_{L}} \overline{e_{L}})$
& \#12
& $({\bf 6}, {\bf 1})_{+4/3}$
& $(\overline{\bf 6}, {\bf 3})_{-1/3}$
& $({\bf 1}, {\bf 3})_{-1}$
& $ -\frac{1}{24} (\mathcal{O}_{1})_{\{ LL \}R}  
    + \frac{1}{96} (\mathcal{O}_{2})_{\{ LL \}R}$
\\
&
$(\overline{u_{R}} \overline{u_{R}})
(d_{R} d_{R})
(\overline{e_{R}} \overline{e_{R}})$
& ---
& $({\bf 6}, {\bf 1})_{+4/3}$
& $(\overline{\bf 6}, {\bf 1})_{+2/3}$
& $({\bf 1}, {\bf 1})_{-2}$
& $ \frac{1}{24} (\mathcal{O}_{3})_{\{ RR \}L} $
\\
\hline
4 
&
$(\overline{u_{L}} \overline{u_{L}})
(d_{R} \overline{e_{L}})
(d_{R} \overline{e_{L}})$
& \#11
& $({\bf 6}, {\bf 3})_{+1/3}$
& $(\overline{\bf 3}, {\bf 2})_{-1/6}$
& $(\overline{\bf 3}, {\bf 2})_{-1/6}$
& $\frac{1}{48} (\mathcal{O}_{1})_{\{ RR \}R}  
   - 
   \frac{1}{192} (\mathcal{O}_{2})_{\{ RR \} R}$
\\
&
$(\overline{u_{R}} \overline{u_{R}})
(d_{L} \overline{e_{R}})
(d_{R} \overline{e_{L}})$
& \#20
& $({\bf 6}, {\bf 1})_{+4/3}$
& $(\overline{\bf 3}, {\bf 2})_{-7/6}$
& $(\overline{\bf 3}, {\bf 2})_{-1/6}$
& $- \frac{1}{48 {\rm i}} (\mathcal{O}_{4})_{LR}
   - \frac{1}{48} (\mathcal{O}_{5})_{LR}   
 $
\\
\hline
5
&
$(\overline{u_{L}} \overline{e_{L}})
(\overline{u_{L}} \overline{e_{L}})
(d_{R} d_{R})$
& \#11
& $({\bf 3}, {\bf 1})_{-1/3}$
& $({\bf 3}, {\bf 1})_{-1/3}$
& $(\overline{\bf 6}, {\bf 1})_{+2/3}$
& $\frac{1}{48} (\mathcal{O}_{1})_{\{ RR \} R}  
   - \frac{1}{192} (\mathcal{O}_{2})_{\{ RR \} R}$
\\
&
&
& $({\bf 3}, {\bf 3})_{-1/3}$
& $({\bf 3}, {\bf 3})_{-1/3}$
& $(\overline{\bf 6}, {\bf 1})_{+2/3}$
& s.a.a
\\
&
$(\overline{u_{L}} \overline{e_{L}})
(\overline{u_{R}} \overline{e_{R}})
(d_{R} d_{R})$
& \#19
& $({\bf 3}, {\bf 1})_{-1/3}$
& $({\bf 3}, {\bf 1})_{-1/3}$
& $(\overline{\bf 6}, {\bf 1})_{+2/3}$
 & $ + \frac{1}{96{\rm i}} (\mathcal{O}_{4})_{RR}
     - \frac{1}{48} (\mathcal{O}_{5})_{RR}
   $
\\
&
$(\overline{u_{R}} \overline{e_{R}})
(\overline{u_{R}} \overline{e_{R}})
(d_{R} d_{R})$
& --- 
& $({\bf 3}, {\bf 1})_{-1/3}$
& $({\bf 3}, {\bf 1})_{-1/3}$
& $(\overline{\bf 6}, {\bf 1})_{+2/3}$
& $- \frac{1}{48} (\mathcal{O}_{3})_{\{ RR \} L}$
\\
\hline \hline
\end{tabular}
} 
\end{center}
\caption{\it Decomposition and operator projection for the three-scalar
 case of T-II.
}
\label{Tab:Top2-3S}
\end{table}

\newpage 
\section{Errata for version 1}

We have discovered a few misprints in tables \ref{Tab:Decom-2-iii} 
and \ref{Tab:Decom-345} of the first version of the paper. 
Some quantum numbers for the intermediate scalar or fermion where 
incorrectly given. The correct quantum numbers are given in table 
\ref{Tab:Misprints}, where the corrections are emphasized in 
red color.
\begin{table}[h]
\begin{center}
{\small
\begin{tabular}{cccccc}
\hline \hline
\# & Operators & BL & $S$ & $\psi$ & $S'$  \\
\hline
2-iii-b &
$(d_{R} \overline{e_{L}}) 
(d_{L}) (\overline{u_{R}})
(\overline{u_{R}} \overline{e_{R}})$
& \#20
& $(\overline{\bf 3}, {\bf 2})_{-1/6}$
& $({\bf 3}, {\bf 1})_{-1/3}$
& $(\overline{\bf 3}, {\bf 1})_{\red +1/3}$ 
\\
&&
& $(\overline{\bf 3}, {\bf 2})_{-1/6}$
& $(\overline{\bf 6}, {\bf 1})_{-1/3}$
& $(\overline{\bf 3}, {\bf 1})_{\red +1/3}$
\\
 \hline
5-i 
& 
$(\overline{u_{L}} \overline{e_{L}})
(d_{R}) (d_{R})
(\overline{u_{L}} \overline{e_{L}})
$
& \#11
& $({\bf 3}, {\bf\red 3})_{-1/3}$
& $({\bf 1}, {\bf 3})_{0}$
& $(\overline{\bf 3}, {\bf\red 3})_{+1/3}$
\\
&
&
& $({\bf 3}, {\bf\red 3})_{-1/3}$
& $({\bf 8}, {\bf 3})_{0}$
& $(\overline{\bf 3}, {\bf\red  3})_{+1/3}$
\\
&
$(\overline{u_{R}} \overline{e_{R}})
(d_{R}) (d_{R})
(\overline{u_{L}} \overline{e_{L}})$
& \#19
& $({\bf 3}, {\bf 1})_{-1/3}$
& $({\bf 1}, {\bf \red 1})_{0}$
& $(\overline{\bf 3}, {\bf 1})_{+1/3}$
  \\
&
& 
& $({\bf 3}, {\bf 1})_{-1/3}$
& $({\bf 8}, {\bf \red 1})_{0}$
& $(\overline{\bf 3}, {\bf 1})_{+1/3}$
  \\
&
$(\overline{u_{R}} \overline{e_{R}})
(d_{R}) (d_{R}) 
(\overline{u_{R}} \overline{e_{R}})$
& ---
& $({\bf 3}, {\bf 1})_{-1/3}$
& $({\bf 1}, {\bf \red 1})_{0}$
& $(\overline{\bf 3}, {\bf 1})_{+1/3}$
\\
&
& 
& $({\bf 3}, {\bf 1})_{-1/3}$
& $({\bf 8}, {\bf \red 1})_{0}$
& $(\overline{\bf 3}, {\bf 1})_{+1/3}$
\\
\hline \hline
\end{tabular}
} 
\end{center}
\caption{\it Misprints in tables 8 and 9. The corrected quantum numbers 
are emphasized in red colour.}
\label{Tab:Misprints}
\end{table}

In addition, in table \ref{Tab:Decom-345} 
two possible decompositions were not listed 
by accident. The missing decompositions correspond to two cases in 
decomposition 4-i, BL\#11, and are given in table \ref{Tab:Missing}. 
\begin{table}[h]
\begin{center}
\small{
\begin{tabular}{ccccccp{5cm}}
\hline \hline 
\# & Operators 
& BL
& $S$ & $\psi$ & $S'$ & Basis op.
\\ 
\hline
4-i 
&
$(d_{R} \overline{e_{L}})
(\overline{u_{L}}) (\overline{u_{L}})
(d_{R} \overline{e_{L}})$
& \#11
& $(\overline{\bf 3}, {\bf 2})_{-1/6}$
& $({\bf 1}, {\bf 3})_{0}$
& $({\bf 3}, {\bf 2})_{+1/6}$
& $ \frac{1}{32} (\mathcal{O}_{1})_{\{ RR \} R}  
   - \frac{1}{128} (\mathcal{O}_{2})_{\{ RR \} R}$
\\
&
& 
& $(\overline{\bf 3}, {\bf 2})_{-1/6}$
& $({\bf 8}, {\bf 3})_{0}$
& $({\bf 3}, {\bf 2})_{+1/6}$
& $ \frac{1}{24} (\mathcal{O}_{1})_{\{RR\}R}
    -
    \frac{1}{96} 
    (\mathcal{O}_{2})_{\{RR\}R}$
\\
\hline \hline
\end{tabular}
} 
\end{center}
\caption{\it Decompositions \#4-i accidentally not listed in 
table 9 of the paper.}
\label{Tab:Missing}
\end{table} 

All the corrections listed above are included in tables 
of the current version.


\begin{thebibliography}{100}

\bibitem{Avignone:2007fu}
I.~Avignone, Frank~T., S.~R. Elliott, and J.~Engel,
\newblock Rev. Mod. Phys. {\bf 80}, 481 (2008), arXiv:0708.1033.

\bibitem{GomezCadenas:2011it}
J.~Gomez-Cadenas, J.~Martin-Albo, M.~Mezzetto, F.~Monrabal, and M.~Sorel,
\newblock Riv. Nuovo Cim. {\bf 35}, 29 (2012), arXiv:1109.5515.

\bibitem{Rodejohann:2012xd}
W.~Rodejohann,
\newblock arXiv:1206.2560.

\bibitem{Barabash:1209.4241}
A.~Barabash,
\newblock  arXiv:1209.4241.

\bibitem{Rodejohann:2011mu}
W.~Rodejohann,
\newblock Int.J.Mod.Phys. {\bf E20}, 1833 (2011), arXiv:1106.1334.

\bibitem{Deppisch:2012nb}
F.~F. Deppisch, M.~Hirsch, and H.~Pas,
\newblock arXiv:1208.0727.

\bibitem{Schechter:1981bd}
J.~Schechter and J.~Valle,
\newblock Phys. Rev. {\bf D25}, 2951 (1982).

\bibitem{Nieves:1984sn}
J.~F. Nieves,
\newblock Phys. Lett. {\bf B147}, 375 (1984).

\bibitem{Takasugi:1984xr}
E.~Takasugi,
\newblock Phys. Lett. {\bf B149}, 372 (1984).

\bibitem{Hirsch:2006yk}
M.~Hirsch, S.~Kovalenko, and I.~Schmidt,
\newblock Phys. Lett. {\bf B642}, 106 (2006), arXiv:hep-ph/0608207.

\bibitem{Duerr:2011zd}
M.~Duerr, M.~Lindner, and A.~Merle,
\newblock JHEP {\bf 1106}, 091 (2011), arXiv:1105.0901.

\bibitem{KlapdorKleingrothaus:2000sn}
H.~Klapdor-Kleingrothaus {\em et~al.},
\newblock Eur. Phys. J. {\bf A12}, 147 (2001), arXiv:hep-ph/0103062.

\bibitem{KlapdorKleingrothaus:2006ff}
H.~Klapdor-Kleingrothaus and I.~Krivosheina,
\newblock Mod. Phys. Lett. {\bf A21}, 1547 (2006).

\bibitem{Auger:2012ar}
EXO Collaboration, M.~Auger {\em et~al.},
\newblock Phys. Rev. Lett. {\bf 109}, 032505 (2012), arXiv:1205.5608.

\bibitem{Collaboration:2012zm}
K.-Z. collaboration,
\newblock arXiv:1211.3863.

\bibitem{KamLANDZen:2012aa}
KamLAND-Zen Collaboration, A.~Gando {\em et~al.},
\newblock Phys. Rev. {\bf C85}, 045504 (2012), arXiv:1201.4664.

\bibitem{MacLallen:2012aa}
EXO-200 Collaboration, R.~MacLallen,
\newblock Recontres de Moriond, {\tt http://moriond.in2p3.fr/}  (2012).

\bibitem{Abt:2004yk}
GERDA Collaboration, I.~Abt {\em et~al.},
\newblock arXiv:hep-ex/0404039.

\bibitem{Guiseppe:2011me}
Majorana Collaboration, C.~Aalseth {\em et~al.},
\newblock Nucl. Phys. Proc. Suppl. {\bf 217}, 44 (2011), arXiv:1101.0119.

\bibitem{Barabash:2011fs}
A.~Barabash,
\newblock AIP Conf. Proc. {\bf 1417}, 5 (2011), arXiv:1109.6423.

\bibitem{Fukuda:1998mi}
Super-Kamiokande Collaboration, Y.~Fukuda {\em et~al.},
\newblock Phys. Rev. Lett. {\bf 81}, 1562 (1998), arXiv:hep-ex/9807003.

\bibitem{Ahmad:2002jz}
SNO Collaboration, Q.~Ahmad {\em et~al.},
\newblock Phys. Rev. Lett. {\bf 89}, 011301 (2002), arXiv:nucl-ex/0204008.

\bibitem{Eguchi:2002dm}
KamLAND Collaboration, K.~Eguchi {\em et~al.},
\newblock Phys. Rev. Lett. {\bf 90}, 021802 (2003), arXiv:hep-ex/0212021.

\bibitem{Abe:2011sj}
T2K Collaboration, K.~Abe {\em et~al.},
\newblock Phys. Rev. Lett. {\bf 107}, 041801 (2011), arXiv:1106.2822.

\bibitem{Adamson:2011qu}
MINOS Collaboration, P.~Adamson {\em et~al.},
\newblock Phys. Rev. Lett. {\bf 107}, 181802 (2011), arXiv:1108.0015.

\bibitem{Abe:2011fz}
DOUBLE-CHOOZ Collaboration, Y.~Abe {\em et~al.},
\newblock Phys. Rev. Lett. {\bf 108}, 131801 (2012), arXiv:1112.6353.

\bibitem{An:2012eh}
DAYA-BAY Collaboration, F.~An {\em et~al.},
\newblock Phys. Rev. Lett. {\bf 108}, 171803 (2012), arXiv:1203.1669.

\bibitem{Ahn:2012nd}
RENO collaboration, J.~Ahn {\em et~al.},
\newblock Phys. Rev. Lett. {\bf 108}, 191802 (2012), arXiv:1204.0626.

\bibitem{Faessler:2012ku}
A.~Faessler, V.~Rodin, and F.~Simkovic,
\newblock arXiv:1206.0464.

\bibitem{Menendez:2008jp}
J.~Menendez, A.~Poves, E.~Caurier, and F.~Nowacki,
\newblock Nucl. Phys. {\bf A818}, 139 (2009), arXiv:0801.3760.

\bibitem{Menendez:2011zza}
J.~Menendez, A.~Poves, E.~Caurier, and F.~Nowacki,
\newblock J. Phys. Conf. Ser. {\bf 312}, 072005 (2011).

\bibitem{Tortola:2012te}
D.~Forero, M.~Tortola, and J.~Valle,
\newblock arXiv:1205.4018.

\bibitem{Lesgourgues:2006nd}
J.~Lesgourgues and S.~Pastor,
\newblock Phys. Rept. {\bf 429}, 307 (2006), arXiv:astro-ph/0603494.

\bibitem{Hannestad:2010kz}
S.~Hannestad,
\newblock Prog. Part. Nucl. Phys. {\bf 65}, 185 (2010), arXiv:1007.0658.

\bibitem{Wong:2011ip}
Y.~Y. Wong,
\newblock Ann. Rev. Nucl. Part. Sci. {\bf 61}, 69 (2011), arXiv:1111.1436.

\bibitem{Bergstrom:2011dt}
J.~Bergstrom, A.~Merle, and T.~Ohlsson,
\newblock JHEP {\bf 1105}, 122 (2011), arXiv:1103.3015.

\bibitem{Pas:2000vn}
H.~Pas, M.~Hirsch, H.~Klapdor-Kleingrothaus, and S.~Kovalenko,
\newblock Phys. Lett. {\bf B498}, 35 (2001), arXiv:hep-ph/0008182.

\bibitem{Pas:1999fc}
H.~Pas, M.~Hirsch, H.~Klapdor-Kleingrothaus, and S.~Kovalenko,
\newblock Phys. Lett. {\bf B453}, 194 (1999).

\bibitem{Riazuddin:1981hz}
Riazuddin, R.~Marshak, and R.~N. Mohapatra,
\newblock Phys. Rev. {\bf D24}, 1310 (1981).

\bibitem{Rizzo:1982kn}
T.~G. Rizzo,
\newblock Phys. Lett. {\bf B116}, 23 (1982).

\bibitem{Keung:1983uu}
W.-Y. Keung and G.~Senjanovic,
\newblock Phys. Rev. Lett. {\bf 50}, 1427 (1983).

\bibitem{Hirsch:1996qw}
M.~Hirsch, H.~Klapdor-Kleingrothaus, and O.~Panella,
\newblock Phys. Lett. {\bf B374}, 7 (1996), arXiv:hep-ph/9602306.

\bibitem{ATLAS:2012ak}
ATLAS Collaboration, G.~Aad {\em et~al.},
\newblock Eur. Phys. J. {\bf C72}, 2056 (2012), arXiv:1203.5420.

\bibitem{CMS:PAS-EXO-12-017}
CMS Collaboration,
\newblock PAS EXO-12-017.

\bibitem{Babu:2001ex}
K.~Babu and C.~N. Leung,
\newblock Nucl. Phys. {\bf B619}, 667 (2001), arXiv:hep-ph/0106054.

\bibitem{deGouvea:2007xp}
A.~de~Gouvea and J.~Jenkins,
\newblock Phys. Rev. {\bf D77}, 013008 (2008), arXiv:0708.1344.

\bibitem{delAguila:2012nu}
F.~del Aguila, A.~Aparici, S.~Bhattacharya, A.~Santamaria, and J.~Wudka,
\newblock JHEP {\bf 1206}, 146 (2012), arXiv:1204.5986.

\bibitem{Weinberg:1979sa}
S.~Weinberg,
\newblock Phys. Rev. Lett. {\bf 43}, 1566 (1979).

\bibitem{Primakoff:1981sx}
H.~Primakoff and P.~S. Rosen,
\newblock Ann. Rev. Nucl. Part. Sci. {\bf 31}, 145 (1981).

\bibitem{Doi:1985dx}
M.~Doi, T.~Kotani, and E.~Takasugi,
\newblock Prog. Theor. Phys. Suppl. {\bf 83}, 1 (1985).

\bibitem{Missimer:1994xd}
J.~H. Missimer, R.~Mohapatra, and N.~C. Mukhopadhyay,
\newblock Phys. Rev. {\bf D50}, 2067 (1994).

\bibitem{Takasugi:2003ah}
E.~Takasugi,
\newblock Nucl. Instrum. Meth. {\bf A503}, 252 (2003).

\bibitem{Kanemura:2012br}
S.~Kanemura, Y.~Kuno, and T.~Ota,
\newblock arXiv:1205.5681.

\bibitem{Gavela:2008ra}
M.~Gavela, D.~Hernandez, T.~Ota, and W.~Winter,
\newblock Phys. Rev. {\bf D79}, 013007 (2009), arXiv:0809.3451.

\bibitem{Bonnet:2009ej}
F.~Bonnet, D.~Hernandez, T.~Ota, and W.~Winter,
\newblock JHEP {\bf 0910}, 076 (2009), arXiv:0907.3143.

\bibitem{Bonnet:2011yx}
F.~Bonnet, M.~Gavela, T.~Ota, and W.~Winter,
\newblock Phys. Rev. {\bf D85}, 035016 (2012), arXiv:1105.5140.

\bibitem{Bonnet:2012kz}
F.~Bonnet, M.~Hirsch, T.~Ota, and W.~Winter,
\newblock JHEP {\bf 1207}, 153 (2012), arXiv:1204.5862.

\bibitem{Mohapatra:1986su}
R.~Mohapatra,
\newblock Phys. Rev. {\bf D34}, 3457 (1986).

\bibitem{Hirsch:1995zi}
M.~Hirsch, H.~Klapdor-Kleingrothaus, and S.~Kovalenko,
\newblock Phys. Rev. Lett. {\bf 75}, 17 (1995).

\bibitem{Hirsch:1995ek}
M.~Hirsch, H.~Klapdor-Kleingrothaus, and S.~Kovalenko,
\newblock Phys. Rev. {\bf D53}, 1329 (1996), arXiv:hep-ph/9502385.

\bibitem{Goswami:2005ng}
S.~Goswami and W.~Rodejohann,
\newblock Phys. Rev. {\bf D73}, 113003 (2006), arXiv:hep-ph/0512234.

\bibitem{Blennow:2010th}
M.~Blennow, E.~Fernandez-Martinez, J.~Lopez-Pavon, and J.~Menendez,
\newblock JHEP {\bf 1007}, 096 (2010), arXiv:1005.3240.

\bibitem{Ibarra:2010xw}
A.~Ibarra, E.~Molinaro, and S.~Petcov,
\newblock JHEP {\bf 1009}, 108 (2010), arXiv:1007.2378.

\bibitem{Choubey:2012ux}
S.~Choubey, M.~Duerr, M.~Mitra, and W.~Rodejohann,
\newblock JHEP {\bf 1205}, 017 (2012), arXiv:1201.3031.

\bibitem{Hirsch:1996qy}
M.~Hirsch, H.~Klapdor-Kleingrothaus, and S.~Kovalenko,
\newblock Phys. Lett. {\bf B378}, 17 (1996), arXiv:hep-ph/9602305.

\bibitem{Hirsch:1996ye}
M.~Hirsch, H.~Klapdor-Kleingrothaus, and S.~Kovalenko,
\newblock Phys.Rev. {\bf D54}, 4207 (1996), arXiv:hep-ph/9603213.

\bibitem{Cuypers:1996ia}
F.~Cuypers and S.~Davidson,
\newblock Eur. Phys. J. {\bf C2}, 503 (1998), arXiv:hep-ph/9609487.

\bibitem{Buchmuller:1986zs}
W.~Buchmuller, R.~Ruckl, and D.~Wyler,
\newblock Phys. Lett. {\bf B191}, 442 (1987).

\bibitem{Mohapatra:1981pm}
R.~N. Mohapatra and J.~Vergados,
\newblock Phys. Rev. Lett. {\bf 47}, 1713 (1981).

\bibitem{Gu:2011ak}
P.-H. Gu,
\newblock Phys. Rev. {\bf D85}, 093016 (2012), arXiv:1101.5106.

\bibitem{Kohda:2012sr}
M.~Kohda, H.~Sugiyama, and K.~Tsumura,
\newblock arXiv:1210.5622.

\bibitem{Wolfenstein:1982bf}
L.~Wolfenstein,
\newblock Phys. Rev. {\bf D26}, 2507 (1982).

\bibitem{Haxton:1982ff}
W.~Haxton, S.~P. Rosen, and G.~Stephenson,
\newblock Phys. Rev. {\bf D26}, 1805 (1982).

\bibitem{Allanach:2009iv}
B.~Allanach, C.~Kom, and H.~Pas,
\newblock Phys. Rev. Lett. {\bf 103}, 091801 (2009), arXiv:0902.4697.

\bibitem{Han:2010rf}
T.~Han, I.~Lewis, and Z.~Liu,
\newblock JHEP {\bf 1012}, 085 (2010), arXiv:1010.4309.

\bibitem{Chatrchyan:2011ns}
CMS Collaboration, S.~Chatrchyan {\em et~al.},
\newblock Phys. Lett. {\bf B704}, 123 (2011), arXiv:1107.4771.

\bibitem{Aad:2012xz}
ATLAS Collaboration, G.~Aad {\em et~al.},
\newblock arXiv:1210.1718.

\bibitem{Aad:2012yz}
ATLAS Collaboration, G.~Aad {\em et~al.},
\newblock ATLAS CONF-2012-088.

\bibitem{Dorsner:2009mq}
I.~Dorsner, S.~Fajfer, J.~F. Kamenik, and N.~Kosnik,
\newblock Phys. Rev. {\bf D81}, 055009 (2010), arXiv:0912.0972.

\bibitem{Dorsner:2011ai}
I.~Dorsner, J.~Drobnak, S.~Fajfer, J.~F. Kamenik, and N.~Kosnik,
\newblock JHEP {\bf 1111}, 002 (2011), arXiv:1107.5393.

\bibitem{Kosnik:2011jr}
N.~Kosnik, I.~Dorsner, J.~Drobnak, S.~Fajfer, and J.~F. Kamenik,
\newblock arXiv:1111.0477.

\bibitem{Aad:2012em}
ATLAS Collaboration, G.~Aad {\em et~al.},
\newblock arXiv:1209.6593.

\bibitem{Cacciapaglia:2010vn}
G.~Cacciapaglia, A.~Deandrea, D.~Harada, and Y.~Okada,
\newblock JHEP {\bf 1011}, 159 (2010), arXiv:1007.2933.

\bibitem{Cacciapaglia:2011fx}
G.~Cacciapaglia {\em et~al.},
\newblock JHEP {\bf 1203}, 070 (2012), arXiv:1108.6329.

\bibitem{Okada:2012gy}
Y.~Okada and L.~Panizzi,
\newblock arXiv:1207.5607.

\bibitem{Cacciapaglia:2012dd}
G.~Cacciapaglia, A.~Deandrea, S.~Perries, V.~Sordini, and L.~Panizzi,
\newblock arXiv:1211.4034.

\bibitem{Azatov:2012rj}
A.~Azatov {\em et~al.},
\newblock arXiv:1204.0455.

\bibitem{Bonne:2012im}
N.~Bonne and G.~Moreau,
\newblock arXiv:1206.3360.

\bibitem{Batell:2012ca}
B.~Batell, S.~Gori, and L.-T. Wang,
\newblock arXiv:1209.6382.

\bibitem{Bertuzzo:2012bt}
E.~Bertuzzo, P.~A. Machado, and R.~Z. Funchal,
\newblock arXiv:1209.6359.

\bibitem{McKeen:2012av}
D.~McKeen, M.~Pospelov, and A.~Ritz,
\newblock arXiv:1208.4597.

\bibitem{Bae:2012ir}
K.~J. Bae, T.~H. Jung, and H.~D. Kim,
\newblock arXiv:1208.3748.

\bibitem{Davoudiasl:2012ig}
H.~Davoudiasl, H.-S. Lee, and W.~J. Marciano,
\newblock arXiv:1208.2973.

\bibitem{Batell:2012mj}
B.~Batell, D.~McKeen, and M.~Pospelov,
\newblock arXiv:1207.6252.

\bibitem{An:2012vp}
H.~An, T.~Liu, and L.-T. Wang,
\newblock arXiv:1207.2473.

\bibitem{Martin:2012dg}
S.~P. Martin and J.~D. Wells,
\newblock Phys. Rev. {\bf D86}, 035017 (2012), arXiv:1206.2956.

\bibitem{Wang:2012gm}
L.~Wang and X.-F. Han,
\newblock arXiv:1206.1673.

\bibitem{Iwamoto:2012hh}
S.~Iwamoto,
\newblock AIP Conf. Proc. {\bf 1467}, 57 (2012), arXiv:1206.0161.

\bibitem{Endo:2011xq}
M.~Endo, K.~Hamaguchi, S.~Iwamoto, and N.~Yokozaki,
\newblock Phys. Rev. {\bf D85}, 095012 (2012), arXiv:1112.5653.

\bibitem{Aad:2012gk}
ATLAS Collaboration, G.~Aad {\em et~al.},
\newblock Phys. Lett. {\bf B716}, 1 (2012), arXiv:1207.7214.

\bibitem{CMS:2012gu}
CMS Collaboration, S.~Chatrchyan {\em et~al.},
\newblock Phys. Lett. {\bf B716}, 30 (2012), arXiv:1207.7235.

\bibitem{Aad:2011ch}
ATLAS Collaboration, G.~Aad {\em et~al.},
\newblock Phys. Lett. {\bf B709}, 158 (2012), arXiv:1112.4828.

\bibitem{Abramowicz:2012tg}
ZEUS Collaboration, H.~Abramowicz {\em et~al.},
\newblock arXiv:1205.5179.

\bibitem{Osipowicz:2001sq}
KATRIN Collaboration, {\tt http://www.katrin.kit.edu}, A.~Osipowicz {\em et~al.},
\newblock (2001), arXiv:hep-ex/0109033.

\bibitem{Dreiner:2008tw}
H.~K. Dreiner, H.~E. Haber, and S.~P. Martin,
\newblock Phys. Rept. {\bf 494}, 1 (2010), arXiv:0812.1594.

\end{thebibliography}

\end{document}